\newcommand\papertitle{\textit{Planck} 2015 results. XIV. Dark energy and
 modified gravity}

\pdfoutput=1

\documentclass[twocolumn,traditabstract,longauth]{aa}
\usepackage{graphicx}

\usepackage{url}
\usepackage[nonamebreak]{natbib}
\usepackage{float}

\usepackage{amsmath}

\usepackage{amssymb}% http://ctan.org/pkg/amssymb

\usepackage{pifont}% http://ctan.org/pkg/pifont
\usepackage{booktabs}
\usepackage{color}
\usepackage{txfonts}
\usepackage{natbib}
\usepackage[labelfont=bf]{caption}
\usepackage[table,usenames,dvipsnames]{xcolor}

\usepackage{hyperref}
\usepackage{ifthen}
\hypersetup{breaklinks,colorlinks,citecolor=blue}

\bibpunct{(}{)}{;}{a}{}{,}  % to follow the A&A style
\usepackage{etex}

\usepackage{placeins}
\usepackage{graphicx}
\usepackage{epstopdf}
\usepackage{ifthen}

\providecommand{\sorthelp}[1]{}

\pdfmapfile{+txfonts.map}

\newcommand{\mksym}[1]{\ifmmode {\rm #1}\else #1\fi}

\newcommand{\lensing}{\mksym{lensing}}

\newcommand{\planckonly}{\planck}

\newcommand{\TT}{\mksym{TT}}
\newcommand{\TTTEEE}{\mksym{TT,TE,EE}}
\newcommand{\planckTTonly}{\planck\ \TT}
\newcommand{\lowTEB}{\mksym{lowP}}

\newcommand{\planckTT}{\planck\ \TT+\lowTEB}
\newcommand{\planckall}{\planck\ \TTTEEE+\lowTEB}
\newcommand{\shortTT}{\TT+\lowTEB}

\newcommand{\lcdm}{$\Lambda$CDM}

\providecommand{\Planck}{\textit{Planck}}
\providecommand{\planck}{\Planck}

\providecommand{\text}[1]{\rm{#1}}

\newcommand{\Mpc}{\text{Mpc}}

\newcommand{\Hunit}{\,\text{km}\,\text{s}^{-1}\,\Mpc^{-1}}

\providecommand{\muK}{\mu\rm{K}}

\newcommand{\redmpl}{M_{\text{P}}}

\providecommand{\Omm}{\Omega_{\mathrm{m}}}

\providecommand{\Omde}{\Omega_{\mathrm{de}}}
\providecommand{\epss}{\epsilon_{\rm s}}
\providecommand{\epsv}{\epsilon_{\rm V}}
\providecommand{\zetas}{\zeta_{\rm s}}

\providecommand{\EFTCAMB}{{\tt EFTcamb}}
\providecommand{\MGCAMB}{{\tt MGcamb}}
\providecommand{\COSMOMC}{{\tt CosmoMC}}

\providecommand{\LCDM}{{$\rm{\Lambda CDM}$}}

\providecommand{\alphaM}{\alpha_{\mathrm{M}}}
\providecommand{\alphaMtoday}{\alpha_{\mathrm{M0}}}
\providecommand{\alphaK}{\alpha_{\mathrm{K}}}
\providecommand{\alphaH}{\alpha_{\mathrm{H}}}
\providecommand{\alphaB}{\alpha_{\mathrm{B}}}
\providecommand{\alphaT}{\alpha_{\mathrm{T}}}

\newcommand{\begm}{\begin{pmatrix}}
\newcommand{\enm}{\end{pmatrix}}

\newcommand\ba{\begin{eqnarray}}
\newcommand\ea{\end{eqnarray}}
\newcommand\bea{\begin{eqnarray}}
\newcommand\eea{\end{eqnarray}}

\newcommand\be{\begin{equation}}
\newcommand\ee{\end{equation}}

\def\pmb#1{\setbox0=\hbox{#1}%
    \kern-.025em\copy0\kern-\wd0
    \kern.05em\copy0\kern-\wd0
    \kern-.025em\raise.0433em\box0}
\def\ltsima{$\; \buildrel < \over \sim \;$}
\def\gtsima{$\; \buildrel > \over \sim \;$}
\def\simlt{\lower.5ex\hbox{\ltsima}}
\def\simgt{\lower.5ex\hbox{\gtsima}}

\def\p2Y{\;_2Y}
\def\m2Y{\;_{-2}Y}
\def\beglet{
  \addtocounter{equation}{1}%
  \setcounter{parentequation}{\value{equation}}%
  \setcounter{equation}{0}%
  \def\theequation{\arabic{parentequation}\alph{equation}}%
  \ignorespaces
}
\def\endlet{
  \setcounter{equation}{\value{parentequation}}%
  \def\theequation{\arabic{equation}}%
}
\providecommand{\beglet}{\begin{subequations}}
\providecommand{\endlet}{\end{subequations}}

\newcommand\beq{\begin{equation}}
\newcommand\eeq{\end{equation}}

\newcommand{\eqn}[1]{(#1)}

\newcommand{\tbl}[1]{Table~#1}
\newcommand{\Tbl}[1]{Table~#1}
\newcommand{\fig}[1]{Fig.~#1}
\newcommand{\Fig}[1]{Fig.~#1}
\newcommand{\sect}[1]{Sect.\ #1}
\newcommand{\eq}[1]{Eq.\ (#1)}
\newcommand{\eqs}[1]{Eqs.\ (#1)}

\newcommand{\hub}{\ensuremath{H}}

\newcommand{\de}{{\rm de}}
\newbox\tablebox    \newdimen\tablewidth
\def\leaderfil{\leaders\hbox to 5pt{\hss.\hss}\hfil}
\def\endtable{\tablewidth=\columnwidth
    $$\hss\copy\tablebox\hss$$
    \vskip-\lastskip\vskip -2pt}

\def\tablenote#1 #2\par{\begingroup \parindent=0.8em
    \abovedisplayshortskip=0pt\belowdisplayshortskip=0pt
    \noindent
    $$\hss\vbox{\hsize\tablewidth \hangindent=\parindent \hangafter=1 \noindent
    \hbox to \parindent{\sup{\rm #1}\hss}\strut#2\strut\par}\hss$$
    \endgroup}
\def\doubleline{\vskip 3pt\hrule \vskip 1.5pt \hrule \vskip 5pt}

\def\setsymbol#1#2{\expandafter\def\csname #1\endcsname{#2}}
\def\getsymbol#1{\csname #1\endcsname}

\def\Planck{\textit{Planck}}

\def\alltwentythirteenresultspapers{\nocite{planck2013-p01, planck2013-p02, planck2013-p02a, planck2013-p02d, planck2013-p02b, planck2013-p03, planck2013-p03c, planck2013-p03f, planck2013-p03d, planck2013-p03e, planck2013-p01a, planck2013-p06, planck2013-p03a, planck2013-pip88, planck2013-p08, planck2013-p11, planck2013-p12, planck2013-p13, planck2013-p14, planck2013-p15, planck2013-p05b, planck2013-p17, planck2013-p09, planck2013-p09a, planck2013-p20, planck2013-p19, planck2013-pipaberration, planck2013-p05, planck2013-p05a, planck2013-pip56, planck2013-p06b, planck2013-p01a}}

\def\alltwentyfifteenresultspapers{\nocite{planck2014-a01, planck2014-a03, planck2014-a04, planck2014-a05, planck2014-a06, planck2014-a07, planck2014-a08, planck2014-a09, planck2014-a11, planck2014-a12, planck2014-a13, planck2014-a14, planck2014-a15, planck2014-a16, planck2014-a17, planck2014-a18, planck2014-a19, planck2014-a20, planck2014-a22, planck2014-a24, planck2014-a26, planck2014-a28, planck2014-a29, planck2014-a30, planck2014-a31, planck2014-a35, planck2014-a36, planck2014-a37, planck2014-ES}}

\newbox\tablebox    \newdimen\tablewidth
\def\leaderfil{\leaders\hbox to 5pt{\hss.\hss}\hfil}
\def\endPlancktable{\tablewidth=\columnwidth 
    $$\hss\copy\tablebox\hss$$
    \vskip-\lastskip\vskip -2pt}

\def\tablenote#1 #2\par{\begingroup \parindent=0.8em
    \abovedisplayshortskip=0pt\belowdisplayshortskip=0pt
    \noindent
    $$\hss\vbox{\hsize\tablewidth \hangindent=\parindent \hangafter=1 \noindent
    \hbox to \parindent{$^#1$\hss}\strut#2\strut\par}\hss$$
    \endgroup}
\def\doubleline{\vskip 3pt\hrule \vskip 1.5pt \hrule \vskip 5pt}

\def\L2{\ifmmode L_2\else $L_2$\fi}

\def\DeltaT{\ifmmode \Delta T\else $\Delta T$\fi}
\def\deltat{\ifmmode \Delta t\else $\Delta t$\fi}
\def\fknee{\ifmmode f_{\rm knee}\else $f_{\rm knee}$\fi}
\def\Fmax{\ifmmode F_{\rm max}\else $F_{\rm max}$\fi}
\def\solar{\ifmmode{\rm M}_{\mathord\odot}\else${\rm M}_{\mathord\odot}$\fi}
\def\Msolar{\ifmmode{\rm M}_{\mathord\odot}\else${\rm M}_{\mathord\odot}$\fi}
\def\Lsolar{\ifmmode{\rm L}_{\mathord\odot}\else${\rm L}_{\mathord\odot}$\fi}
\def\inv{\ifmmode^{-1}\else$^{-1}$\fi}
\def\mo{\ifmmode^{-1}\else$^{-1}$\fi}
\def\sup#1{\ifmmode ^{\rm #1}\else $^{\rm #1}$\fi}
\def\expo#1{\ifmmode \times 10^{#1}\else $\times 10^{#1}$\fi}
\def\,{\thinspace}
\def\lsim{\mathrel{\raise .4ex\hbox{\rlap{$<$}\lower 1.2ex\hbox{$\sim$}}}}
\def\gsim{\mathrel{\raise .4ex\hbox{\rlap{$>$}\lower 1.2ex\hbox{$\sim$}}}}

\def\simprop{\mathrel{\raise .4ex\hbox{\rlap{$\propto$}\lower 1.2ex\hbox{$\sim$}}}}
\def\deg{\ifmmode^\circ\else$^\circ$\fi}
\def\pdeg{\ifmmode $\setbox0=\hbox{$^{\circ}$}\rlap{\hskip.11\wd0 .}$^{\circ}
          \else \setbox0=\hbox{$^{\circ}$}\rlap{\hskip.11\wd0 .}$^{\circ}$\fi}
\def\arcs{\ifmmode {^{\scriptstyle\prime\prime}}
          \else $^{\scriptstyle\prime\prime}$\fi}
\def\arcm{\ifmmode {^{\scriptstyle\prime}}
          \else $^{\scriptstyle\prime}$\fi}
\newdimen\sa  \newdimen\sb
\def\parcs{\sa=.07em \sb=.03em
     \ifmmode \hbox{\rlap{.}}^{\scriptstyle\prime\kern -\sb\prime}\hbox{\kern -\sa}
     \else \rlap{.}$^{\scriptstyle\prime\kern -\sb\prime}$\kern -\sa\fi}
\def\parcm{\sa=.08em \sb=.03em
     \ifmmode \hbox{\rlap{.}\kern\sa}^{\scriptstyle\prime}\hbox{\kern-\sb}
     \else \rlap{.}\kern\sa$^{\scriptstyle\prime}$\kern-\sb\fi}
\def\ra[#1 #2 #3.#4]{#1\sup{h}#2\sup{m}#3\sup{s}\llap.#4}
\def\dec[#1 #2 #3.#4]{#1\deg#2\arcm#3\arcs\llap.#4}
\def\deco[#1 #2 #3]{#1\deg#2\arcm#3\arcs}
\def\rra[#1 #2]{#1\sup{h}#2\sup{m}}

\def\dots{\relax\ifmmode \ldots\else $\ldots$\fi}
\def\WHzsr{\ifmmode $W\,Hz\mo\,sr\mo$\else W\,Hz\mo\,sr\mo\fi}
\def\mHz{\ifmmode $\,mHz$\else \,mHz\fi}
\def\GHz{\ifmmode $\,GHz$\else \,GHz\fi}
\def\mKs{\ifmmode $\,mK\,s$^{1/2}\else \,mK\,s$^{1/2}$\fi}
\def\muKs{\ifmmode \,\mu$K\,s$^{1/2}\else \,$\mu$K\,s$^{1/2}$\fi}
\def\muKRJs{\ifmmode \,\mu$K$_{\rm RJ}$\,s$^{1/2}\else \,$\mu$K$_{\rm RJ}$\,s$^{1/2}$\fi}
\def\muKHz{\ifmmode \,\mu$K\,Hz$^{-1/2}\else \,$\mu$K\,Hz$^{-1/2}$\fi}
\def\MJysr{\ifmmode \,$MJy\,sr\mo$\else \,MJy\,sr\mo\fi}
\def\MJysrmK{\ifmmode \,$MJy\,sr\mo$\,mK$_{\rm CMB}\mo\else \,MJy\,sr\mo\,mK$_{\rm CMB}\mo$\fi}
\def\microns{\ifmmode \,\mu$m$\else \,$\mu$m\fi}

\def\muK{\ifmmode \,\mu$K$\else \,$\mu$\hbox{K}\fi}
\def\microK{\ifmmode \,\mu$K$\else \,$\mu$\hbox{K}\fi}
\def\muW{\ifmmode \,\mu$W$\else \,$\mu$\hbox{W}\fi}
\def\kms{\ifmmode $\,km\,s$^{-1}\else \,km\,s$^{-1}$\fi}
\def\kmsMpc{\ifmmode $\,\kms\,Mpc\mo$\else \,\kms\,Mpc\mo\fi}
\providecommand{\sorthelp}[1]{}

\begin{document}

\title{\papertitle}

%This author list corresponds to \title{Author list for A16\_Dark\_energy\_and\_modified\_gravity}
%Prepared by M. Lopez-Caniego (Marcos.Lopez.Caniego@sciops.esa.int), ESAC/ESA
%This version is from Tue Feb  9 10:24:47 2016 CET
%\subtitle{There are 236 co-authors in this list}
\author{\small
Planck Collaboration: P.~A.~R.~Ade\inst{95}
\and
N.~Aghanim\inst{63}
\and
M.~Arnaud\inst{79}
\and
M.~Ashdown\inst{75, 7}
\and
J.~Aumont\inst{63}
\and
C.~Baccigalupi\inst{93}
\and
A.~J.~Banday\inst{107, 11}
\and
R.~B.~Barreiro\inst{70}
\and
N.~Bartolo\inst{35, 71}
\and
E.~Battaner\inst{109, 110}
% ...
}
\institute{\small
APC, AstroParticule et Cosmologie, Universit\'{e} Paris Diderot, CNRS/IN2P3, CEA/lrfu, Observatoire de Paris, Sorbonne Paris Cit\'{e}, 10, rue Alice Domon et L\'{e}onie Duquet, 75205 Paris Cedex 13, France\goodbreak
\and
Aalto University Mets\"{a}hovi Radio Observatory and Dept of Radio Science and Engineering, P.O. Box 13000, FI-00076 AALTO, Finland\goodbreak
\and
African Institute for Mathematical Sciences, 6-8 Melrose Road, Muizenberg, Cape Town, South Africa\goodbreak
\and
Agenzia Spaziale Italiana Science Data Center, Via del Politecnico snc, 00133, Roma, Italy\goodbreak
\and
Aix Marseille Universit\'{e}, CNRS, LAM (Laboratoire d'Astrophysique de Marseille) UMR 7326, 13388, Marseille, France\goodbreak
\and
Aix Marseille Universit\'{e}, Centre de Physique Th\'{e}orique, 163 Avenue de Luminy, 13288, Marseille, France\goodbreak
\and
Astrophysics Group, Cavendish Laboratory, University of Cambridge, J J Thomson Avenue, Cambridge CB3 0HE, U.K.\goodbreak
\and
Astrophysics \& Cosmology Research Unit, School of Mathematics, Statistics \& Computer Science, University of KwaZulu-Natal, Westville Campus, Private Bag X54001, Durban 4000, South Africa\goodbreak
\and
CGEE, SCS Qd 9, Lote C, Torre C, 4$^{\circ}$ andar, Ed. Parque Cidade Corporate, CEP 70308-200, Bras\'{i}lia, DF, Brazil\goodbreak
\and
CITA, University of Toronto, 60 St. George St., Toronto, ON M5S 3H8, Canada\goodbreak
\and
CNRS, IRAP, 9 Av. colonel Roche, BP 44346, F-31028 Toulouse cedex 4, France\goodbreak
\and
CRANN, Trinity College, Dublin, Ireland\goodbreak
\and
California Institute of Technology, Pasadena, California, U.S.A.\goodbreak
\and
Centre for Theoretical Cosmology, DAMTP, University of Cambridge, Wilberforce Road, Cambridge CB3 0WA, U.K.\goodbreak
\and
Centro de Estudios de F\'{i}sica del Cosmos de Arag\'{o}n (CEFCA), Plaza San Juan, 1, planta 2, E-44001, Teruel, Spain\goodbreak
\and
Computational Cosmology Center, Lawrence Berkeley National Laboratory, Berkeley, California, U.S.A.\goodbreak
\and
Consejo Superior de Investigaciones Cient\'{\i}ficas (CSIC), Madrid, Spain\goodbreak
\and
DSM/Irfu/SPP, CEA-Saclay, F-91191 Gif-sur-Yvette Cedex, France\goodbreak
\and
DTU Space, National Space Institute, Technical University of Denmark, Elektrovej 327, DK-2800 Kgs. Lyngby, Denmark\goodbreak
\and
D\'{e}partement de Physique Th\'{e}orique, Universit\'{e} de Gen\`{e}ve, 24, Quai E. Ansermet,1211 Gen\`{e}ve 4, Switzerland\goodbreak
\and
Departamento de Astrof\'{i}sica, Universidad de La Laguna (ULL), E-38206 La Laguna, Tenerife, Spain\goodbreak
\and
Departamento de F\'{\i}sica, Universidad de Oviedo, Avda. Calvo Sotelo s/n, Oviedo, Spain\goodbreak
\and
Department of Astronomy and Astrophysics, University of Toronto, 50 Saint George Street, Toronto, Ontario, Canada\goodbreak
\and
Department of Astrophysics/IMAPP, Radboud University Nijmegen, P.O. Box 9010, 6500 GL Nijmegen, The Netherlands\goodbreak
\and
Department of Physics \& Astronomy, University of British Columbia, 6224 Agricultural Road, Vancouver, British Columbia, Canada\goodbreak
\and
Department of Physics and Astronomy, Dana and David Dornsife College of Letter, Arts and Sciences, University of Southern California, Los Angeles, CA 90089, U.S.A.\goodbreak
\and
Department of Physics and Astronomy, University College London, London WC1E 6BT, U.K.\goodbreak
\and
Department of Physics and Astronomy, University of Sussex, Brighton BN1 9QH, U.K.\goodbreak
\and
Department of Physics, Florida State University, Keen Physics Building, 77 Chieftan Way, Tallahassee, Florida, U.S.A.\goodbreak
\and
Department of Physics, Gustaf H\"{a}llstr\"{o}min katu 2a, University of Helsinki, Helsinki, Finland\goodbreak
\and
Department of Physics, Princeton University, Princeton, New Jersey, U.S.A.\goodbreak
\and
Department of Physics, University of California, Berkeley, California, U.S.A.\goodbreak
\and
Department of Physics, University of California, Santa Barbara, California, U.S.A.\goodbreak
\and
Department of Physics, University of Illinois at Urbana-Champaign, 1110 West Green Street, Urbana, Illinois, U.S.A.\goodbreak
\and
Dipartimento di Fisica e Astronomia G. Galilei, Universit\`{a} degli Studi di Padova, via Marzolo 8, 35131 Padova, Italy\goodbreak
\and
Dipartimento di Fisica e Scienze della Terra, Universit\`{a} di Ferrara, Via Saragat 1, 44122 Ferrara, Italy\goodbreak
\and
Dipartimento di Fisica, Universit\`{a} La Sapienza, P. le A. Moro 2, Roma, Italy\goodbreak
\and
Dipartimento di Fisica, Universit\`{a} degli Studi di Milano, Via Celoria, 16, Milano, Italy\goodbreak
\and
Dipartimento di Fisica, Universit\`{a} degli Studi di Trieste, via A. Valerio 2, Trieste, Italy\goodbreak
\and
Dipartimento di Matematica, Universit\`{a} di Roma Tor Vergata, Via della Ricerca Scientifica, 1, Roma, Italy\goodbreak
\and
Discovery Center, Niels Bohr Institute, Blegdamsvej 17, Copenhagen, Denmark\goodbreak
\and
Discovery Center, Niels Bohr Institute, Copenhagen University, Blegdamsvej 17, Copenhagen, Denmark\goodbreak
\and
European Space Agency, ESAC, Planck Science Office, Camino bajo del Castillo, s/n, Urbanizaci\'{o}n Villafranca del Castillo, Villanueva de la Ca\~{n}ada, Madrid, Spain\goodbreak
\and
European Space Agency, ESTEC, Keplerlaan 1, 2201 AZ Noordwijk, The Netherlands\goodbreak
\and
Gran Sasso Science Institute, INFN, viale F. Crispi 7, 67100 L'Aquila, Italy\goodbreak
\and
HGSFP and University of Heidelberg, Theoretical Physics Department, Philosophenweg 16, 69120, Heidelberg, Germany\goodbreak
\and
Helsinki Institute of Physics, Gustaf H\"{a}llstr\"{o}min katu 2, University of Helsinki, Helsinki, Finland\goodbreak
\and
INAF - Osservatorio Astronomico di Padova, Vicolo dell'Osservatorio 5, Padova, Italy\goodbreak
\and
INAF - Osservatorio Astronomico di Roma, via di Frascati 33, Monte Porzio Catone, Italy\goodbreak
\and
INAF - Osservatorio Astronomico di Trieste, Via G.B. Tiepolo 11, Trieste, Italy\goodbreak
\and
INAF/IASF Bologna, Via Gobetti 101, Bologna, Italy\goodbreak
\and
INAF/IASF Milano, Via E. Bassini 15, Milano, Italy\goodbreak
\and
INFN, Sezione di Bologna, Via Irnerio 46, I-40126, Bologna, Italy\goodbreak
\and
INFN, Sezione di Roma 1, Universit\`{a} di Roma Sapienza, Piazzale Aldo Moro 2, 00185, Roma, Italy\goodbreak
\and
INFN, Sezione di Roma 2, Universit\`{a} di Roma Tor Vergata, Via della Ricerca Scientifica, 1, Roma, Italy\goodbreak
\and
INFN/National Institute for Nuclear Physics, Via Valerio 2, I-34127 Trieste, Italy\goodbreak
\and
IPAG: Institut de Plan\'{e}tologie et d'Astrophysique de Grenoble, Universit\'{e} Grenoble Alpes, IPAG, F-38000 Grenoble, France, CNRS, IPAG, F-38000 Grenoble, France\goodbreak
\and
IUCAA, Post Bag 4, Ganeshkhind, Pune University Campus, Pune 411 007, India\goodbreak
\and
Imperial College London, Astrophysics group, Blackett Laboratory, Prince Consort Road, London, SW7 2AZ, U.K.\goodbreak
\and
Infrared Processing and Analysis Center, California Institute of Technology, Pasadena, CA 91125, U.S.A.\goodbreak
\and
Institut N\'{e}el, CNRS, Universit\'{e} Joseph Fourier Grenoble I, 25 rue des Martyrs, Grenoble, France\goodbreak
\and
Institut Universitaire de France, 103, bd Saint-Michel, 75005, Paris, France\goodbreak
\and
Institut d'Astrophysique Spatiale, CNRS, Univ. Paris-Sud, Universit\'{e} Paris-Saclay, B\^{a}t. 121, 91405 Orsay cedex, France\goodbreak
\and
Institut d'Astrophysique de Paris, CNRS (UMR7095), 98 bis Boulevard Arago, F-75014, Paris, France\goodbreak
\and
Institut f\"ur Theoretische Teilchenphysik und Kosmologie, RWTH Aachen University, D-52056 Aachen, Germany\goodbreak
\and
Institute for Space Sciences, Bucharest-Magurale, Romania\goodbreak
\and
Institute of Astronomy, University of Cambridge, Madingley Road, Cambridge CB3 0HA, U.K.\goodbreak
\and
Institute of Theoretical Astrophysics, University of Oslo, Blindern, Oslo, Norway\goodbreak
\and
Instituto de Astrof\'{\i}sica de Canarias, C/V\'{\i}a L\'{a}ctea s/n, La Laguna, Tenerife, Spain\goodbreak
\and
Instituto de F\'{\i}sica de Cantabria (CSIC-Universidad de Cantabria), Avda. de los Castros s/n, Santander, Spain\goodbreak
\and
Istituto Nazionale di Fisica Nucleare, Sezione di Padova, via Marzolo 8, I-35131 Padova, Italy\goodbreak
\and
Jet Propulsion Laboratory, California Institute of Technology, 4800 Oak Grove Drive, Pasadena, California, U.S.A.\goodbreak
\and
Jodrell Bank Centre for Astrophysics, Alan Turing Building, School of Physics and Astronomy, The University of Manchester, Oxford Road, Manchester, M13 9PL, U.K.\goodbreak
\and
Kavli Institute for Cosmological Physics, University of Chicago, Chicago, IL 60637, USA\goodbreak
\and
Kavli Institute for Cosmology Cambridge, Madingley Road, Cambridge, CB3 0HA, U.K.\goodbreak
\and
Kazan Federal University, 18 Kremlyovskaya St., Kazan, 420008, Russia\goodbreak
\and
LAL, Universit\'{e} Paris-Sud, CNRS/IN2P3, Orsay, France\goodbreak
\and
LERMA, CNRS, Observatoire de Paris, 61 Avenue de l'Observatoire, Paris, France\goodbreak
\and
Laboratoire AIM, IRFU/Service d'Astrophysique - CEA/DSM - CNRS - Universit\'{e} Paris Diderot, B\^{a}t. 709, CEA-Saclay, F-91191 Gif-sur-Yvette Cedex, France\goodbreak
\and
Laboratoire Traitement et Communication de l'Information, CNRS (UMR 5141) and T\'{e}l\'{e}com ParisTech, 46 rue Barrault F-75634 Paris Cedex 13, France\goodbreak
\and
Laboratoire de Physique Subatomique et Cosmologie, Universit\'{e} Grenoble-Alpes, CNRS/IN2P3, 53, rue des Martyrs, 38026 Grenoble Cedex, France\goodbreak
\and
Laboratoire de Physique Th\'{e}orique, Universit\'{e} Paris-Sud 11 \& CNRS, B\^{a}timent 210, 91405 Orsay, France\goodbreak
\and
Lawrence Berkeley National Laboratory, Berkeley, California, U.S.A.\goodbreak
\and
Lebedev Physical Institute of the Russian Academy of Sciences, Astro Space Centre, 84/32 Profsoyuznaya st., Moscow, GSP-7, 117997, Russia\goodbreak
\and
Max-Planck-Institut f\"{u}r Astrophysik, Karl-Schwarzschild-Str. 1, 85741 Garching, Germany\goodbreak
\and
McGill Physics, Ernest Rutherford Physics Building, McGill University, 3600 rue University, Montr\'{e}al, QC, H3A 2T8, Canada\goodbreak
\and
National University of Ireland, Department of Experimental Physics, Maynooth, Co. Kildare, Ireland\goodbreak
\and
Nicolaus Copernicus Astronomical Center, Bartycka 18, 00-716 Warsaw, Poland\goodbreak
\and
Niels Bohr Institute, Blegdamsvej 17, Copenhagen, Denmark\goodbreak
\and
Niels Bohr Institute, Copenhagen University, Blegdamsvej 17, Copenhagen, Denmark\goodbreak
\and
Nordita (Nordic Institute for Theoretical Physics), Roslagstullsbacken 23, SE-106 91 Stockholm, Sweden\goodbreak
\and
Optical Science Laboratory, University College London, Gower Street, London, U.K.\goodbreak
\and
SISSA, Astrophysics Sector, via Bonomea 265, 34136, Trieste, Italy\goodbreak
\and
SMARTEST Research Centre, Universit\`{a} degli Studi e-Campus, Via Isimbardi 10, Novedrate (CO), 22060, Italy\goodbreak
\and
School of Physics and Astronomy, Cardiff University, Queens Buildings, The Parade, Cardiff, CF24 3AA, U.K.\goodbreak
\and
School of Physics and Astronomy, University of Nottingham, Nottingham NG7 2RD, U.K.\goodbreak
\and
Sorbonne Universit\'{e}-UPMC, UMR7095, Institut d'Astrophysique de Paris, 98 bis Boulevard Arago, F-75014, Paris, France\goodbreak
\and
Space Research Institute (IKI), Russian Academy of Sciences, Profsoyuznaya Str, 84/32, Moscow, 117997, Russia\goodbreak
\and
Space Sciences Laboratory, University of California, Berkeley, California, U.S.A.\goodbreak
\and
Special Astrophysical Observatory, Russian Academy of Sciences, Nizhnij Arkhyz, Zelenchukskiy region, Karachai-Cherkessian Republic, 369167, Russia\goodbreak
\and
Stanford University, Dept of Physics, Varian Physics Bldg, 382 Via Pueblo Mall, Stanford, California, U.S.A.\goodbreak
\and
Sub-Department of Astrophysics, University of Oxford, Keble Road, Oxford OX1 3RH, U.K.\goodbreak
\and
The Oskar Klein Centre for Cosmoparticle Physics, Department of Physics,Stockholm University, AlbaNova, SE-106 91 Stockholm, Sweden\goodbreak
\and
Theory Division, PH-TH, CERN, CH-1211, Geneva 23, Switzerland\goodbreak
\and
UPMC Univ Paris 06, UMR7095, 98 bis Boulevard Arago, F-75014, Paris, France\goodbreak
\and
Universit\"{a}t Heidelberg, Institut f\"{u}r Theoretische Astrophysik, Philosophenweg 12, 69120 Heidelberg, Germany\goodbreak
\and
Universit\'{e} de Toulouse, UPS-OMP, IRAP, F-31028 Toulouse cedex 4, France\goodbreak
\and
Universities Space Research Association, Stratospheric Observatory for Infrared Astronomy, MS 232-11, Moffett Field, CA 94035, U.S.A.\goodbreak
\and
University of Granada, Departamento de F\'{\i}sica Te\'{o}rica y del Cosmos, Facultad de Ciencias, Granada, Spain\goodbreak
\and
University of Granada, Instituto Carlos I de F\'{\i}sica Te\'{o}rica y Computacional, Granada, Spain\goodbreak
% ...
}

\abstract{
We study the implications of \Planck\ data for models of dark energy (DE)
and modified gravity (MG), beyond the standard cosmological constant scenario.
We start with cases where the DE only directly affects the background
evolution, considering Taylor expansions of the equation of state $w(a)$,
as well as principal component analysis and parameterizations related to the
potential of a minimally coupled DE scalar field. When estimating the
density of DE at early times, we significantly improve present constraints and find that it has to be below $\approx 2\,\%$
(at $95\%$ confidence) of the critical density even when forced to play a
role for $z < 50$ only.
We then move to general parameterizations of the DE or MG perturbations that
encompass both effective field theories and the phenomenology of gravitational
potentials in MG models.  Lastly,
we 
test a range of specific models, such as k-essence,
$f(R)$ theories and coupled DE.
In addition to the latest \Planck\ data, for our main analyses we use
background constraints from baryonic acoustic oscillations, type-Ia supernovae
and local measurements of the Hubble constant.  We 
further show the impact of
measurements of the cosmological perturbations, such as redshift-space
distortions and weak gravitational lensing.
These additional probes are important tools for testing MG models and for
breaking degeneracies that are still present in the combination of \Planck\
and background data sets.

All results that include only background parameterizations (expansion
of the equation of state, early DE, general potentials in minimally-coupled
scalar fields or principal component analysis) are in agreement with
\LCDM.  When testing models that also change perturbations (even when the background is fixed to \LCDM), some tensions
appear in a few scenarios: the maximum one found is $\sim 2 \sigma$ for \planckTT\ when parameterizing observables related to
the gravitational potentials with a chosen time dependence; the tension increases to at most $3 \sigma$ when external data sets are included. It however disappears when including CMB lensing.}

\date{\vspace{-0.23in} February 5, 2015}
\titlerunning{\papertitle}
\authorrunning{Planck Collaboration}

\maketitle
\providecommand{\sorthelp}[1]{}
\alltwentythirteenresultspapers
\alltwentyfifteenresultspapers

\section{Introduction}
\label{sec:introduction}

The cosmic microwave background (CMB) is a key probe of our cosmological model
\citep{planck2014-a15}, providing information on the primordial Universe
and its physics, including inflationary models \citep{planck2014-a24}
and constraints on primordial non-Gaussianities \citep{planck2014-a19}.
In this paper we use the 2015 data release from \planck \footnote{\Planck\ (\url{http://www.esa.int/Planck}) is a project of the European Space Agency  (ESA) with instruments provided by two scientific consortia funded by ESA member states and led by Principal Investigators from France and Italy, telescope reflectors provided through a collaboration between ESA and a scientific consortium led and funded by Denmark, and additional contributions from NASA (USA).}
\citep{planck2014-a01} to perform a systematic analysis of a large set
of dark energy and modified gravity theories.

Observations have long shown that only a small fraction 
of the total energy density in the Universe (around 5\,\%)
is in the form of baryonic matter, with the dark matter
needed for structure formation accounting for about another 26\,\%.
In one scenario the dominant component,
generically referred to as dark energy (hereafter DE), brings the total
close to the critical density and is responsible for the recent phase of
accelerated expansion.
In another scenario the accelerated expansion arises, partly or fully, due
to a modification of gravity on cosmological scales.
Elucidating the nature of this DE and testing General Relativity (GR)
on cosmological
scales are major challenges for contemporary cosmology, both on the theoretical
and experimental sides
\citep[e.g.,][]{LSSTAbell,2012arXiv1206.1225A, Clifton:2011jh, 2014arXiv1407.0059J, Huterer2015}.

In preparation for future experimental investigations
of DE and modified gravity (hereafter MG), it is
important to determine what we already know about these models
at different epochs in redshift and different length scales.  
CMB anisotropies fix the cosmology at early times, while
additional cosmological data sets further constrain on how DE or MG evolve at
lower redshifts.  The aim of this paper is to investigate models for dark energy and modified gravity using
 \Planck\ data in combination with other data sets.

The simplest model for DE is a cosmological constant, $\Lambda$, first
introduced by \cite{1917SPAW.......142E} in order to keep the Universe static,
but soon dismissed when the Universe was found to be expanding
\citep{1927ASSB...47...49L, Hubble:1929ig}.
This constant has been reintroduced several times over the years
in attempts to explain several astrophysical phenomena,
including most recently the flat spatial geometry implied by the
CMB and supernova observations of a recent phase
of accelerated expansion \citep{Riess1998, Perlmutter:1999}.
A cosmological constant is described by a single parameter,
the inclusion of which brings the model
($\Lambda$CDM) into excellent agreement with the data.
$\Lambda$CDM still represents a good fit to a wide range of
observations, more than 20 years after it was introduced.
Nonetheless, theoretical estimates for the vacuum density are many orders of
magnitude larger than its observed value.
In addition, $\Omega_\Lambda$ and $\Omega_{\rm m}$ are of the same
order of magnitude only at present, which
marks our epoch as a special time in the
evolution of the Universe (the ``coincidence problem'').
This lack of a clear theoretical understanding has motivated the development
of a wide variety of alternative models.  Those models which are close to
$\Lambda$CDM are in broad agreement with current constraints on the background
cosmology, but the perturbations may still evolve differently, and hence it
is important to test their predictions against CMB data. 

There are at least three difficulties we had to face within this paper.
First, there appears to be a vast array of possibilities in the literature and no agreement yet in the scientific community on a comprehensive 
framework for discussing the landscape of models.
A second complication is that robust constraints on DE come from a combination
of different data sets working in concert.  Hence we have to be careful in
the choice of the data sets so that we do not find apparent
hints for non-standard models that are in fact due to systematic errors.
A third area of concern is the fact that numerical codes available at present for DE and MG are not as well tested in these scenarios as for \LCDM, especially given the accuracy reached by the data. 
Furthermore, in some cases, we need to rely on stability routines that deserve further investigation to assure that they are not excluding more models than required.

In order to navigate the range of modelling possibilities, we adopt
the following three-part approach.
\begin{enumerate}
\item {\bf Background parameterizations}.  Here we consider only
parameterizations of background-level cosmological
quantities.  Perturbations are always included, but their evolution depends only on the background.
This set includes models involving 
expansions,
parameterizations or principal component analyses of the equation of state
$w \equiv p/ \rho$ of a DE fluid with pressure $p$ and energy density $\rho$.
Early DE also belongs to this class.
\item {\bf Perturbation parameterizations}.  Here the
perturbations themselves are parameterized or modified explicitly, not only as a
consequence of a change in background quantities.
There are two main branches we consider: firstly, effective field theory for DE
\citep[EFT, e.g.\ ][]{Gubitosi:2012hu}, which has a clear theoretical
motivation, since it includes all theories derived when accounting for
all symmetry operators in the Lagrangian, written in unitary gauge, i.e.~in terms of metric perturbations only. This is a very general
classification that has the advantage of providing a broad overview of (at least) all
universally coupled DE models.  However, a clear disadvantage is that the number of free parameters
is large and the constraints are consequently weak.
Moreover, in currently available numerical codes one needs to rely on
stability routines which are not fully tested and may discard more models
than necessary.

As a complementary approach, we include a more phenomenological class of models
obtained by directly parameterizing two independent functions of the
gravitational potentials.  This approach can in principle probe
{\em all\/} degrees of freedom at the background and perturbation level
\citep[e.g.][]{Kunz:2012aw} and is easier to handle in numerical codes.
While the connection to physical models is less obvious here than in EFT,
this approach allows us to gain a more intuitive understanding of the general
constraining power of the data.
\item {\bf Examples of particular models}.  Here we focus on a selection of
theories that have already been discussed in the li\-terature and are better
understood theoretically; these 
can partly be considered as applications of
previous cases for which the CMB constraints are more informative, because
there is less freedom in any particular theory
than in a more general one.
\end{enumerate}

The CMB is the cleanest probe of large scales, which are of particular
interest for modifications to gravity.
We will investigate the constraints coming from \Planck\ data
in combination with other data sets, addressing strengths and potential
weaknesses of different analyses. 
Before describing in detail the models and data sets that correspond to our
requirements, in Sect.~\ref{sec:CMBrelevant} we first address the main
question that motivates our paper, discussing why CMB is relevant for DE.  We then present the specific model
parameterizations in Sect.~\ref{sec:models}.
The choice of data sets is discussed in detail in Sec. \ref{sec:data}
before we present results in Sect.~\ref{sec:results} and discuss 
conclusions in Sect.~\ref{sec:conclusions}.

\section{Why is the CMB relevant for dark energy?}
\label{sec:CMBrelevant}

The CMB anisotropies are largely generated at the last-scattering epoch, and
hence can be used to pin down the theory at early times.  In fact many forecasts
of future DE or MG experiments are for new data {\it plus\/} constraints from
\Planck.  However, there are also several effects that DE and MG models can
have on the CMB, some of which are to:

\begin{enumerate}
\item change the expansion history and hence distance to the last scattering
surface, with a shift in the peaks, sometimes referred to as a geometrical
projection effect \citep{1996ApJ...471...30H};
\item cause the decay of gravitational potentials at late times,
affecting the low-multipole CMB anisotropies through 
the integrated Sachs-Wolfe (or ISW) effect \citep{SachsWolfe,Kofman1985};
\item enhance the cross-correlation between the CMB and
large-scale structure, through the ISW effect \citep{Giannantonio:2008zi};
\item change the lensing potential, through additional DE perturbations or
modifications of GR \citep{acquaviva_baccigalupi_2006, 2013JCAP...09..004C};
\item change the growth of structure \citep{Peebles1984,Barrow1993}
leading to a mismatch between the CMB-inferred amplitude of the fluctuations
$A_{\rm s}$ and late-time measurements of $\sigma_8$
\citep{Kunz:2003iz, Baldi:2010td};
\item impact small scales, modifying the damping tail in $C_\ell^{TT}$,
giving a measurement of the abundance of DE at different redshifts
\citep{2011PhRvD..83l3504C, reichardt_etal_2011};
\item affect the ratio between odd and even peaks if modifications of gravity
treat baryons and cold dark matter differently \citep{amendola_etal_2012};
\item modify the lensing $B$-mode contribution, through changes in the
lensing potential \citep{Amendola:2014wma};
\item modify the primordial $B$-mode amplitude and scale dependence,
by changing the sound speed of gravitational waves
\citep{2013JCAP...02..024A, Amendola:2014wma,2014arXiv1405.7974R}.
\end{enumerate}

\begin{figure}[htb!]
\begin{center}
\hspace*{-1cm}
\begin{tabular}{ccc}
\includegraphics[width=.45\textwidth]{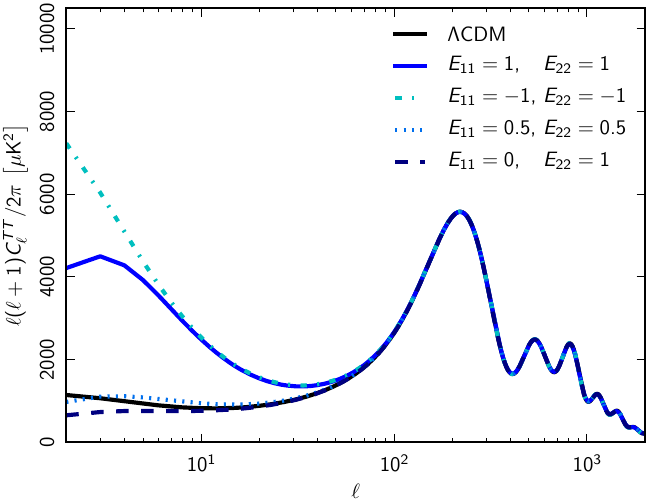} \\
\includegraphics[width=.45\textwidth]{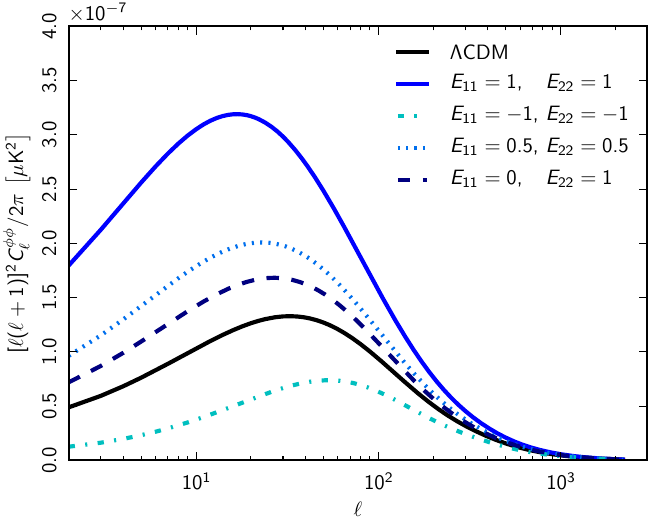} 
 \end{tabular}
\caption{Typical effects of modified gravity on theoretical CMB temperature (top panel)
and lensing potential (bottom panel) power spectra.  
An increase (or decrease)
of $E_{22}$ with respect to zero introduces a gravitational slip,
higher at present, when $\Omde$ is higher (see \eq{\ref{eq:mudef}} and \eq{\ref{eq:etadef}}); this in turns changes
the Weyl potential and leads to a higher (or lower) lensing potential.
On the other hand, whenever $E_{11}$ and $E_{22}$ are different from zero
(quite independently of their sign) $\mu$ and $\eta$ change in time: as the
dynamics in the gravitational potential is increased, this leads to an
enhancement in the ISW effect.  Note also that even when the temperature
spectrum is very close to \LCDM\ (as for $E_{11} = E_{22} = 0.5$) the lensing potential is still different with respect to \LCDM,
shown in black.
}
\label{fig:MGcls}
\end{center}
\end{figure}

In this paper we restrict our analysis to scalar perturbations.  The dominant
effects on the temperature power spectrum are due to lensing and the ISW
effect, as can be seen in \fig{\ref{fig:MGcls}}, which shows typical
power spectra of temperature anisotropies and lensing potential for modified
gravity models.
Different curves correspond to different
choices of the $\mu$ and $\eta$ functions, which change the relation between
the metric potentials and the sources, as well as introducing a gravitational slip; we will define these functions in \sect{\ref{subsubsec:mgcamb}},
\eq{\ref{eq:mudef}} and \eq{\ref{eq:etadef}}, respectively. Spectra are obtained using 
a scale-independent evolution for both $\mu$ and $\eta$.  
The two parameters in the figure then determine the change in amplitude of
$\mu$ and $\eta$ with respect to the \lcdm\ case, in which
$E_{11} = E_{22} = 0$ and $\mu = \eta = 1$.

\section{Models and parameterizations} \label{sec:models}

We now provide an overview of the models addressed in this paper. 
Details on the specific parameterizations will be discussed in
\sect{\ref{sec:results}}, where we also present the results for each
specific  method.

We start by noticing that one can generally follow two different approaches:
(1) given a theoretical set up, one can specify the action (or Lagrangian)
of the theory and derive background and perturbation equations in that
framework; or (2) more phenomenologically, one can construct functions
that map closely onto cosmological observables, probing the geometry of
spacetime and the growth of perturbations.  Assuming
spatial flatness for simplicity, the geometry is given  by the expansion rate
$H$ and perturbations to the metric.  If we
consider only scalar-type components
the metric perturbations can be written in terms of the gravitational potentials
$\Phi$ and $\Psi$ (or equivalently by any two independent combinations of
these potentials).
Cosmological observations thus constrain one ``background'' function of time
$H(a)$ and two ``perturbation'' functions of scale and time
$\Phi(k,a)$ and $\Psi(k,a)$ \citep[e.g.,][]{Kunz:2012aw}.
These functions fix the metric, and thus the Einstein tensor
$G_{\mu\nu}$.  Einstein's equations link this tensor to the energy-momentum
tensor $T_{\mu\nu}$, which in turn can be related to DE or MG properties.

Throughout this paper we will adopt the metric given by the line element
\begin{equation}
ds^2 = a^2 \left[ - (1+2 \Psi) d\tau^2 + (1-2 \Phi) dx^2 \right] \, .
 \label{eq:metric}
\end{equation}
The gauge invariant potentials $\Phi$ and $\Psi$ are related to the \cite{PhysRevD.22.1882} 
potentials 
$\Phi_{\rm{A}}$ and $\Phi_H$ 
and to the
\cite{Kodama:1985bj} potentials 
${\Psi_{\rm{KS}}}$ and ${\Phi_{\rm{KS}}}$ in the following way:
$\Psi = \Phi_{\rm{A}} = \Psi_{\rm{KS}}$
and $\Phi = -\Phi_{\rm{H}} = -\Phi_{\rm{KS}}$.
Throughout the paper we
use a metric signature $(-,+++)$ and follow the notation of 
\cite{Ma:1995ey}; the speed of light is set to $c=1$, except where explicitly stated otherwise.

We define the equation of state $\bar p(a) = w(a) \bar \rho(a)$, where
$\bar p$ and $\bar \rho$ are the average pressure
and energy density.  The sound speed $c_{\rm{s}}$ is defined in the fluid
rest frame 
in terms of pressure and density perturbations as
$\delta p(k,a) = c_{\rm{s}}^2(k,a) \delta\rho(k,a)$.
The anisotropic stress $\sigma(k,a)$ \citep[equivalent to $\pi_{\rm T}$
in the notation of][]{Kodama:1985bj} is the scalar part of the off-diagonal
space-space stress energy tensor perturbation.  The set of functions
$\{H,\Phi,\Psi\}$ describing the metric is formally equivalent to the set of
functions $\{w,c_{\rm{s}}^2,\sigma\}$
\citep{Ballesteros:2011cm}. 

Specific theories typically cover only subsets of this function space and
thus make specific
predictions for their form.  In the following sections we will
discuss the particular theories that we consider in this paper.

\subsection{Background parameterizations\label{sec:bgparam}}

The first main `category' of theories we describe includes parameterizations of background quantities. 
Even when we are only interested in constraints on background parameters, we are implicitly assuming a prescription for Dark Energy fluctua\-tions.  The conventional approach, that we adopt also here, is to choose a minimally-coupled scalar field model \citep{wetterich_1988, ratra_peebles_1988}, also known as quintessence, which corresponds to the choice of a rest-frame sound speed 
$c_{\rm s}^2  = 1$
(i.e., equal to the speed of light) and $\sigma = 0$ (no scalar anisotropic stress). In this case the relativistic sound speed suppresses the dark energy perturbations on sub-horizon scales, preventing it from contributing significantly to clustering.

Background parameterizations discussed in this paper include:
\begin{itemize}
\item ($w_0$, $w_a$) Taylor expansion at first order (and potentially higher orders);
\item Principal Component Analysis of $w(a)$ \citep{huterer_starkman_2003}, that allows to estimate constraints on $w$ in independent redshift bins;
\item general parameterization of any minimally coupled scalar field in terms of three parameters $\epsilon_s$, $\zeta_s$, $\epsilon_\infty$. This is a novel way to describe minimally coupled scalar field models without explicitly specifying the form of the potential \citep{huang_bond_kofman:2011};
\item Dark Energy density as a function of $z$ (including parameterizations such as early Dark Energy).
\end{itemize}

The specific implementation for each of them is discussed in \sect{\ref{sec:bkg_res}} together with corresponding results. 
We will conclude the background investigation by describing, in \sect{\ref{sec:compress}}, a compressed Gaussian likelihood that captures most of the constraining power of the \planck\ data applied to smooth Dark Energy or curved models \citep[following][]{Mukherjee:2008kd}. The compressed likelihood is useful for example to include more easily the \planckonly\ CMB data in Fisher-forecasts for future large-scale structure surveys.

\subsection{Perturbation parameterizations}

Modified gravity models (in which gravity is modified with respect to
GR) in general affect both the background and the perturbation equations.  In this
subsection we go beyond background parameterizations and identify two different
approaches to constrain MG models, one more theoretically motivated and a
second more phenomenological one.
We will not embark on a full-scale survey of DE and MG models here, but
refer the reader to e.g. \cite{Amendola:2012ys} for more details.

\subsubsection{Modified gravity and effective field theory}
 \label{subsubsec:EFT}

The first approach starts from a Lagrangian, derived from an effective field
theory (EFT) expansion \citep{2008JHEP...03..014C}, discussed in \cite{2009JCAP...02..018C} and \cite{Gubitosi:2012hu} in the context of DE. Specifically, EFT describes the space of  (universally coupled) scalar field
theories, 
with a Lagrangian written in unitary gauge that preserves isotropy and homogeneity at the background level,
assumes the weak equivalence principle, and has only one extra dynamical field
besides the matter fields conventionally considered in cosmology. 
The action reads:
\begin{eqnarray} \label{Eqn:EFTLag}
 S &=& \int d^4x\sqrt{-g}\left\{\frac{m_0^2}{2}\left[1+\Omega(\tau)\right]R
 + \Lambda(\tau) - a^2c(\tau) \delta g^{00} \right. \nonumber \\  
  &\quad&\left. +\ \frac{M_2^4 (\tau)}{2} \left(a^2\delta g^{00}\right)^2 
 - \bar{M}_1^3 (\tau){2}a^2 \delta g^{00} \delta {K}^\mu_\mu \right.
  \nonumber \\ 
  &\quad& \left. -\ \frac{\bar{M}_2^2 (\tau)}{2} \left(\delta K^\mu_\mu\right)^2
 - \frac{\bar{M}_3^2 (\tau)}{2} \delta K^\mu_\nu \delta K^\nu_\mu
 +\frac{a^2\hat{M}^2(\tau)}{2}\delta g^{00}\delta R^{(3)} \right. \nonumber \\ 
  &\quad&\left. +\ m_2^2 (\tau) \left(g^{\mu \nu} + n^\mu n^\nu\right)
  \partial_\mu \left(a^2g^{00}\right) \partial_\nu\left(a^2 g^{00}\right)
  \vphantom{\frac{m_0^2}{2}}\right\} \nonumber \\
  &\quad& +\ S_{\rm{m}} \big[\chi_i ,g_{\mu \nu}\big].
\end{eqnarray} 
Here $R$ is the Ricci scalar, $\delta R^{(3)}$ is its spatial perturbation,
${K}^\mu_\nu$ is the extrinsic curvature, and $m_0$ is the bare (reduced)
Planck mass.  The matter part of the action, $S_{\rm{m}}$, includes all fluid
components except dark energy, i.e., baryons, cold dark matter, radiation,
and neutrinos.  The action in Eq.~(\ref{Eqn:EFTLag}) depends on nine
time-dependent functions \citep{bloomfield_etal_2013}, here
$\{\Omega,c,\Lambda,\bar{M}_1^3,\bar{M}_2^4,\bar{M}_3^2,M_2^4,\hat{M}^2,
 m_2^2\}$, whose choice specifies the theory. 
In this way, EFT provides a direct link to any scalar field theory.
A particular subset of EFT theories are the \cite{Horndeski:1974wa}
models, which include (almost) all stable scalar-tensor theories, universally coupled to gravity, with
second-order equations of motion in the fields and depend on five functions
of time \citep{2013JCAP...08..025G, Bellini:2014fua, 2014JCAP...05..043P}.

Although the EFT approach has the advantage of being very versatile,
in practice it is necessary to choose suitable parameterizations for the free functions
listed above, in order to compare the action with the data.  We will describe
our specific choices, together with results for each of them,
in \sect{\ref{results:modifiedgravity}}.

\subsubsection{MG and phenomenological parameterizations}
 \label{subsubsec:mgcamb}
The second approach adopted in this paper to test MG is more phenomenological
and starts from the consideration that cosmological observations probe
quantities related to the metric perturbations, in addition to the expansion
rate.  Given the line element of Eq.~(\ref{eq:metric}),
the metric perturbations are determined by the two potentials $\Phi$ and
$\Psi$, so that we can model all observationally relevant degrees of freedom
by parameterizing these two potentials (or, equivalently, two independent
combinations of them) as functions of time and scale.  Since a 
non-vanishing anisotropic stress (proportional to $\Phi-\Psi$) is a generic
signature of modifications of GR \citep{Mukhanov:1990me,Saltas:2014dha},
the parameterized potentials will correspond to predictions of MG models.

Various parameterizations have been considered in the literature.
Some of the more popular (in longitudinal gauge) are:
\begin{enumerate}
\item {$Q(a,\vec{k})$}, which modifies the relativistic Poisson
equation through extra DE clustering according to 
\begin{equation} \label{eq:Qdef}
 -k^2 \Phi \equiv 4\pi G a^2 Q(a,\vec{k}) \rho \Delta,
\end{equation}
where $\Delta$ is the comoving density perturbation;
\item $\mu(a,\vec{k})$ (sometimes also called $Y(a,\vec{k})$), which modifies the equivalent equation for $\Psi$
 rather than $\Phi$:
\begin{equation} \label{eq:mudef}
 -k^2 \Psi \equiv 4\pi G a^2 \mu(a,\vec{k}) \rho \Delta;
\end{equation}
\item $\Sigma(a,\vec{k})$, which modifies lensing (with the lensing/Weyl
potential being $\Phi+\Psi$), such that 
\begin{equation} \label{eq:sigmadef}
 -k^2 (\Phi+\Psi) \equiv 8\pi G a^2 \Sigma(a,\vec{k}) \rho  \Delta;
\end{equation}
\item $\eta(a,\vec{k})$, which reflects the presence of a non-zero anisotropic
stress, the difference between $\Phi$ and $\Psi$ being equivalently written as
a deviation of the ratio\footnote{This parameter is called $\gamma$ in the code
{\tt MGCAMB}, but since $\gamma$ is also often used for the growth index, we
prefer to use the symbol $\eta$.}
\begin{equation} \label{eq:etadef}
 \eta(a,\vec{k}) \equiv \Phi/\Psi.
\end{equation}
\end{enumerate}
In the equations above,
$\rho\Delta = \rho_{\rm m} \Delta_{\rm m} + \rho_{\rm r} \Delta_{\rm r}$
so that the parameters $Q$, $\mu$, or $\Sigma$ quantify the deviation of the
gravitational potentials from the value expected in GR due to perturbations
of matter and relativistic particles.
At low redshifts, where most DE mo\-dels become relevant, we can neglect the
relativistic contribution.  The same is true for $\eta$, where we can neglect
the contribution of relativistic particles to the anisotropic stress at late
times.

The four functions above are certainly not independent.  It is sufficient to
choose two independent functions of time and scale to describe all
modifications with respect to General Relativity
\citep[e.g.][]{Zhang:2007nk,Amendola:2007rr}.
Popular choices include: $(\mu, \eta)$, which have a simple functional form for
many theories; $(\mu,\Sigma)$, which is more closely related
to what we actually observe, given that CMB lensing, weak galaxy lensing
and the ISW effect measure a projection or derivative of the Weyl
potential $\Phi+\Psi$.  Furthermore, redshift space distortions constrain the
velocity field, which is linked to $\Psi$ through the Euler equation of motion.

All four quantities, $Q$, $\mu$, $\Sigma$, and $\eta$, are free functions of
time and scale.  Their parameterization in terms of the scale factor $a$ and
momentum $\vec{k}$ will be specified in Section \ref{sec:genpar},
together with results obtained by confronting this class of models with data.

\subsection{Examples of particular models}
 \label{subsubsec:particularmodels}
The last approach is to consider particular models.
Even though these are in principle included in the case described in
Sect.~\ref{subsubsec:EFT}, it is nevertheless still useful to highlight
some well known examples of specific interest, which we list below.
\begin{itemize}
\item Minimally-coupled models beyond simple quintessence.  Specifically, we
consider ``k-essence'' models, which are defined by an arbitrary sound speed
$c_{\rm s}^2$ in addition to a free equation of state parameter $w$
\citep{ArmendarizPicon:2000dh}.
\item An example of a generalized scalar field model \citep{Deffayet:2010qz} and
of Lorentz-violating massive gravity \citep{2004JHEP...10..076D, 2008PhyU...51..759R}, both in the `equation of state'
formalism of \cite{Battye:2012eu}.
\item Universal ``fifth forces.''  We will show results for $f(R)$ theories
\citep{Wetterich1994, Capozziello:2002rd, PhysRevD.75.083504, DeFelice:2010aj},
which form a subset of all models contained in the EFT approach. 
\item Non-universal fifth forces.  We will illustrate results for coupled
DE models \citep{amendola_2000},
in which dark matter particles feel a force mediated by the DE scalar field.
\end{itemize}
All these particular models are based on specific actions, ensuring full
internal consistency. The reviews by \cite{Amendola:2012ys}, \cite{Clifton:2011jh}, \cite{2014arXiv1407.0059J} and 
\cite{Huterer2015} contain detailed descriptions of a large number of models
discussed in the literature.

\section{Data\label{sec:data}} 
We now discuss the data sets we use, both from \Planck\ and in combination
with other experiments.
As mentioned earlier, if we combine many different data sets (not all of which
will be equally reliable) and take them all at face value, we risk attributing
systematic problems between data sets to genuine physical effects
in DE or MG models.  On the other hand, we need to avoid bias in confirming
$\Lambda$CDM, and remain open to the possibility that some tensions may be
providing hints that point towards DE or MG models. 
While discussing results in Sect.~\ref{sec:results}, we will try to assess
the impact of additional data sets, separating them from the 
\Planck\ baseline choice, keeping in mind caveats that might appear when
considering some of them.  For a more detailed discussion of the data sets
we refer to \cite{planck2014-a15}.

\subsection{\Planck\ data sets}
\subsubsection{\Planck\ low-$\ell$ data}

The 2013 papers used WMAP polarization measurements \citep{bennett2012} at
multipoles $\ell \le 23$ to constrain the optical depth parameter $\tau$.
The corresponding likelihood was denoted ``WP'' in the 2013 papers. 

For the present release, we use in its place a \Planck\ polarization likelihood
that is built through low-resolution maps of Stokes $Q$ and $U$ polarization
measured by LFI at 70\,GHz
(excluding data from Surveys~2 and 4), foreground-cleaned with the LFI 30\,GHz
and HFI 353\,GHz maps, used as polarized 
synchrotron and dust templates, respectively (see \cite{planck2014-a13}).

The foreground-cleaned LFI 70\,GHz polarization maps are processed, together
with the temperature map from the {\tt Commander} component
separation algorithm over 94\,\% of the sky
\citep[see][for further details]{planck2014-a11}, using the
low-$\ell$ \planck\ temperature-polarization likelihood.
This likelihood is pixel-based, extends up to multipoles $\ell=29$
and masks the polarization maps with a specific polarization mask,
which uses 46\,\% of the sky. 
Use of this likelihood is denoted as ``lowP'' hereafter.

The \Planck\ lowP likelihood, when combined with the high-$\ell$ \Planck\
temperature one, provides a best fit value for the optical depth
$\tau = 0.078 \pm 0.019$, which is about $1\,\sigma$ lower than the value
inferred from the WP polarization likelihood, i.e., $\tau = 0.089 \pm 0.013$,
in the \Planck\ 2013 papers \citep[see also][]{planck2014-a15}.
However, we find that the LFI 70\,GHz and WMAP polarization maps
are extremely consistent when both are cleaned with the HFI 353\,GHz polarized
dust template, as discussed in more detail in \cite{planck2014-a13}.
\subsubsection{\Planck\ high-$\ell$ data}
\label{sec:hi-l}

Following \cite{planck2013-p08}, the high-$\ell$ part of the likelihood
($30<\ell<2500$) uses a Gaussian approximation, 
\begin{equation}
-{\rm log}{\cal L}(\hat{C} | C(\theta))
 = \frac{1}{2} (\hat{C} - C(\theta))^{\rm T}\cdot\tens{C}^{-1}\cdot 
 (\hat{C} - C(\theta)) + {\rm const.} \,\, ,
 \label{eq:basic-likelihood}
 \end{equation}
with $\hat{C}$ the data vector, $C(\theta)$ the model with parameters
$\theta$ and $\tens{C}$ the covariance matrix.
The data vector consists of the temperature power spectra of the best CMB
frequencies of the HFI instrument.  Specifically, as discussed in \cite{planck2014-a13}, we use 100\,GHz, 143\,GHz
and 217\,GHz half-mission cross-spectra, measured on the cleanest 
part of the sky, avoiding the Galactic plane, as well as the brightest point
sources and regions where the CO emission is the strongest.
The point source masks are specific to each frequency.  We retain, 66\,\% 
of the sky for the 100\,GHz map, 57\,\% for 143\,GHz, and 47\,\% for
217\,GHz.  All the spectra are corrected for beam and pixel window functions.
Not all cross-spectra and multipoles are included in the data vector;
specifically, the $TT$ $100\times143$ and $100\times217$ cross-spectra,
which do not bring much extra information, are discarded.  Similarly, we only
use multipoles in the range $30<\ell<1200$ for $100\times100$ and
$30<\ell<2000$ for $143\times143$, discarding modes where the S/N is too low.
We do not co-add the different cross-frequency spectra, since, even after
masking the highest dust-contaminated regions, each cross-frequency spectrum
has a different, frequency-dependent residual foreground contamination,
which we deal with in the model part of the likelihood function.

The model, $C(\theta)$ can be rewritten as 
\begin{equation}
 C_{\mu,\nu}(\theta) = \frac{ C^{\rm cmb}
 + C^{\mathrm{fg}}_{\mu,\nu}(\theta) }{\sqrt{A_\mu A_\nu}},
\end{equation}
where $C^{\rm cmb}$ is the set of CMB $C_\ell$s, which is independent of
frequency, $C^{\mathrm fg}_{\mu,\nu}(\theta)$ is the foreground model
contribution to the cross-frequency spectrum $\mu\times\nu$, and $A_\mu$
the calibration factor for the $\mu\times\mu$ spectrum.  We retain the
following contributions in our foreground modelling: dust; clustered
cosmic infrared background (CIB); thermal Sunyaev-Zeldovich (tSZ) effect;
kinetic Sunyaev-Zeldovich (kSZ) effect; tSZ-CIB cross-correlations;
and point sources.  The dust, CIB and point source contributions are the
dominant contamination.  Specifically, dust is the dominant foreground at
$\ell<500$, while the diffuse point source term (and CIB for the
$217\times217$) dominates the small scales.
All our foreground models are based upon smooth $C_\ell$ templates with free
amplitudes.  All templates but the dust are based on analytical models,
as described in \cite{planck2014-a13}.  The dust is based on a mask difference
of the 545\,GHz map and is well described by a power law of index $n=-2.63$,
with a wide bump around $\ell=200$.  A prior for the dust amplitude is
computed from the cross-spectra with the 545\,GHz map.  We refer the reader to
\cite{planck2014-a13} for a complete description of the foreground model.
The overall calibration for the $100\times100$ and $217\times217$ power spectra
free to vary within a prior measured on a small fraction of the sky near the
Galactic pole.

The covariance matrix ${\bf C}$ accounts for the correlations due to the mask
and is computed following the equations in \cite{planck2013-p08}.  The fiducial
model used to compute the covariance is based on a joint fit of $\Lambda$CDM
and nuisance parameters.  The covariance includes the non-Gaussianity of the
noise, but assumes Gaussian statistics for the dust.  The non-whiteness of the
noise is estimated from the difference between the cross- and auto-half mission
spectra and accounted for in an approximate manner in the covariance.
Different Monte Carlo based corrections are applied to the covariance matrix
calculation to account for inaccuracies in the analytic formulae at
large scales ($\ell<50$) and when dealing with the point source mask.
Beam-shape uncertainties are folded into the covariance matrix.  A complete
description of the computation and its validation is discussed in
\cite{planck2014-a13}.

The $TT$ unbinned covariance matrix is of size about $8000\times8000$.
When adding the polarization, the matrix has size $23000\times23000$,
which translates into a significant memory requirement and slows the
likelihood computation considerably.  We thus bin the data and covariance
matrix, using a variable bin-size scheme, to reduce the 
data vector dimension by about a factor of ten.  We checked that for the
$\Lambda$CDM model, including single parameter classical extensions, the
cosmological and nuisance parameter fits are identical with or without binning.

\subsubsection{\Planck\ CMB lensing}

Gravitational lensing by large-scale structure introduces dependencies in CMB
observables on the late-time geometry and clustering, which otherwise would be
degenerate in the primary
anisotropies~\citep{2002PhRvD..65b3003H, lewis_challinor_2006}. This provides
some sensitivity to dark energy and late-time modifications of gravity from
the CMB alone. The source plane for CMB lensing is the last-scattering surface,
so the peak sensitivity is to lenses at $z\approx 2$ (i.e., half-way to the
last-scattering surface) with typical sizes of order $10^2\,{\rm Mpc}$.
Although this peak lensing redshift is rather high for constraining simple
late-time dark energy models, CMB lensing deflections at angular multipoles
$\ell \la 60$ have sources extending to low enough redshift that DE
becomes dynamically important (e.g.,~\citealt{2014MNRAS.445.2941P}).

The main observable effects of CMB lensing are a smoothing of the acoustic
peaks and troughs in the temperature and polarization power spectra, the
generation of significant non-Gaussianity in the form of a non-zero connected
4-point function, and the conversion of $E$-mode to $B$-mode polarization.
The smoothing effect on the power spectra is included routinely in all results
in this paper. We additionally include measurements of the power spectrum
$C_\ell^{\phi\phi}$ of the CMB lensing potential $\phi$, which are extracted
from the \planck\ temperature and polarization 4-point functions, as presented
in~\citet{planck2014-a17} and discussed further below. Lensing also produces
3-point non-Gaussianity, which peaks in squeezed configurations, due to the
correlation between the lensing potential and the ISW effect in the
large-angle temperature anisotropies. This effect has been measured at around
$3\,\sigma$ with the full-mission \planck\
data~\citep{planck2014-a17,planck2014-a26}. Although in principle this is
a further probe of DE~\citep{2002PhRvD..65d3007V}
and MG~\citep{2004PhRvD..70b3515A},
we do not include these $T$--$\phi$ correlations in this paper
as the likelihood was not readily available.
We plan however to test this effect in future work.

The construction of the CMB lensing likelihood we use in this paper is
described fully in~\citet{planck2014-a17}; see also~\citet{planck2014-a15}.
It is a simple Gaussian approximation in the estimated $C_\ell^{\phi\phi}$
bandpowers, covering the multipole range
$40\,{\leq}\,{\ell}\,{\leq}\,400$.  The $C_\ell^{\phi\phi}$ are estimated from
the full-mission temperature and polarization 4-point functions, using the
{\tt SMICA} component-separated maps~\citep{planck2014-a11} over approximately
70\,\% of the sky.  A large number of tests of internal consistency of the
estimated $C_\ell^{\phi\phi}$ to different data cuts
(e.g., whether polarization is included, or whether individual frequency bands
are used in place of the {\tt SMICA} maps) are reported
in~\citet{planck2014-a17}. All such tests are passed for the conservative
multipole range $40 \leq \ell \leq 400$ that we adopt in this paper.  For
multipoles $\ell > 400$, there is marginal evidence of systematic effects in
reconstructions of the lensing deflections from temperature anisotropies alone,
based on curl-mode tests. Reconstructing the lensing deflections on large
angular scales is very challenging because of the large ``mean-field'' due to
survey anisotropies, which must be carefully subtracted with simulations.
We conservatively adopt a minimum multipole of $\ell = 40$ here, although the
results of the null tests considered in~\citet{planck2014-a17} suggest that
this could be extended down to $\ell = 8$.  For \planck, the multipole range
$40 \leq \ell \leq 400$ captures the majority of the S/N on
$C_\ell^{\phi\phi}$ for \lcdm\ models, although this restriction may be more
lossy in extended models.  The \planck\ 2014 lensing measurements are the most
significant to date (the amplitude of $C_\ell^{\phi\phi}$ is measured at
greater than $40\,\sigma$), and we therefore choose not to include lensing
results from other CMB experiments in this paper.

\subsubsection{\Planck\ CMB polarization}

The $TE$ and $EE$ likelihood follows the same principle as the $TT$ likelihood
described in Sect.~\ref{sec:hi-l}.  The data vector is extended to contain the
$TE$ and $EE$ cross-half-mission power spectra of the same 100\,GHz, 143\,GHz,
and 217\,GHz frequency maps.  Following \cite{planck2014-XXX}, we mask the
regions where the dust intensity is important, and retain 70\,\%, 50\,\%,
and 41\,\% of the sky for our three frequencies.  We ignore any other polarized
galactic emission and in particular synchrotron, which has been shown to be
negligible, even at 100\,GHz.  We use all of the cross-frequency spectra,
using the multipole range $30<\ell<1000$ for the 100\,GHz 
cross-spectra and $500<\ell<2000$ for the 217\,GHz cross-spectra.
Only the $143\times143$ spectrum covers the full $30<\ell<2000$ range.
We use the same beams as for the $TT$ spectra and do not correct for leakage
due to beam mismatch.  A complete description of the beam mismatch effects
and correction is described in \cite{planck2014-a13}. 

The model is similar to the $TT$ one.  We retain a single foreground component
accounting for the polarized emission of the dust.  Following
\cite{planck2014-XXX}, the dust $C_\ell$ template is a power law with index
$n=-2.4$.  A prior for the dust amplitude is measured in the cross-correlation
with the 353\,GHz maps.  The calibration parameters are fixed to unity.

The covariance matrix is extended to polarization, as described in
\cite{planck2014-a13}, using the correlation between the $TT$, $TE$,
and $EE$ spectra.  It is computed similarly to the $TT$ covariance matrix,
as described in Sect.~\ref{sec:hi-l}.

In this paper we will only show results that include CMB high-$\ell$
polarization data where we find that it has a significant impact.
DE and MG can in principle also affect the $B$-mode power spectrum through 
lensing of $B$-modes (if the lensing Weyl potential is modified)
or by changing the position and amplitude of the primordial peak
\citep{2013JCAP...02..024A, 2014arXiv1408.2224P}, including modifications of the sound speed of
gravitational waves \citep{Amendola:2014wma, 2014arXiv1405.7974R}.  Due to
the unavailability of the likelihood, results from $B$-mode polarization
are left to future work.

\subsection{Background data combination} \label{subsec:priority} 
We identify a first basic combination of data sets that we mostly rely on,
for which we have a high confidence that systematics are under control.
Throughout this paper, we indicate for simplicity with ``BSH'' the combination
BAO + SN-Ia + $H_0$, which we now discuss in detail.

\subsubsection{Baryon acoustic oscillations}
\label{sec:BAO}

Baryon acoustic oscillations (BAO) are the imprint of oscillations
in the baryon-photon plasma on the matter power spectrum
and can be used as a standard ruler,
calibrated to the CMB-determined sound horizon at the end of the drag epoch.
Since the acoustic scale is so large, BAO are largely unaffected by nonlinear
evolution.
As in the cosmological parameter paper, \cite{planck2014-a15}, BAO is
considered as the primary data set to break parameter degeneracies from
CMB measurements and offers constraints on the background evolution of
MG and DE models.
The BAO data can be used to measure both the angular diameter distance
$D_{\rm A} (z)$, and the expansion rate of the Universe $H(z)$ either separately or through the combination
\begin{equation}
 D_{\rm V}(z) = \left [ (1+z)^2 D^2_{\rm A}(z) {cz \over H(z)} \right ]^{1/3}.  \label{BAO1}
\end{equation}

As in \citet{planck2014-a15} we use: the SDSS Main Galaxy Sample at
$z_{\rm eff}=0.15$ \citep{Ross:14}; the Baryon Oscillation
Spectroscopic Survey (BOSS) ``LOWZ'' sample at $z_{\rm eff} = 0.32$
\citep{Anderson:14}; the BOSS CMASS (i.e. ``constant mass'' sample)
at $z_{\rm eff} = 0.57$ of
\citet{Anderson:14}; and the six-degree-Field Galaxy survey (6dFGS)
at $z_{\rm eff} = 0.106$ \citep{Beutler:11}.
The first two measurements are based on peculiar
velocity field reconstructions to sharpen the BAO feature and reduce
the errors on the quantity $D_{\rm V}/r_{\rm s}$; the analysis in
\citet{Anderson:14} provides constraints on both
$D_A(z_{\rm eff})$ and $H(z_{\rm eff})$.
In all cases considered here the BAO observations are modelled as
distance ratios, and therefore provide no direct measurement of $H_0$.
However, they provide a link between the expansion rate at low
redshift and the constraints placed by \Planck\ at $z\approx 1100$.

\subsubsection{Supernovae}

Type-Ia supernovae (SNe) are among the most important probes of
expansion and historically led to the general acceptance that a DE
component is needed \citep{Riess:1998,Perlmutter:1999}. 
Supernovae are considered as ``standardizable candles'' and
so provide a measurement of the luminosity distance as a function
of redshift. However, the absolute luminosity of SNe is considered
uncertain and is marginalized out, which also removes any constraints
on $H_0$.

Consistently with \cite{planck2014-a15}, we use here the analysis by
\citet{Betoule:2013} 
of the ``Joint Light-curve Analysis'' (JLA) sample. JLA is constructed
from the SNLS and SDSS SNe data, together with several samples of low
redshift SNe.  Cosmological constraints from the JLA
sample\footnote{A \COSMOMC\ likelihood module for the JLA sample
is available at
\url{http://supernovae.in2p3.fr/sdss_snls_jla/ReadMe.html}.}
are discussed by \citet{Betoule:2014},
and as mentioned in \cite{planck2014-a15} the constraints are
consistent with the 2013 and 2104 \Planck\ values for standard \LCDM.

\subsubsection{The Hubble constant}
\label{sec:hubble}
The CMB measures mostly physics at the epoch of recombination,
and so provides only weak direct constraints about low-redshift quantities
through the integrated Sachs-Wolfe effect and CMB lensing.
The CMB-inferred constraints
on the local expansion rate $H_0$ are model dependent, and this makes
the comparison to direct measurements interesting, since any mismatch
could be evidence of new physics.

Here, we rely on the re-analysis of the \citet{Riess:11}
(hereafter R11) Cepheid data made by \citet{Efstathiou:14} (hereafter
E14). By using a revised geometric maser distance to NGC 4258 from
\citet{Humphreys:13}, E14 obtains the following value for the Hubble
constant:
\begin{equation}
H_0 =  (70.6 \pm 3.3)\,{\rm km}\,{\rm s}^{-1}\,{\rm Mpc}^{-1},
\label{H0prior1}
\end{equation}
which is within $1\,\sigma$ of the \planckTT\ estimate.  In this
paper we use \eq{\ref{H0prior1}} as a conservative $H_0$ prior.  We note
that the 2015 \planckTT\ value is perfectly consistent with the 2013
\Planck\ value \citep{planck2013-p11} and so the tension with the R11 $H_0$
determination is still present at about $2.4\,\sigma$.  We refer to the
cosmological parameter paper \cite{planck2014-a15} for a more comprehensive
discussion of the different values of $H_0$ present in the
literature.

\subsection{Perturbation data sets}
The additional freedom present in MG models can be calibrated using external
data that test perturbations in particular.  In the following we describe
other available data sets that we included in the grid of runs for this paper.

\subsubsection{Redshift space distortions} 
Observations of the anisotropic clustering of galaxies in
redshift space permit the measurement of their peculiar velocities,
which are related to the Newtonian potential $\Psi$ via the Euler equation.
This, in turn, allows us to break a degeneracy with gravitational lensing
that is sensitive to the combination $\Phi+\Psi$.
Galaxy redshift surveys now provide very precise constraints on
redshift-space clustering.
The difficulty in using these data
is that much of the signal currently comes from scales where
nonlinear effects and galaxy bias are significant and must be accurately
modelled \citep[see, e.g., the discussions in][]{bianchi_etal_2012,
 gilmarin_etal_2012}.
Moreover, adopting the wrong fiducial cosmological model to convert angles
and redshifts into distances can bias measurements of the rate-of-growth of
structure \citep{Reid:12,Howlett:14}.  Significant progress in the
modelling has been achieved in the last few years, so
we shall focus here on the most recent (and relatively conservative) studies.
A compilation of earlier measurements can be found in the references above.

In linear theory, anisotropic clustering along the line of sight and in
the transverse directions measures the combination $f(z)\sigma_8(z)$,
where the growth rate is defined by
\beq 
f(z)=\frac{\rm{d} \ln \sigma_8}{\rm{d} \ln a}\,.
\eeq
where $\sigma_8$ is calculated including all matter and neutrino density perturbations.
Anisotropic clustering also contains geometric information from the
Alcock-Paczynski (AP) effect~\citep{Alcock:79}, which is sensitive to 
\beq
 F_{\rm AP}(z)= (1+z) D_{\rm A}(z)H (z) \,.
\eeq
In addition, fits which constrain RSD frequently also measure the
BAO scale, $D_V(z)/r_{\rm s}$, where $r_{\rm s}$ is the comoving sound
horizon at the drag epoch, and $D_V$ is given in \eq{\ref{BAO1}}.
As in \citet{planck2014-a15} we consider only analyses which solve
simultaneously for the acoustic scale, $F_{\rm AP}$ and $f\sigma_8$.

The Baryon Oscillation Spectroscopic Survey (BOSS) collaboration have
measured the power spectrum of their CMASS galaxy sample
\citep{Beutler:2013yhm} in the range
$k=0.01$--$0.20\,{\rm h}\,\mathrm{Mpc}^{-1}$.
\citet{Samushia:2013yga} have estimated the multipole moments of
the redshift-space correlation function of CMASS galaxies on scales
$>25\,{\rm h}^{-1}$Mpc.
Both papers provide tight constraints on the quantity $f\sigma_8$, and the
constraints are consistent.
The \citet{Samushia:2013yga} result was shown to behave marginally better
in terms of small-scale bias compared to mock simulations, so we choose to
adopt this as our baseline result.
Note that when we use the data of \citet{Samushia:2013yga}, we
exclude the measurement of the BAO scale, $D_{\rm V}/r_{\rm s}$,
from~\citet{2013arXiv1312.4877A}, to avoid double counting.

The \citet{Samushia:2013yga} results are expressed as a $3\times3$
covariance matrix for the three parameters $D_{\rm V}/r_{\rm s}$,
$F_{\rm AP}$ and
$f\sigma_8$, evaluated at an effective redshift of $z_{\rm eff}=0.57$.
Since \citet{Samushia:2013yga} do not apply a density field
reconstruction in their analysis, the BAO constraints are slightly weaker
than, though consistent with, those of \citet{Anderson:14}.

\subsubsection{Galaxy weak lensing}
\label{sec:gallenssec}
The distortion of the shapes of distant galaxies by large-scale structure
along the line of sight (weak gravitational lensing or cosmic shear) is 
particularly important for
constraining DE and MG, due to its dependence on the
growth of fluctuations and the two scalar metric potentials.

Currently the largest weak lensing (WL) survey is the Canada France Hawaii
Telescope Lensing Survey (CFHTLenS), and we make use of two data sets from
this survey:
\begin{enumerate}
\item 2D CFHTLenS data \citep{2013MNRAS.430.2200K}, whose shear correlation
functions $\xi^{\pm}$ are estimated in the angular range 0.9 to 296.5 arcmin;
\item the tomographic CFHTLenS blue galaxy sample \citep{Heymans:2013fya},
whose data have an intrinsic alignment signal consistent with zero,
eliminating the need to marginalize over any additional nuisance parameters,
and where the shear correlation functions are estimated in six redshift bins,
each with an angular range $1.7 < \theta < 37.9$ arcmin. 
\end{enumerate}
Since these data are not independent we do not combine them, but rather check
the consistency of our results with each.  The galaxy lensing convergence power
spectrum, $P_{ij}^{\kappa}(\ell)$, can be written in terms of the Weyl
potential, $P_{\Phi+\Psi}$, by
\bea
P_{ij }^{\kappa}(\ell) \approx  {2\pi^2\ell}
  \int \frac{ {\rm d}  \chi}{\chi} g_{i} \left(\chi \right)
  g_{j} \left(\chi \right) P_{\Phi+\Psi}  \left(\ell/ \chi, \chi \right)\,,
\eea
where we have made use of the Limber approximation in flat space, and $\chi$
is the comoving distance. The lensing efficiency is given by
\bea
g_i (\chi) = \int_{\chi}^{\infty} {\rm d} \chi^{\prime} n_{i}
 \left( \chi^{\prime} \right) \frac{\chi^{\prime}-\chi}{\chi^{\prime}}\,,
\eea
where $n_{i} \left( \chi \right)$ is the radial distribution of source galaxies
in bin $i$.  In the case of no anisotropic stress and no additional clustering
from the DE, the convergence power spectrum can be written in the usual form 
\bea \label{eqn:convergencePk}
P_{ij}^{\kappa}(\ell) &=&\frac{9}{4}\Omega_{\rm m}^2 H_0^4
 \int_0^{\infty}\frac{g_i(\chi)g_j(\chi)}{a^2(\chi)}
 P(\ell/\chi, \chi) \rm{d} \chi\,. 
\eea
However, in this paper we always use the full Weyl potential to compute the
theoretical WL predictions.  The convergence can also be written in terms
of the correlation functions $\xi^{\pm}$ via
\bea \label{eqn:xi}
 \xi_{i,j}^{\pm}(\theta) = \frac{1}{2\pi}
 \int {\rm d}\ell \, \ell\, P_{ij}^{\kappa}(\ell) J_{\pm}(\ell \theta),
\eea
where the Bessel functions are $J_{+} = J_{0}$ and $J_{-} = J_{4}$.   

\begin{figure}[!t]
\begin{center}
\includegraphics[width=0.45\textwidth]{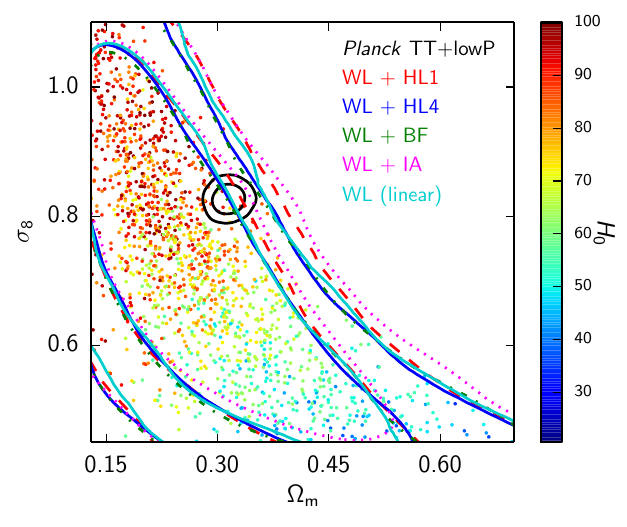}
\caption{$\Omega_{\rm m}$--$\sigma_8$  constraints for tomographic lensing
from \citet{Heymans:2013fya}, using a very conservative angular cut, as
described in the text (see \sect{\ref{sec:gallenssec}}).  We show results using linear theory, nonlinear
corrections from {\tt Halofit} (HL) versions 1, 4, marginalization over
baryonic AGN feedback (BF), and intrinsic alignment (IA) (the latter two
using nonlinear corrections and {\tt Halofit} 4).  Coloured points indicate
$H_0$ values from WL+HL4.}
\label{fig:wl_comparison}
\end{center}
\end{figure}

In this paper we need to be particularly careful about the contribution of
nonlinear scales to $\xi^{\pm}$, since the behaviour of MG models in the
nonlinear regime is not known very precisely. 
The standard approach is to correct the power spectrum on nonlinear scales
using the {\tt Halofit} fitting function. Since its inception, there have
been several revisions to improve the agreement with $N$-body simulations.
We use the following convention to label the particular {\tt Halofit} model:
\begin{enumerate}
\item the original model of~\citet{Smith:2002dz};
\item an update from higher resolution $N$-body simulations, to include the
effect of massive neutrinos \citep{Bird:2011rb};
\item an update to improve the accuracy on small
scales\footnote{\url{http://www.roe.ac.uk/~jap/haloes/}};
\item an update from higher resolution $N$-body simulations, including
DE cosmologies with constant equation of state \citep{Takahashi:2012em}.
\end{enumerate}
Given this correction, one can scale the Weyl potential transfer functions
by the ratio of the nonlinear to linear matter power spectrum
\beq
T_{\Phi+\Psi} (k,z) \rightarrow T_{\Phi+\Psi} (k,z)
 \sqrt{\frac{P_{\delta}^{\rm nonlin} (k,z)}{P_{\delta}^{\rm lin} (k,z)}}\,.
\eeq

Both \citep{2013MNRAS.430.2200K} and \citep{Heymans:2013fya} quote a
``conservative'' set of cuts to mitigate uncertainty over the nonlinear
modelling scheme.  For the 2D analysis of \cite{2013MNRAS.430.2200K} angular
scales $\theta<17\arcmin$ are excluded for $\xi^{+}$, and $\theta<54\arcmin$
for $\xi^{-}$.  For the tomographic analysis of \cite{Heymans:2013fya},
angular scales $\theta<3 \arcmin$ are excluded for $\xi^{+}$ for any bin
combination involving the two lowest redshift bins, and no cut is applied for
the highest four redshift bins.  For $\xi^{-}$, angular scales
$\theta<30\arcmin$ are excluded for any bin combination involving the four
lowest redshift bins, and $\theta<16\arcmin$ for the highest two bins. 

These cuts, however, may be insufficient for our purposes, since we are
interested in extensions to \lcdm. We therefore choose a very conservative
set of cuts to mitigate the total contribution from nonlinear scales.
In order to select these cuts we choose the baseline \Planck\ TT+lowP \lcdm\
cosmology as described in \cite{planck2014-a15}, for which one can use
Eq.~(\ref{eqn:convergencePk}).  The cuts are then chosen by considering
$\Delta \chi^2= |\chi^2_{\rm lin}-\chi^2_{\rm nonlin}| $ of the WL likelihood
as a function of angular cut.  In order for this to remain $\Delta \chi^2<1$
for each of the {\tt Halofit} versions, we find it necessary to remove
$\xi^{-}$ entirely from each data set, and exclude $\theta<17\arcmin$ for
$\xi^{+}$ for both the 2D and tomographic bins. We note that a similar
approach to~\citet{Kitching:2014dtq} could also be followed using 3D
CFHTLenS data, where the choice of cut is more well defined in $k$-space, however the likelihood for this was not available at the time of this paper.

On small scales, the effects of intrinsic alignments and baryonic feedback
can also become significant.  In order to check the robustness of our cuts to
these effects we adopt the same methodology of~\citet{MacCrann:2014wfa}.
Using the same baseline model and choosing {\tt Halofit} version 4, we scale
the matter power spectrum by an active galactic nuclei (AGN) component,
derived from numerical simulations~\citep{vanDaalen:2011xb}, marginalizing
over an amplitude $\alpha_{\rm AGN}$.  The AGN baryonic feedback model has been shown by  \citet{2014arXiv1407.4301H} to provide the best fit to small-scale CFHTLens data. For intrinsic alignment we adopt the
model of~\cite{Bridle:2007ft}, including the additional nonlinear alignment
contributions to $\xi^{\pm}$, and again marginalizing over an amplitude
$\alpha_{\rm IA}$.  For more details on this procedure, we refer the reader
to~\citet{MacCrann:2014wfa}. 

The robustness of our ultra-conservative cuts to nonlinear modelling, baryonic
feedback and intrinsic alignment margina\-lization, is illustrated in
Fig.~\ref{fig:wl_comparison} for the tomographic data, with similar constraints
obtained from 2D data.  Assuming the same base \lcdm\ cosmology, and applying
priors of $\Omega_{\rm b} h^2 = 0.0223 \pm 0.0009$,
$n_{\rm s} = 0.96 \pm 0.02$, and $40 \Hunit <H_0<100 \Hunit$ to avoid
over-fitting the model, we find that the WL likelihood is insensitive to
nonlinear physics.  We therefore choose to adopt the tomographic data with
the ultra-conservative cuts as our baseline data set.

\begin{table*}[!tbh]
\begingroup
\newdimen\tblskip \tblskip=5pt
\caption{Table of models tested in this paper.  We have tested all models for
the combinations: \planckonly, \planckonly+BSH, \planckonly+WL,
\planckonly+BAO/RSD and \planckonly+WL+BAO/RSD.  Throughout the text, unless
otherwise specified, \planckonly\ refers to the baseline \planckTT\
combination.  The effects of CMB \lensing\ and \planckonly\ \TTTEEE\
polarization have been tested on all runs above and are, in particular, used
to constrain the amount of DE at early times.}
    \label{tbl:runsTable}
\footnotesize
\setbox\tablebox=\vbox{
\newdimen\digitwidth % These five lines change what an asterisk
\setbox0=\hbox{\rm 0} % means to TeX. Instead of meaning
\digitwidth=\wd0 % "print an '*' here", it now means "leave
\catcode`*=\active % as much blank space as a single number
\def*{\kern\digitwidth} % takes up".
\newdimen\signwidth % These five lines change the meaning of an
\setbox0=\hbox{{\rm +}} % exclamation mark in the same way, so that it
\signwidth=\wd0 % leaves as much space as a plus or minus sign.
\catcode`!=\active % These definitions will disappear at the end of
\def!{\kern\signwidth} % the \vbox.
\newdimen\pointwidth % These five lines change the meaning of a
\setbox0=\hbox{\rm .} % question mark in the same way, so that it
\pointwidth=\wd0 % leaves as much space as a period.
\catcode`?=\active % These definitions will disappear at the end of
\def?{\kern\pointwidth} % the \vbox.
\halign{\tabskip=0pt\hbox to 1.75in{#\leaderfil}\tabskip=4.em&
 #\hfil\tabskip=0.0pt\cr
\noalign{\doubleline}
\noalign{\vskip -3pt}
\omit\hfil Model\hfil& Section\cr
\noalign{\vskip 3pt\hrule\vskip 4pt} 
\LCDM& \cite{planck2014-a15}\cr
\noalign{\vskip 3pt\hrule\vskip 4pt} 
\omit{\bf Background parameterizations:}\hfil& \omit\cr 
${w}$&   \cite{planck2014-a15}\cr
$w_0$, $w_a$& Sect.~\ref{sec:w0wa}: Figs.~\ref{fig:w1D}, \ref{fig:w2D_wwa},
 \ref{fig:w_reconstructed}\cr
$w$ higher order expansion& Sect.~\ref{sec:w0wa}\cr
1-parameter $w(a)$& Sect.~\ref{GSw}: \Fig \ref{fig:gs}\cr
$w$ PCA& Sect.~\ref{PCA}: Fig.~\ref{fig:weights}\cr
$\epsilon_{\rm s}$, $\zeta_{\rm s}$, $\epsilon_\infty$&
 Sect.~\ref{subset:weaklypar}: Figs.~\ref{fig:epss}, \ref{fig:epsinf}\cr
Early DE& Sect.~\ref{subsub:ede}: Figs.~\ref{fig:jla}, \ref{fig:ede3}\cr
\noalign{\vskip 3pt\hrule\vskip 4pt} 
\omit{\bf Perturbation parameterizations:}\hfil& \omit\cr 
EFT exponential& Sect.~\ref{EFT}: \Fig{\ref{fig:EXPom}}\cr
EFT linear& Sect.~\ref{EFT}: \Fig{\ref{fig:linearEFT}}\cr
\omit $\mu, \eta$ scale-independent: \hfil& \omit\cr 
 \,\,\,\,	DE-related& Sect.~\ref{sec:genpar}: Figs.~\ref{fig:MGcls},
 \ref{fig:mueta}, \ref{fig:musigma}, \ref{fig:muetarecon}, \ref{fig:lenscomp}\cr
 \,\,\,\,	time related& Sect.~\ref{sec:genpar}: Figs.~\ref{fig:mueta},
 \ref{fig:muetarecon}\cr
$\mu, \eta$ scale-dependent:& \omit\cr 
 \,\,\,\,	DE-related& Sect.~\ref{sec:genpar}: Fig.~\ref{fig:mueta_scale}\cr
 \,\,\,\,	time related& Sect.~\ref{sec:genpar}\cr
\noalign{\vskip 3pt\hrule\vskip 4pt} 
\omit{\bf Other particular examples:}\hfil& \omit\cr 
DE sound speed and k-essence & Sect.~\ref{sec:kessence} \cr
Equation of state approach: & \cr  
 \,\,\,\, Lorentz-violating massive gravity & Sect.~\ref{sec:EOS}\cr  
 \,\,\,\, Generalized scalar fields & Sect.~\ref{sec:EOS}\cr  
$f(R)$& Sect.~\ref{sec:fr}: Figs.~{\ref{fig:logB0-tau}}, \ref{fig:logB0}\cr
Coupled DE& Sect.~\ref{sec:cq}: Figs.~\ref{fig:cq1D}, \ref{fig:cq2D_H0}\cr
\noalign{\vskip 3pt\hrule\vskip 4pt}
}}
\endPlancktable
\endgroup
\end{table*} 

\subsection{Combining data sets} 

We show for convenience in Table~\ref{tbl:runsTable} the schematic summary
of models. All models have been tested for
the combinations: \planckonly, \planckonly+BSH, \planckonly+WL,
\planckonly+BAO/RSD and \planckonly+WL+BAO/RSD.  Throughout the text, unless
other\-wise specified, \planckonly\ refers to the baseline \planckTT\
combination. The effects of CMB \lensing\ and \planckonly\ \TTTEEE\
polarization have been tested on all runs above and are, in parti\-cular, used
to constrain the amount of DE at early times.
For each of them we indicate the section in which the model is described
and the corresponding figures. 
In addition, all combinations in the table have been tested with and without
CMB lensing.  The impact of \Planck\ high-$\ell$ polarization has been tested
on all mo\-dels for the combination \Planck+BAO+SNe+$H_0$.

\section{Results} 
\label{sec:results}
We now proceed by illustrating in detail the models and parame\-terizations
described in \sect\ref{sec:models}, through presenting results for each of
them.  The structure of this section is as follows.  We start in
\sect{\ref{sec:bkg_res}} with smooth dark energy models that are effectively
parameterized by the expansion history of the Universe alone.
In \sect{\ref{results:modifiedgravity}} we study the constraints on the
presence of non-negligible dark energy perturbations, both in the context of
ge\-neral modified gravity models described through effective field theories
and with phenomenological parameterizations of the gravitational potentials
and their combinations, as illustrated in \sect{\ref{subsubsec:mgcamb}}.
The last part, \sect{\ref{sec:examplemodels}}, illustrates results for a range
of particular examples often considered in the literature.

\subsection{Background parameterizations} \label{sec:bkg_res}
In this section, we consider models where DE is a generic quintessence-like component
with equation of state $w \equiv p/\rho$, where $p$ and $\rho$ are the
spatially averaged (background) DE pressure and density.
Although it is important to include, as we do, DE perturbations, models in
this section have a sound speed that is equal to the
speed of light, which means that they are smooth on sub-horizon scales
(see \sect{\ref{sec:bgparam}} for more details).
We start with Taylor expansions and a principal component analysis of $w$
in a fluid formalism, then consider actual
quintessence mo\-dels parameterized through their potentials and finally study
the li\-mits that can be put on the abundance of
DE density at early times.  At the end of the sub-section we provide the
necessary information to compress the \planck\ CMB power spectrum 
into a 4-parameter Gaussian likelihood for applications where the full
likelihood is too unwieldy.

\subsubsection{Taylor expansions of $w$ and $w_0, w_a$ parameterization}
\label{sec:w0wa}
\begin{figure*}[!t]
\begin{center}
\includegraphics[width=1\textwidth]{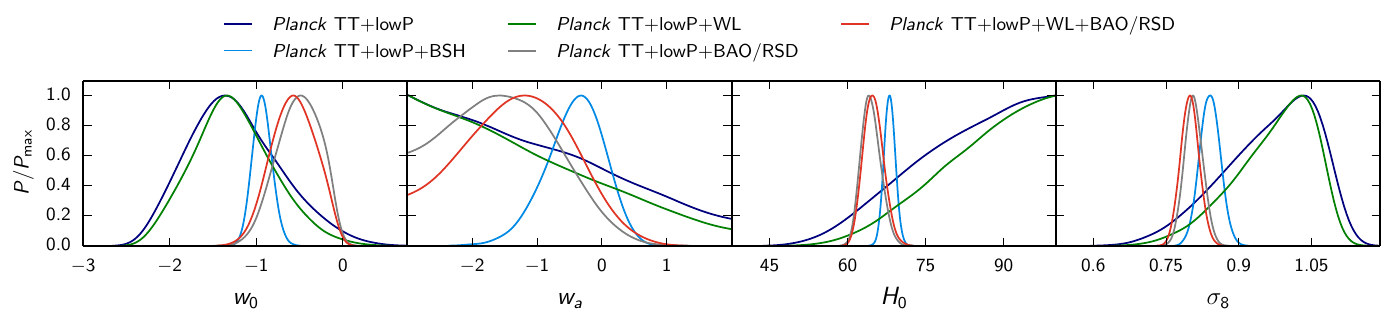}
\caption{Parameterization $\{w_0, w_a\}$ (see \sect{\ref{sec:w0wa}}).
Marginalized posterior distributions for $w_0$, $w_a$, $H_0$ and $\sigma_8$
for various data combinations.  The tightest constraints come from the
\planckTTonly+\lowTEB+BSH combination, which indeed tests background
observations, and is compatible with $\Lambda$CDM.}
\label{fig:w1D}
\end{center}
\end{figure*}
\begin{figure*}[!t]
\begin{center}
\includegraphics[height=21em]{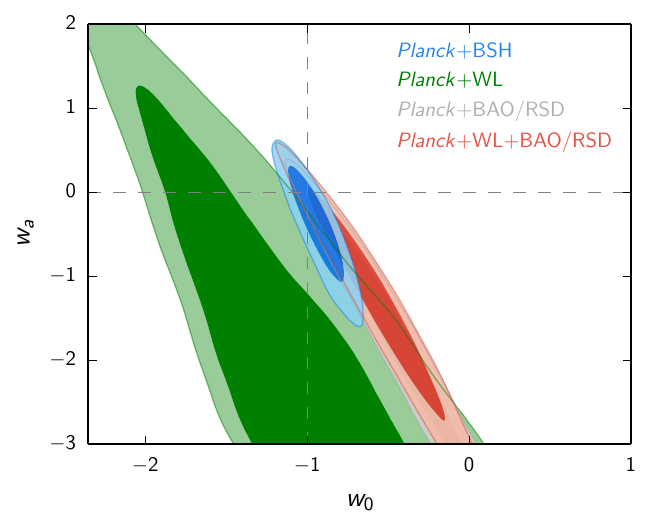}
\includegraphics[height=21em]{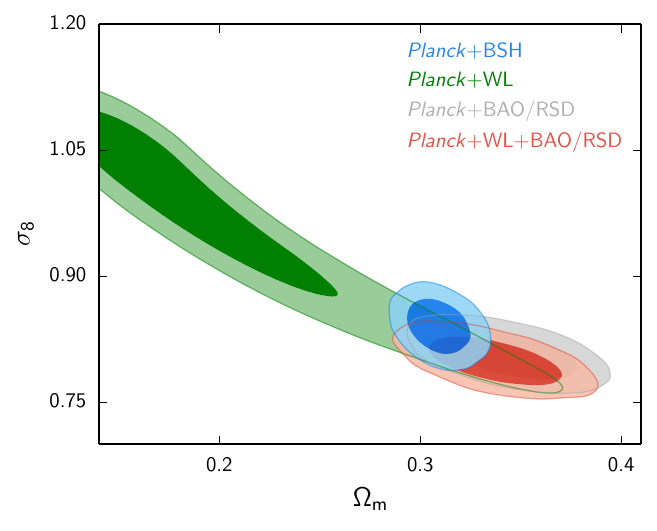}
\caption{Marginalized posterior distributions of the ($w_0, w_a$)
parameterization (see \sect{\ref{sec:w0wa}}) for various data combinations.
The best constraints come from the priority combination and are compatible
with $\rm{\Lambda}$CDM.  The dashed lines indicate the point in parameter
space $(-1,0)$ corresponding to the $\Lambda$CDM model.  CMB lensing and
polarization do not significantly change the constraints.
Here \planckonly\ indicates \planckTT.}
\label{fig:w2D_wwa}
\end{center}
\end{figure*}

If the dark energy is not a cosmological constant with $w=-1$ then there is no
reason why $w$ should remain constant. 
In order to test a time-varying equation of state, we expand $w(a)$ in a
Taylor series. The first order corresponds to the $\{w_0,w_a\}$ case, also discussed in \cite{planck2014-a15}: 
\begin{equation} \label{w0wapar}
w(a) = w_0 + (1-a) w_a \, .
\end{equation} 
We use the parameterized post-Friedmann (PPF) model of \cite{Hu:2007pj}
and \cite{Fang:2008sn} to allow for values $w < -1$ (note that there is another PPF formalism discussed in \cite{2014PhRvD..89b4026B}).  Marginalized posterior
distributions for $w_0$, $w_a$, $H_0$ and $\sigma_8$ are shown in
Fig.~\ref{fig:w1D} and the corresponding 2D contours can be found in
Fig.~\ref{fig:w2D_wwa} for $w_a$ vs $w_0$ and for $\sigma_8$ vs $\Omm$.
Results from \planckTTonly +\lowTEB+BSH data are shown in blue and corresponds
to the combination we consider the most secure, which in this case also gives
the strongest constraints.  This is expected, since the BAO and SNe data
included in the BSH combination provide the best constraints on the background
expansion rate.
\begin{figure}[!bH]
\begin{center}
\includegraphics[height=20em]{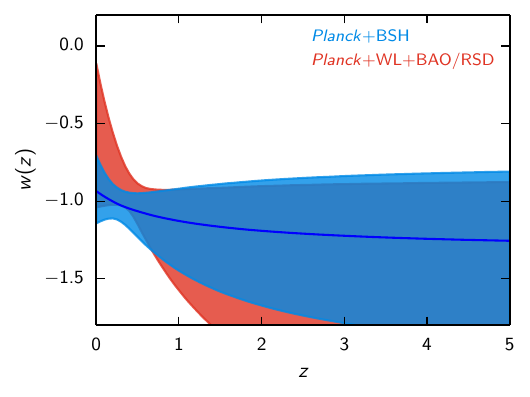}
\caption{Reconstructed equation of state $w(z)$ as a function of redshift
(see \sect{\ref{sec:w0wa}}), when assuming a Taylor expansion of $w(z)$ to
first-order ($N = 1$ in Eq.~\ref{eq:w_expansion}), for different combinations
of the data sets.  The coloured areas show the regions which contain 95\,\%
of the models.  The central blue line is the median line for \planckTT+BSH.
Here \planckonly\ indicates \planckTT.}
\label{fig:w_reconstructed}
\end{center}
\end{figure}
Results for weak lensing (WL) and redshift space distortions (RSD) are also
shown, both separately and combined. 
The constraints from these probes are weaker, since we are considering a
smooth dark energy model where the perturbations are suppressed on small
scales.  While the WL data appear to be in slight tension with \lcdm,
according to the green contours shown in \fig{\ref{fig:w2D_wwa}},
the difference in total $\chi^2$ between the best-fit in the $\{w_0,w_a\}$
model and in \lcdm\ for \planckTT +WL is $\Delta \chi^2 = -5.6$, which is not
very significant for 2 extra parameters (for normal errors a $2\,\sigma$
deviation corresponds to a $\chi^2$ absolute difference of 6.2).
The WL contributes a $\Delta\chi^2$ of $-2.0$ and the
$\Delta\chi^2_{\rm{CMB}} = -3.3$ (virtually the same as when using \planckTT\ alone,
for which $\Delta\chi^2_{\rm{CMB}} = -3.2$, which seems to indicate that WL is not in tension with \planckTT\ within a ($w_0, w_a$) cosmology).  
However, as also discussed in \cite{planck2014-a15}, these data combinations
prefer very high values of $H_0$, which is visible also in the third panel
of \fig{\ref{fig:w1D}}.
The combination \planckTT+BSH, on the other hand, is closer to \lcdm, with a
total $\chi^2$ difference between ($w_0, w_a$) and \lcdm\  of only $-0.8$.  
We also show in Fig.~\ref{fig:w_reconstructed} the equation of state
reconstructed as a function of redshift from the linear expansion in the
scale factor $a$ for different combinations of data.

One might wonder whether it is reasonable to stop at first order in $w(a)$.
We have therefore tested a generic expansion in powers of the scale factor
up to order $N$:
\begin{equation} 
\label{eq:w_expansion}
w(a) = w_0 + \sum_{i=1}^N (1-a)^i w_i \, . 
\end{equation}
We find that all parameters are very stable when allowing higher order
polynomials; the $w_i$ parameters are weakly constrained and going from $N=1$
(the linear case) to $N = 2$ (quadratic case) to $N=3$ (cubic expansion) does not improve the goodness of
fit and stays compatible with \LCDM, which indicates that a linear parameterization is sufficient.

\subsubsection{1-parameter varying $w$} \label{GSw}
\begin{figure}[!t]
\begin{center}
\includegraphics[height=21em]{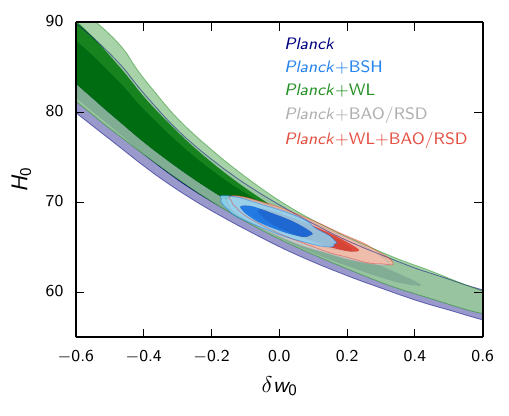}
\caption{Marginalized posterior contours in the $h$--$\delta w_0$ plane are
shown for 1-parameter varying $w$ models (see \sect{\ref{GSw}}) for
different data combinations.  Here \planckonly\ indicates \planckTT.}
\label{fig:gs}
\end{center}
\end{figure}

A simple example of a varying $w$ model that can be written in terms of one
extra parameter only (instead of $w_0, w_a$) was proposed in
\cite{2011MNRAS.416..907G, 2014MNRAS.438.1948S}, motivated in connection to a DE minimally-coupled
scalar field, slowly rolling down a potential $\frac{1}{2}m^2\phi^2$,
analogous to the one predicted in chaotic inflation
\citep{1983PhLB..129..177L}.
More generally, one can fully characte\-rize the background by expanding a
varying equation of state
$w(z)\equiv-1+\delta w(z)\approx-1+\delta w_0\times H_0^2/H^2(z)$, where:
\begin{equation}
\frac{H^2(z)}{H_0^2} \approx \Omm(1+z)^3 + \Omde
 \left[ \frac{(1+z)^3}{\Omm(1+z)^3+\Omde} \right]^{\delta w_0/\Omde} \, , 
\end{equation}
at first order in $\delta w_0$, which is then the only extra parameter.
Marginalized posterior contours in the plane $h$--$\delta w_0$ are shown in
\fig{\ref{fig:gs}}.
The tightest constraints come from the combination \planckTT+\lensing+BSH
that gives $\delta w_0 = -0.008\pm 0.068$ at 68\,\% confidence level,
which slightly improves constraints found by \cite{2014arXiv1411.1074A}. 

\subsubsection{Principal Component Analysis on \textit{w(z)}} \label{PCA}

A complementary way to measure the evolution of the equation of state,
which is better able to model rapid variations, proceeds
by choosing $w$ in $N$ fixed bins in redshift and by performing
a principal component analysis to uncorrelate
the constraints. We consider $N = 4$ different bins in $z$ and
assume that $w$ has a constant value $p_i$ in each of them.  We then smooth
the transition from one bin to the other such that:
\begin{eqnarray}
w(z) &=& p_{i-1} + \Delta w \left(\tanh\left[\frac{z-z_i}{s}\right] + 1\right)
 \ \textrm{for} \, z < z_i \, , \ \textrm{i $\epsilon$ \{1, 4\}},
\end{eqnarray}
with $\Delta w \equiv (p_i - p_{i-1})/{2}$, a smoothing scale $s = 0.02$, and a
binning $z_i = (0, 0.2, 0.4, 0.6, 1.8)$.  We have tested also a larger number
of bins (up to $N=18$) and have found no improvement in the goodness of fit.

\begin{figure}[!t]
\begin{center}
\includegraphics[width=0.45\textwidth]{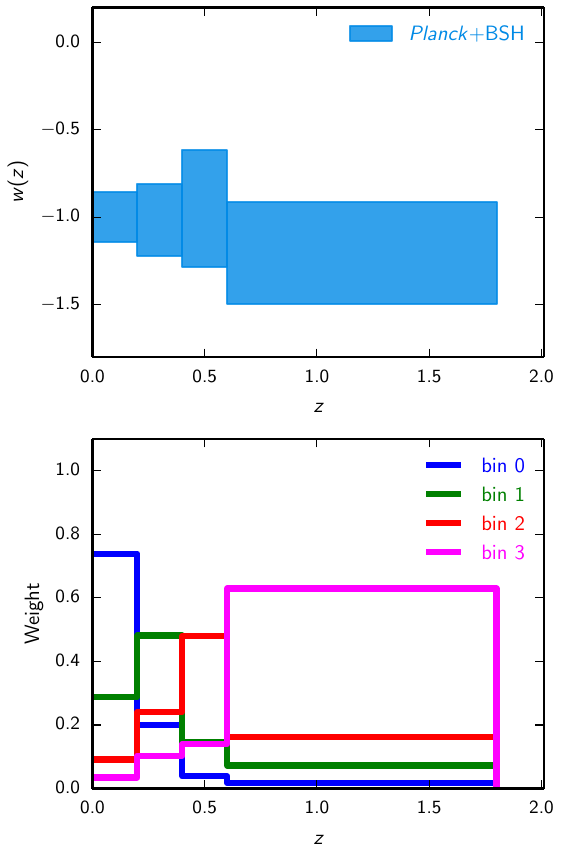}
\caption{PCA analysis constraints (described in \sect{\ref{PCA}}).
The top panel shows the reconstructed equation of state $w(z)$ after the PCA
analysis. Vertical error bars correspond to mean and standard deviations of
the $q$ vector parameters, while horizontal error bars are the amplitude of
the original binning. The bins are not exactly independent but are rather smeared out as illustrated in the bottom panel. The bottom panel shows the PCA corresponding weights
on $w(z)$ as a function of redshift for the combination \planckTT+BSH. In other words, error bars in the top panel correspond therefore to the errors in the $q$ parameters, which are linear combinations of the $p$ parameters, i.e. a smeared out distribution with weights shown in the lower panel.}
\label{fig:weights}
\end{center}
\end{figure}

The constraints on the vector $\vec{p}_i (i = 1, \dots, N)$ of values that
$w(z)$ can assume in each bin is difficult to interpret, due to the
correlations between bins.  To uncorrelate the bins, we perform a principal
component analysis
\citep{huterer_starkman_2003, huterer_cooray_2005, Said_etal_2013}.
We first run {\tt COSMOMC} \citep{Lewis:2002ah} on the original binning values
$\vec{p}_i$; then extract the covariance matrix that refers to the parameters
we want to constrain:
\begin{equation}
  C \equiv \left\langle pp^{\rm T}\right\rangle -
  \langle p\rangle\langle p^{\rm T}\rangle ,
\end{equation} 
where $p$ is the vector of parameters $p_i$ and $p^{\rm T}$ is its transpose.
We calculate the Fisher matrix, $F = C^{-1}$, and diagonalize it,
$F = O^{\rm T} \Lambda O$, where $\Lambda$ is diagonal and $O$ is the
orthogonal matrix whose columns are the eigenvectors of the Fisher matrix. 
We then define $\widetilde{W} = O^{\rm T} \Lambda^{1/2} O$
\citep[e.g.,][]{huterer_cooray_2005}
and normalize this such that its rows sum up to unity; this matrix can be used
to find the new vector $q = \widetilde{W} p$ of uncorrelated parameters that
describe $w(z)$.  This choice of $\tilde{W}$ has been shown to be convenient,
since most of the weights (i.e., the rows of $\widetilde{W}$) are found to be
positive and fairly well localized in redshift.  In \fig{\ref{fig:weights}}
(lower panel) we show the weights for each bin as a function of redshift.
Because they overlap only partially, we can assume the binning to be the same
as the original one and attach to each of them error bars corresponding to the
mean and standard deviations of the $q$ values.  The result is shown in
\fig{\ref{fig:weights}}, top panel.  The equation of state is compatible with
the $\Lambda$CDM value $w = -1$.  Note however that this plot contains more
information than a Taylor expansion to first order. 

\subsubsection{Parameterization for a weakly-coupled canonical scalar field.}
\label{subset:weaklypar}
We continue our investigation of background parameterizations by considering a
slowly rolling scalar field.  In this case, as in inflation, we can avoid writing down an
explicit potential $V(\phi)$ and instead parameterize $w(a)$ at late times,
in the presence of matter, as \citep{huang_bond_kofman:2011}
\begin{equation}
w = -1 + \frac{2}{3} \epss F^2\left(\frac{a}{a_{\rm de}}\right) \, ,
 \label{eq:w_qcdm}
\end{equation}
where the ``slope parameter'' $\epss$ is defined as:
\begin{equation} \label{def_eps_s_quint}
\epss \equiv \left.\epsv\right\vert_{a=a_{\rm de}} \, ,
\end{equation}
with $\epsv \equiv (\frac{d \ln V}{d\phi})^2\redmpl^2/2$
being a function of the slope of the potential.  Here
$\redmpl \equiv {1}/{\sqrt{8\pi G}}$ is the reduced Planck mass
and $a_{\rm de}$ is the scale factor where the total matter and DE densities
are equal.  The function $F(x)$ in Eq.~(\ref{eq:w_qcdm}) is defined as:
\begin{equation}
F(x) \equiv \frac{\sqrt{1+x^3}}{x^{3/2}}
 - \frac{\ln{\left(x^{3/2}+\sqrt{1+x^3}\right)}}{x^3}\, .
\end{equation}
\eq{\ref{eq:w_qcdm}} parameterizes $w(a)$ with one parameter $\epss$,
while $a_{\rm de}$ depends on $\Omm$ and $\epss$ and can be derived using an
approxi\-mated fitting formula that facilitates numerical computation
\citep{huang_bond_kofman:2011}.  
Positive (negative) values of $\epss$ correspond to quintessence (phantom)
models.

\eq{\ref{eq:w_qcdm}} is only valid for late-Universe slow-roll
($\epsv \lesssim 1$ and $\eta_V \equiv \redmpl^2 {V''}/{V}\ll 1$) or the
moderate-roll ($\epsv\lesssim 1$ and $\eta_V \lesssim 1$) regime.
For quintessence models, where the scalar field rolls down from a very steep
potential, at early times $\epsv(a) \gg 1$, however  the fractional density
$\Omega_\phi(a)\rightarrow 0$ and the combination $\epsv(a)\Omega_\phi(a)$
aprroaches a constant, defined to be a second parameter
$\epsilon_{\infty}\equiv \lim_{a\rightarrow 0} \epsv(a)\Omega_\phi(a) $.

One could also add a third parameter $\zetas$ to capture the time-dependence
of $\epsv$ via corrections to the functional dependence of $w(a)$ at late time.
This parameter is defined as the relative difference of
$d\sqrt{\epsv\Omega_\phi}/d y$ at $a = a_{\rm de}$ and at $a \rightarrow 0$,
where $y \equiv (a/a_{\rm de})^{3/2}/\sqrt{1+(a/a_{\rm de})^3}$.
If $\epsilon_\infty \ll 1$, $\zetas$ is proportional to the second derivative
of $\ln V(\phi)$, but for large $\epsilon_\infty$, the dependence is more
complicated \citep{huang_bond_kofman:2011}.
In other words, while $\epss$ is sensitive to the late time evolution of
$1+w(a)$, $\epsilon_\infty$ captures its early time behaviour.
Quintessence/phantom models can be mapped into $\epss$--$\epsilon_\infty$
space and the classification can be further refined with $\zetas$.
For $\Lambda$CDM, all three parameters are zero.

\begin{figure}
\includegraphics[height=20em]{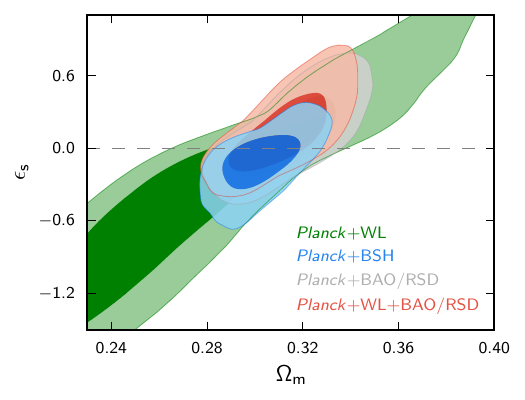}
\caption{Marginalized posterior distributions showing 68\,\%  and 95\,\% C.L.
constraints on $\Omm$ and $\epss$ for scalar field models
(see \sect{\ref{subset:weaklypar}}).  The dashed line for $\epss = 0$ is
the $\Lambda$CDM model.  Here \planckonly\ indicates \planckTT.}
\label{fig:epss}
\end{figure}

\begin{figure}
  \centering
\includegraphics[  height=20em]{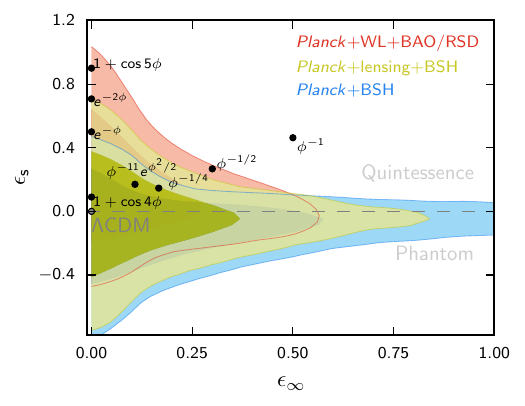} \,\,\,\,\,\,\,\,\,\,
\caption{Marginalized posterior distributions at 68\,\% C.L. and 95\,\% C.L. in
the parameter space of $\epss$ and $\epsilon_\infty$ for scalar field modes
(see \sect{\ref{subset:weaklypar}}).  We have computed $\epss$ and
$\epsilon_\infty$ for various quintessence potentials $V(\phi)$, with the
functional forms of $V(\phi)$ labelled on the figure.  The field $\phi$ is in
reduced Planck mass $\redmpl$ units.  The normalization of $V(\phi)$ is
computed using $\Omm = 0.3$.  Here \planckonly\ indicates \planckTT.
\label{fig:epsinf}}
\end{figure}

In \fig{\ref{fig:epss}} we show the marginalized posterior distributions at
68.3\,\% and 95.4\,\% confidence levels in the parameter space $\epss$--$\Omm$,
marginalizing over the other parameters. 
In \fig{\ref{fig:epsinf}} we show the current constraints on quintessence
models projected in $\epss$--$\epsilon_\infty$ space.  The constraints are
obtained by marginalizing over all other cosmological parameters.  The models
here include exponentials $V=V_0\exp(-\lambda\phi/\redmpl)$
\citep{wetterich_1988}, cosines from pseudo-Nambu Goldstone bosons
(pnGB) $V=V_0[1+\cos({\lambda\phi/\redmpl})]$
\citep{frieman_hill_stebbins_waga:1995, kaloper_sorbo_2006}, power laws
$V=V_0(\phi/\redmpl)^{-n}$ \citep{ratra_peebles_1988}, and models motivated
by supergravity (SUGRA)
$V=V_0(\phi/\redmpl)^{-\alpha} \exp{[(\phi/\redmpl)^2]}$
\citep{brax_martin_1999}.  The model projection is done with a fiducial
$\Omm = 0.3$ cosmology.  We have verified that variations of 1\,\% compared
to the fiducial $\Omm$ lead to negligible changes in the constraints.

Mean values and uncertainties for a selection of cosmological parameters are
shown in Table~\ref{tbl:weaklypar}, for both the 1-parameter case
(i.e., $\epss$ only, with $\epsilon_\infty = 0$ and $\zetas = 0$, describing
``thawing'' quintessence/phantom models, where $\dot\phi = 0$ in the early
Universe) and the 3-parameter case (general quintessence/phantom models where
an early-Universe fast-rolling phase is allowed).  When we vary the data sets
and theoretical prior (between the 1-parameter and 3-parameter cases), the
results are all compatible with $\Lambda$CDM and mutually compatible with each
other.  Because $\epss$ and $\epsilon_\infty$ are correlated, caution has to
be taken when looking at the marginalized constraints in the table.  For
instance, the constraint on $\epss$ is tighter for the 3-parameter case,
because in this case flatter potentials are preferred in the late Universe
in order to slow-down larger $\dot{\phi}$ from the early Universe.
A better view of the mutual consistency can be obtained from
\fig{\ref{fig:epsinf}}.  We find that the addition of polarization data does
not have a large impact on these DE parameters.  Adding polarization data to
\Planck+BSH shifts the mean of $\epss$ by $-1/6\,\sigma$ and reduces the
uncertainty of $\epss$ by 20\,\%, while the 95\,\% upper bound on
$\epsilon_\infty$ remains unchanged.

\begin{table*}
\footnotesize 
\nointerlineskip
\setbox\tablebox=\vbox{
\newdimen\digitwidth % These five lines change what an asterisk
\setbox0=\hbox{\rm 0} % means to TeX. Instead of meaning
\digitwidth=\wd0 % "print an '*' here", it now means "leave
\catcode`*=\active % as much blank space as a single number
\def*{\kern\digitwidth} % takes up".
\newdimen\signwidth % These five lines change the meaning of an
\setbox0=\hbox{{\rm +}} % exclamation mark in the same way, so that it
\signwidth=\wd0 % leaves as much space as a plus or minus sign.
\catcode`!=\active % These definitions will disappear at the end of
\def!{\kern\signwidth} % the \vbox.
\newdimen\pointwidth % These five lines change the meaning of a
\setbox0=\hbox{\rm .} % question mark in the same way, so that it
\pointwidth=\wd0 % leaves as much space as a period.
\catcode`?=\active % These definitions will disappear at the end of
\def?{\kern\pointwidth} % the \vbox.
\halign{\hbox to 1.20in{#\leaderfil}\tabskip=2.0em&
\hfil#\hfil\tabskip=2.0em&
\hfil#\hfil\tabskip=2.0em&
\hfil#\hfil\tabskip=2.0em&
\hfil#\hfil\tabskip=0.0em\cr
\noalign{\doubleline}
\omit Parameter\hfil& \planckonly+BSH (1-param.)& \planckonly+BSH (3-param.)&
 \planckonly+WL+BAO/RSD& \planckonly+lensing+BSH\cr
\noalign{\vskip 3pt\hrule\vskip 4pt}  
$\epss$& $-0.08^{+0.32}_{-0.32}$&  $-0.11^{+0.16}_{-0.12}$&
 $0.14^{+0.17}_{-0.25}$ & $-0.03^{+0.16}_{-0.17}$\cr
$\epsilon_\infty$& fixed $=0$& $\le 0.76$ (95\,\% CL)&
 $\le 0.38$ (95\,\% CL)& $\le 0.52$ (95\,\% CL)\cr
$\zetas$& fixed $=0$& not constrained& not constrained& not constrained\cr
\noalign{\vskip 3pt\hrule\vskip 4pt} 
}}
\endtable
\caption{Marginalized mean values and 68\,\% CL errors for a selection of
cosmological parameters for the weakly-coupled scalar field parameterization
described in the text (\sect{\ref{subset:weaklypar}}).  Here ``1-param'' in the first
column refers to the priors $\epsilon_\infty = 0$ and $\zetas = 0$
(slow- or moderate-roll ``thawing'' models). \label{tbl:weaklypar} }
\end{table*}

\subsubsection{Dark energy density at early times} \label{subsub:ede}
\begin{figure}
\includegraphics[width=0.45\textwidth]{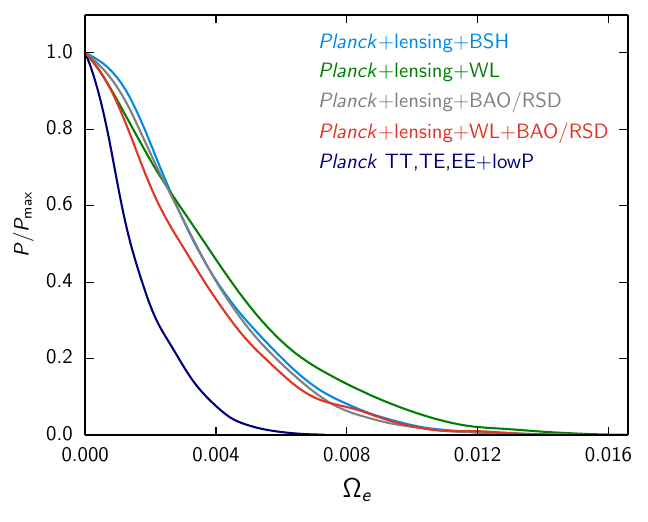}
\caption{Marginalized posterior distributions for $\Omega_{\rm{e}}$ for the
early DE parameterization in \eqn{\ref{ede1_par}} and for different
combinations of data (see \sect{\ref{subsub:ede}}).  Here \planckonly\
indicates \planckTT. \label{fig:jla}}
\end{figure}

Quintessence models can be divided into two classes, namely cosmologies with
or without DE at early times.  Although the equation of state and the DE
density are related to each other, it is often convenient to think directly
in terms of DE density rather than the equation of state.  In this section we
provide a more direct estimate of how much DE is allowed by the data as a
function of time. 
A key parameter for this purpose is $\Omega_{\rm{e}}$, which measures the
amount of DE present at early times (``early dark energy,'' EDE)
\citep{Wetterich:2004pv}. 
Early DE parameterizations encompass features of a large class of dynamical
DE mo\-dels.  The amount of early DE influences CMB peaks
and can be strongly constrained when including small-scale measurements and
CMB lensing.
Assuming a constant fraction of $\Omega_{\rm{e}}$ until recent times
\citep{doran_robbers_2006}, the DE density is parameterized as:
\begin{equation} \label{ede1_par}
\Omde(a) = \frac{\Omde^0 - \Omega_{\rm{e}}(1-a^{-3 w_0})}{\Omde^0
 + \Omm^0 a^{3w_0}} + \Omega_{\rm{e}}(1-a^{-3w_0}) \, .
\end{equation}
This expression requires two parameters in addition to those of
$\rm{\Lambda}$CDM, namely $\Omega_{\rm{e}}$ and $w_0$,
while $\Omm^0=1-\Omde^0$ is the present matter abundance.
The strongest constraints to date were discussed in \citep{planck2013-p11},
finding $\Omega_{\rm{e}} < 0.010$ at 95\,\% CL using \Planck\ combined with
WMAP polarization.  Here we update the analysis using \Planck\ 2015 data.
In \fig{\ref{fig:jla}} we show marginalized posterior distributions for
$\Omega_{\rm{e}}$ for different combination of data sets; the corresponding marginalized limits are shown in Table~{\ref{tab:ede1}}, improving substantially current constraints, especially when the \planckall\ polarization is included, leading to $\Omega_{\rm{e}} < 0.0036$ at $95\%$ confidence level for \planckall +BSH.

\begin{table*}
\footnotesize 
\nointerlineskip
\setbox\tablebox=\vbox{
\newdimen\digitwidth % These five lines change what an asterisk
\setbox0=\hbox{\rm 0} % means to TeX. Instead of meaning
\digitwidth=\wd0 % "print an '*' here", it now means "leave
\catcode`*=\active % as much blank space as a single number
\def*{\kern\digitwidth} % takes up".
\newdimen\signwidth % These five lines change the meaning of an
\setbox0=\hbox{{\rm +}} % exclamation mark in the same way, so that it
\signwidth=\wd0 % leaves as much space as a plus or minus sign.
\catcode`!=\active % These definitions will disappear at the end of
\def!{\kern\signwidth} % the \vbox.
\newdimen\pointwidth % These five lines change the meaning of a
\setbox0=\hbox{\rm .} % question mark in the same way, so that it
\pointwidth=\wd0 % leaves as much space as a period.
\catcode`?=\active % These definitions will disappear at the end of
\def?{\kern\pointwidth} % the \vbox.
\halign{\hbox to 1.0in{#\leaderfil}\tabskip=0.0em&
\hfil#\hfil\tabskip=1.em&
\hfil#\hfil\tabskip=1.em&
\hfil#\hfil\tabskip=1.em&
\hfil#\hfil\tabskip=1.em&
\hfil#\hfil\tabskip=0.em&
\hfil#\hfil\tabskip=0.em\cr
\noalign{\doubleline}
\omit Parameter\hfil& \planckTT& \planckTT& \planckTT& \planckTT&
 \planckall\cr
\omit& +\lensing+BSH& +\lensing+WL& +\lensing+BAO/RSD& +\lensing+WL+BAO/RSD&
 + BSH\cr
\noalign{\vskip 3pt\hrule\vskip 4pt} 
$\Omega_{\rm{e}}$& $<0.0071$& $<0.0087$& $<0.0070$& $<0.0070$& $<0.0036$\cr
$w_0$& $<-0.93$& $<-0.76$& $<-0.90$& $<-0.90$& $<-0.94$\cr
\noalign{\vskip 3pt\hrule\vskip 4pt}
}}
\endtable
\caption{Marginalized 95\,\% limits on $\Omega_{\rm{e}}$ and $w_0$ for the
early DE parameterization in \eq{\ref{ede1_par}} and different combinations of data (see
\sect{\ref{subsub:ede}}).  Including high-$\ell$ polarization significantly
tightens the bounds.}
\label{tab:ede1}
\end{table*} 

\begin{figure}
\includegraphics[width=0.45\textwidth]{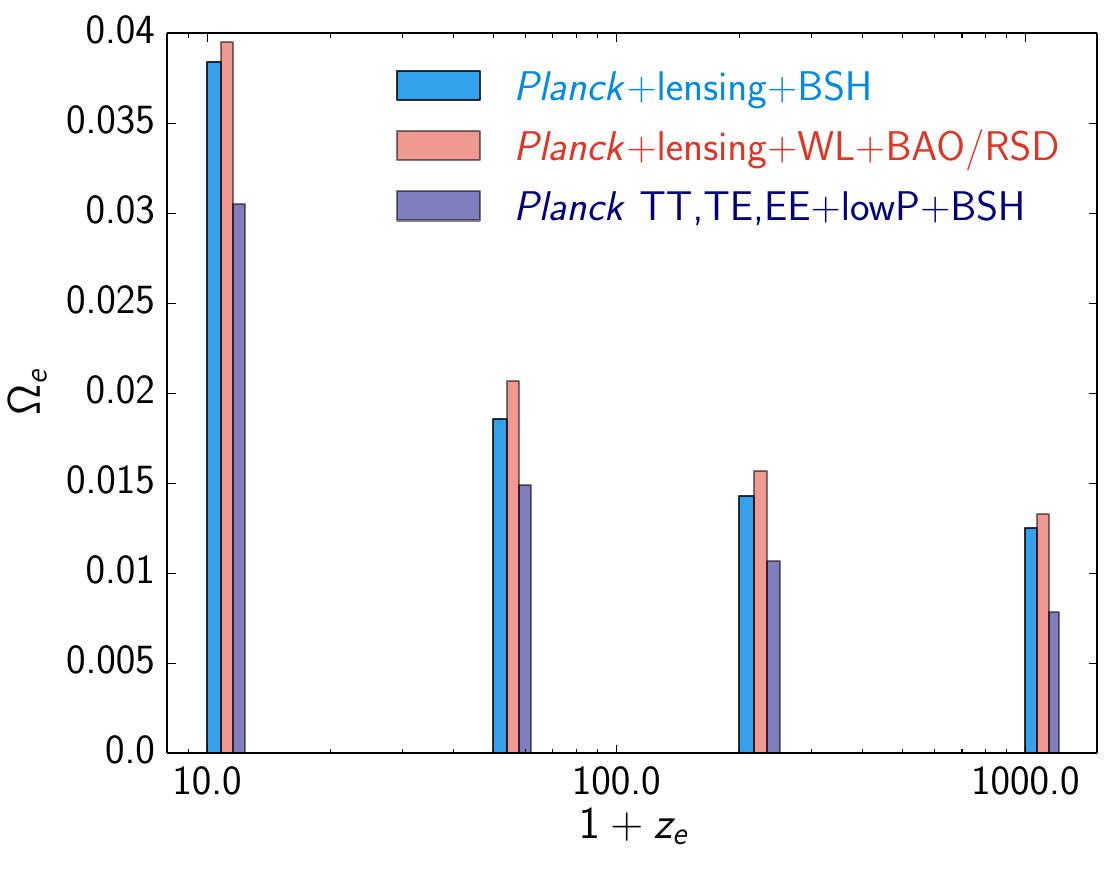}
\caption{Amount of DE at early times $\Omega_e$ as a function of the redshift
$z_e$ after which early DE is non-negligible (see \eq{\ref{param:ede3}}, \sect{\ref{subsub:ede}})
for different combinations of data sets.  The heights of the columns give the limit at
95\,\% CL on $\Omega_e$, as obtained from Monte Carlo runs for the values
$z_e = 10$, $50$, $200$ and $1000$.  The width of the columns has no physical meaning and is just due to plotting purposes.  Here \planckonly\ indicates \planckTT. \label{fig:ede3}}
\end{figure}

As first shown in \cite{pettorino_amendola_wetterich_2013}, bounds on
$\Omega_{\rm{e}}$ can be weaker if DE is present only over a limited range
of redshifts.  In particular, EDE reduces structure growth in the period after
last scattering, implying a smaller number of
clusters as compared to $\Lambda$CDM, and therefore a weaker lensing potential
to influence the anisotropies at high $\ell$.  It is possible to isolate this
effect by switching on EDE only after last scattering, at a scale factor
$a_{\rm{e}}$ (or equivalently for redshifts smaller than $z_{\rm{e}}$).
Here we adopt the parameterization ``EDE3'' proposed in
\cite{pettorino_amendola_wetterich_2013} to which we refer for more details:

\begin{equation} \label{param:ede3}
\Omde(a) =
\begin{cases}
 \frac{\displaystyle \Omega_{\rm{de}0}}
  {\displaystyle \Omega_{\rm{de0}}+\Omega_{\rm{m0}}
  a^{-3}+\Omega_{\rm{r0}} a^{-4}} &\mbox{for } a \le a_{\rm{e}} \,;\\[10pt]
 \Omega_{\rm{e}} &\mbox{for } a_{\rm{e}} < a < a_{\rm{c}} \,:\\[5pt]
 \frac{\displaystyle \Omega_{\rm{de0}}}
  {\displaystyle \Omega_{\rm{de0}}+\Omega_{\rm{m0}}
  a^{-3}+\Omega_{\rm{r0}} a^{-4}} &\mbox{for } a>a_{\rm{c}} \,.
\end{cases}
\end{equation}
In this case, early dark energy is present in the time interval
$a_{\rm{e}} < a < a_{\rm{c}}$,
 while outside this interval it behaves as in \lcdm, including the radiation contribution, unlike in \eq{\ref{ede1_par}}.
During that interval in time, there is a non-negligible EDE contribution,
parameterized by $\Omega_{\rm{e}}$.
The constant $a_{\rm{c}}$ is fixed by continuity, so that the parameters
\{$\Omega_{\rm{e}}, a_{\rm{e}}$\} fully determine how much EDE
there was and how long its presence lasted. 
We choose four fixed values of $a_{\rm{e}}$ corresponding to $z_{\rm{e}} = 10$,
$50$, $200$ and $1000$ and include $\Omega_{\rm{e}}$ as a free parameter in
MCMC runs for each value of $a_{\rm{e}}$.
Results are shown in \fig{\ref{fig:ede3}} where we plot
$\Omega_{\rm{e}}$ as a function of the redshift $z_{\rm{e}}$ at which DE starts
to be non-negligible.  The smaller the value
of $z_{\rm{e}}$, the weaker are the constraints, though still very tight, with
$\Omega_{\rm{e}} \la 2\,\% \,\,(95\,\%$ CL) for $z_{\rm{e}} \approx 50$.

\subsubsection{Compressed likelihood\label{sec:compress}}
Before concluding the set of results on background parameterizations, we
discuss here how to reduce the full likelihood information to few parameters.
As discussed for example in \cite{Kosowsky:2002zt} and \cite{Wang:2007mza},
it is possible to compress a large part of the information contained
in the CMB power spectrum into just a few numbers\footnote{There are also alternative approaches that compress the power spectra directly, like e.g.\ PICO \citep{Fendt:2006uh}.}: 
here we use specifically the CMB shift
parameter $R$ \citep{Efstathiou:1998xx}, the
angular scale of the sound horizon at last scattering $\ell_{\rm A}$
(or equivalently $\theta_\ast$), as well the baryon density $\omega_{\rm b}$
and the scalar spectral index $n_{\rm s}$.
The first two quantities are defined as
\begin{equation}
R \equiv \sqrt{\Omm H_0^2} \, D_{\rm A}(z_\ast)/c \,\,\,\, , \,\,\,\, \
 \ell_{\rm A} \equiv \pi D_{\rm A}(z_\ast)/r_{\rm s}(z_\ast)
 = \pi/\theta_\ast \, ,
\end{equation}
where $D_{\rm A}(z)$ is the comoving angular diameter distance to redshift
$z$, $z_\ast$ is the redshift for which the optical depth is unity and
$r_{\rm s}(z_\ast)=r_\ast$ is the comoving size of the sound horizon at
$z_\ast$.  These numbers are effectively observables and
they apply to models with either non-zero curvature or a smooth DE component \citep{Mukherjee:2008kd}.
It should be noted, however, that the constraints on these quantities,
especially on $R$, are sensitive to changes in the growth of perturbations.
This can be seen easi\-ly with the help of the
``dark degeneracy'' \citep{Kunz:2007rk}, i.e., the possibility to absorb part
of the dark matter into the dark energy, which changes $\Omm$ without
affecting observables.  For this reason the compressed likelihood presented
here cannot be used for models with low sound speed or modifications of gravity (and is therefore located at the end of this ``background'' section).

\begin{table*}[!htb]
\footnotesize 
\nointerlineskip
\setbox\tablebox=\vbox{
\newdimen\digitwidth % These five lines change what an asterisk
\setbox0=\hbox{\rm 0} % means to TeX. Instead of meaning
\digitwidth=\wd0 % "print an '*' here", it now means "leave
\catcode`*=\active % as much blank space as a single number
\def*{\kern\digitwidth} % takes up".
\newdimen\signwidth % These five lines change the meaning of an
\setbox0=\hbox{{$-$}} % exclamation mark in the same way, so that it
\signwidth=\wd0 % leaves as much space as a plus or minus sign.
\catcode`!=\active % These definitions will disappear at the end of
\def!{\kern\signwidth} % the \vbox.
\newdimen\pointwidth % These five lines change the meaning of a
\setbox0=\hbox{\rm .} % question mark in the same way, so that it
\pointwidth=\wd0 % leaves as much space as a period.
\catcode`?=\active % These definitions will disappear at the end of
\def?{\kern\pointwidth} % the \vbox.
\halign{\hbox to 1.50in{#\leaderfil}\tabskip=0.0em&
\hfil#\hfil\tabskip=4.0em&
\hfil#\hfil\tabskip=2.0em&
\hfil#\hfil\tabskip=2.0em&
\hfil#\hfil\tabskip=2.0em&
\hfil#\hfil\tabskip=0.0em\cr
\noalign{\doubleline}
\noalign{\vskip -3pt} 
\omit Smooth DE models\hfil& \planckTT&  $R$&  $\ell_{\rm A}$&
 $\Omega_{\rm b}h^2$& $n_{\rm s}$\cr
\noalign{\vskip 3pt\hrule\vskip 4pt} 
$R$&      $***1.7488*\pm0.0074$& $!1.0*$& $!0.54$& $-0.63$& $-0.86$\cr
$\ell_{\rm A}$& $301.76**\pm0.14$& $!0.54$& $!1.0*$& $-0.43$& $-0.48$\cr
$\Omega_{\rm b}h^2$ &$****0.02228\pm0.00023$& $-0.63$& $-0.43$& $!1.0*$&
 $!0.58$\cr
$n_{\rm s}$& $***0.9660*\pm0.0061$& $-0.86$& $-0.48$& $!0.58$& $!1.0*$\cr
\noalign{\vskip 3pt\doubleline}
\noalign{\vskip -3pt} 
\omit Marginalized over $A_{\rm L}$\hfil& \planckTT&  $R$& $\ell_{\rm A}$&
 $\Omega_{\rm b}h^2$& $n_{\rm s}$\cr
\noalign{\vskip 3pt\hrule\vskip 4pt}
$R$& $ ***1.7382*\pm0.0088$&
 $!1.0*$& $!0.64$& $-0.75$& $-0.89$\cr
$\ell_{\rm A}$& $301.63**\pm0.15$&
 $!0.64$& $!1.0*$& $-0.55$& $-0.57$\cr
$\Omega_{\rm b}h^2$& $****0.02262\pm0.00029$&
 $-0.75$& $-0.55$& $!1.0*$& $!0.71$\cr
$n_{\rm s}$& $***0.9741*\pm0.0072$&
 $-0.89$& $-0.57$& $!0.71$& $!1.0*$\cr
\noalign{\vskip 3pt\hrule\vskip 4pt}
}}
\endtable
\caption{Compressed likelihood discussed in \sect{\ref{sec:compress}}.  The
left columns give the marginalized mean values and standard deviation for the
parameters of the compressed likelihood for \planckTT, while the right
columns present the normalized covariance or correlation matrix $\tens{D}$
for the parameters of the compressed likelihood for \planckTT.
The covariance matrix $\tens{C}$ is then given by
$\tens{C}_{ij} = \sigma_i \sigma_j \tens{D}_{ij}$ (without summation),
where $\sigma_i$ is the standard deviation
of parameter $i$.  While the upper values were derived for $w$CDM and are
consistent with those of \LCDM\ and the $\{w_0,w_a\}$ model, we marginalized
over the amplitude of the lensing power spectrum for the lower values,
which leads to a more conservative compressed likelihood.}
    \label{tbl:wm2}  \label{tbl:wm1}
\end{table*}

The marginalized mean values and 68\,\% confidence intervals for the
compressed likelihood values are shown in \tbl{\ref{tbl:wm1}} for \planckTT.
The posterior distribution of $\{R, \ell_{\rm A}, \omega_{\rm b}, n_{\rm s}\}$
is approximately Gaussian, which allows us to specify the likelihood easily by
giving the mean values and the covariance matrix, as derived from a Monte Carlo
Markov chain (MCMC) approach, in this case from the grid chains for the $w$CDM
model.  Since these quantities are very close to observables directly derivable
from the data, and since smoothly parameterized DE models are all compatible
with the \Planck\ observations to a comparable degree, they lead to very
similar central values and essentially the same covariance matrix.  The
Gaussian likelihood in $\{R, \ell_{\rm A}, \omega_{\rm b}, n_{\rm s}\}$ given
by \tbl{\ref{tbl:wm1}} is thus useful for combining \Planck\ temperature and
low-$\ell$ polarization data with other data sets and for inclusion in Fisher
matrix forecasts for future surveys.  This is especially useful when
interested in parameters such as $\{w_0,w_a\}$, for which the posterior is very non-Gaussian
and cannot be accurately represented by a direct covariance matrix
(as can be seen in Fig.\ \ref{fig:w2D_wwa}).

The quantities that make up the compressed likelihood are supposed to be
``early universe observables'' that describe the observed power spectrum and
are insensitive to late time physics.  However, lensing by large-scale
structure has an important smoothing effect on the $C_\ell$ and is detected at
over $10\,\sigma$ in the power spectrum (see section~5.2 of
\cite{planck2014-a15}).  We checked by comparing different MCMC chains that the
compressed likelihood is stable for \LCDM, $w$CDM and the $\{w_0,w_a\}$ model.
However, the ``geometric degeneracy'' in curved models is broken significantly
by the impact of CMB lensing on the power spectrum
\citep[see \Fig25 of][]{planck2014-a15} and for non-flat models one needs
to be more careful.  For this reason we also provide the ingredients for the
compressed likelihood marginalized over the amplitude $A_{\rm L}$ of the
lensing power spectrum in the lower part of \tbl{\ref{tbl:wm1}}.
Marginalizing over $A_{\rm L}$ increases the errors in some variables by over
20\,\% and slightly shifts the mean values, giving a more conservative choice
for models where the impact of CMB lensing on the power spectrum is
non-negligible.

We notice that the constraints on $\{R,\ell_{\rm A},\omega_{\rm b},n_{\rm s}\}$
given in \tbl{\ref{tbl:wm2}} for \planckTT\ data are significantly weaker
than those predicted by table~II of \cite{Mukherjee:2008kd}, which were based on
the ``Planck Blue Book'' specifications \cite{planck2005-bluebook}.
This is because these forecasts also used high-$\ell$ polarization.
If we derive the actual \Planck\ covariance matrix for the \planckall\
likelihood then we find constraints that are about 50\,\% smaller than those
given above, and are comparable and even somewhat stronger than those quoted
in \cite{Mukherjee:2008kd}.  The mean values have of course shifted to
represent what \Planck\ has actually measured.

\subsection{Perturbation parameterizations}\label{results:modifiedgravity}
Up to now we have discussed in detail the ensemble of background
parameterizations, in which DE is assumed to be a smooth fluid, minimally
interacting with gravity. General modi\-fications of gravity, however, change
both the background and the perturbation equations, allowing for contribution
to cluste\-ring (via a sound speed different than unity) and anisotropic stress
different from zero.  Here we illustrate results for perturbation degrees of
freedom, approaching MG from two different perspectives, as
discussed in \sect{\ref{sec:models}}. First we discuss results for EFT
cosmologies, with a ``top-down'' approach that starts from the most general
action allowed by symmetry and selects from there interesting classes
belonging to so-called ``Horndeski models'', which, as mentioned in \sect{\ref{subsubsec:EFT}}, include almost all stable scalar-tensor theories, universally coupled, with second-order equations of motion in the fields.
We then proceed by parameterizing directly the gravitational potentials and
their combinations, as illustrated in \sect{\ref{subsubsec:mgcamb}}. In this
way we can test more phenomenologically their effect on lensing and clustering,
in a ``bottom-up'' approach from observations to theoretical mo\-dels.

\subsubsection{Modified gravity: EFT and Horndeski models\label{sec:eftcamb}}
\label{EFT}
The first of the two approaches described in \sect{\ref{subsubsec:EFT}} adopts
effective field theory (EFT) to investigate DE
\citep{2013JCAP...08..025G, Gubitosi:2012hu}, based on the action of Eq.~(\ref{Eqn:EFTLag}).
The parameters that appear in the action, when choosing the nine time-dependent functions
$\{\Omega, c, \Lambda, \bar{M}_1^3,\bar{M}_2^4,\bar{M}_3^2, M_2^4,\hat{M}^2,
m_2^2\}$, describe the effective DE.
The full background and perturbation equations for this action have been
implemented in the publicly available Boltzmann code {\tt EFTCAMB} \citep{PhysRevD.89.103530, PhysRevD.90.043513}
\footnote{\url{http://www.lorentz.leidenuniv.nl/\~hu/codes/}, version 1.1, Oct. 2014.}.
Given an expansion history (which we fix to be \lcdm, i.e., effectively
$w\,{=}\,-1$) and an EFT function
$\Omega(a)$, {\tt EFTCAMB} computes $c$ and $\Lambda$ from the Friedmann
equations and the assumption of spatial flatness \citep{2014arXiv1405.3590H}.
As we have seen in \sect{\ref{sec:bkg_res}}, for smooth DE models the
constraints on the DE equation of state are compatible with
$w\,{=}\,-1$; hence this choice is not a limitation for the following
analysis.  In addition, {\tt EFTCAMB} uses a set of stability criteria in
order to specify whether a given model is stable and ghost-free, i.e. without
negative energy density for the new degrees of freedom.
This will automatically place a
theoretical prior on the parameter space while performing the MCMC analysis. 

The remaining six functions, $\bar{M}_1^3$, $\bar{M}_2^4$, $\bar{M}_3^2$,
$M_2^4$, $\hat{M}^2$, $m_2^2$, are internally redefined in terms of
the dimensionless parameters $\alpha_i$ with $i$ running from $1$
to $6$:
\begin{align}
\alpha_1^4&=\frac{M^4_2}{m_0^2H_0^2}, &
 \alpha_2^3&=\frac{\bar{M}^3_1}{m_0^2H_0}, &
 \alpha_3^2&=\frac{\bar{M}^2_2}{m_0^2}, \nonumber \\ \nonumber
\alpha_4^2&=\frac{\bar{M}^2_3}{m_0^2}, &
 \alpha_5^2&=\frac{\hat{M}^2}{m_0^2}, &
 \alpha_6^2&=\frac{m^2_2}{m_0^2}.
\end{align}
We will always demand that
\begin{eqnarray}
m_2^2 &=& 0\ (\textrm{or equivalently}\ \alpha_6^2 = 0), \\
\bar{M}_3^2 &=& -\bar{M}_2^2\ (\textrm{or equivalently}\
 \alpha_4^2 = -\alpha_3^2), \label{eq:dercondition}
\end{eqnarray}
which eliminates models containing higher-order spatial derivatives
\citep{2013JCAP...08..025G, Gleyzes:2014qga}.
In this case the nine functions of time discussed above reduce to a minimal
set of five functions of time that can be labelled
\{$\alphaM, \alphaK, \alphaB, \alphaT, \alphaH$\},
in addition to the Planck mass $M_\ast^2$ (the evolution of which is
determined by $H$ and $\alpha_M$), and an additional function of time
describing the background evolution, e.g., $H(a)$.
The former are related to the EFT functions via the following relations:
\begin{eqnarray}
M_\ast^2 &=& m_0^2\Omega + \bar{M}_2^2 \label{eq:Mast}; \\
M_\ast^2 H \alphaM &=& m_0^2 \dot{\Omega} + \dot{\bar{M}}_2^2;
  \label{eq:alphaM} \\
M_\ast^2 H^2 \alphaK &=& 2c + 4 M_2^4; \\ \label{eq:alphaK}
M_\ast^2 H \alphaB &=& -m_0^2 \dot{\Omega} -\bar{M}_1^3; \label{eq:alphaB} \\
M_\ast^2 \alphaT &=& -\bar{M}_2^2; \\
M_\ast^2 \alphaH &=& 2 \hat{M}^2 - \bar{M}_2^2.
\end{eqnarray}

These five $\alpha$ functions
are closer to a physical description of the theories under investigation  \citep{Bellini:2014fua}.
For example: $\alpha_T$ enters in the equation for gravitational waves,
affecting their speed and the position of the primordial peak in $B$-mode
polarization; $\alpha_M$ affects the lensing potential, but also the
amplitude of the primordial polarization peak in $B$-modes
\citep{Amendola:2014wma,2014arXiv1405.7974R,2014arXiv1408.2224P}.
It is then 
possible to relate the desired choice for the Horndeski variables to an
appropriate choice of the EFT functions,
\begin{eqnarray}
\partial_\tau (M_\ast^2)  &=& {\cal H} M_\ast^2 \alphaM \label{eq:Mstar},\\
m_0^2(\Omega+1) &=& (1 + \alphaT) M_\ast^2 \label{eq:Omega_alphaT}, \\
\bar{M}_2^2 &=& - \alphaT M_\ast^2 \label{eq:alphaT}, \\
4 M_2^4 &=& M_\ast^2 {\cal H}^2 \alphaK - 2c, \\
\bar{M}_1^3 &=& -M_\ast^2 {\cal H} \alphaB + m_0^2 \dot{\Omega},
 \label{eq:alphaB} \\
2 \hat{M}^2 &=& M_\ast^2 (\alphaH - \alphaT), \label{eq:hatM}
\end{eqnarray}
where ${\cal {H}}$ is the conformal Hubble function, $m_0$ the bare Planck
mass and $M_\ast$ the effective Plank mass.
Fixing $\alphaM$ corresponds to fixing $M_\ast$ through \eq{\ref{eq:Mstar}}.
Once $\alphaT$ has been chosen, $\Omega$ is obtained from
\eq{\ref{eq:Omega_alphaT}}.
Finally, $\alphaB$ determines $\bar{M}_1^3$ via \eq{\ref{eq:alphaB}},
while the choice of $\alphaH$ fixes $\hat{M}^2$ via \eq{\ref{eq:hatM}}. 
In this way, our choice of the EFT functions can be guided by the selection
of different ``physical'' scenarios, corresponding
to turning on different Horndeski functions.

To avoid possible consistency issues with higher derivatives, we
set\footnote{Because of the way {\tt EFTCAMB} currently implements these
equations internally, it is not possible to satisfy
Eq.~(\ref{eq:dercondition}) otherwise.} 
$\bar{M}_3^2 = \bar{M}_2^2 = 0$ in order to satisfy Eq.~(\ref{eq:dercondition}).
From \eq{\ref{eq:alphaT}} and \eq{\ref{eq:Mast}} this implies $\alphaT = 0$,
so that tensor waves move with the speed of light. In addition, we set
$\alphaH = 0$ so as to remain in the original class of Horndeski
theories.
As a consequence, $\hat{M}^2 = 0$ from Eq.~(\ref{eq:hatM}) and
$M_*^2 = m_0^2 (1+\Omega)$ from \eq{\ref{eq:Mast}}.
For simplicity we also turn off all other higher-order EFT operators
and set $\bar{M}_1^3 = M_2^4 = 0$. Comparing \eq{\ref{eq:alphaM}} and
\eq{\ref{eq:alphaB}}, this implies $\alphaB = -\alphaM$.

In summary, in the following we consider Horndeski models in which
$\alphaM = - \alphaB$, $\alphaK$ is fixed by \eq{\ref{eq:alphaK}},
with $M_2$ = 0 as a function of $c$ and $\alphaT = \alphaH = 0$. We are thus
considering non-minimally coupled ``K-essence'' type models, similar to the
ones discussed in \cite{Sawicki:2012re}.

The only free function in this case is $\alphaM$, which is linked to
$\Omega$ through:
\begin{equation}
\alpha_M = \frac{a}{\Omega+1} \frac{\rm{d} \Omega}{\rm{d} a}  \, .  \label{eq:alphaMomega}
\end{equation}
By choosing a non-zero $\alphaM$ (and therefore a time evolving $\Omega$) we
introduce a non-minimal coupling in the action (see Eq.~\ref{Eqn:EFTLag}),
which will lead to non-zero anisotropic stress and to modifications of the
lensing potential, typical signatures of MG models. Here we will use a scaling
ansatz, $\alphaM = \alphaMtoday a^\beta$, where $\alphaMtoday$ is the value of
$\alphaM$ today, and $\beta>0$ determines how quickly the modification of
gravity decreases in the past. 

\begin{figure}[htb!]
\begin{center}
\hspace*{-1cm}
\begin{tabular}{cc}
\includegraphics[width=.45\textwidth]{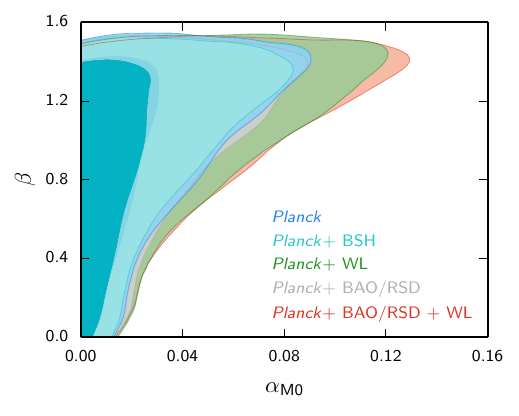}
 \end{tabular}
\caption{Marginalized posterior distributions at $68\,\%$ and $95\,\%$ C.L.
for the two parameters $\alphaMtoday$ and $\beta$ of the
exponential evolution, $\Omega(a) = \exp(\Omega_0 \, a^\beta) - 1.0$, see \sect{\ref{sec:eftcamb}}.
Here $\alphaMtoday$ is defined as $\Omega_0 \beta$ and the background is fixed
to $\Lambda$CDM. $\Omega_{\rm{M0}} = 0$
corresponds to the $\Lambda$CDM model also at perturbation level.
Note that \textit{Planck} means \planckTT. Adding WL to the data sets results
in broader contours, as a
consequence of the slight tension between the \Planck\ and WL data sets.}
\label{fig:EXPom}
\end{center}
\end{figure}

\begin{figure}[htb!]
\begin{center}
\hspace*{-1cm}
\begin{tabular}{cc}
\includegraphics[width=.45\textwidth]{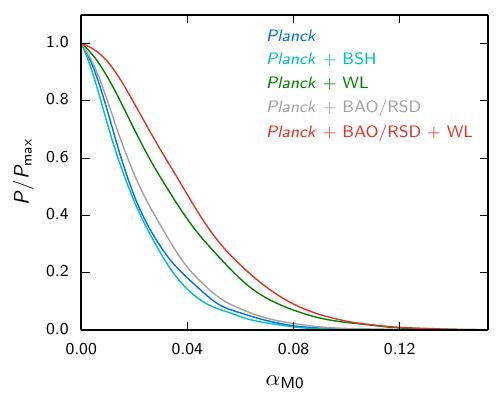}
\end{tabular}\caption{Marginalized posterior distribution of the linear EFT
model background parameter, $\Omega$, with $\Omega$ parameterized as a linear
function of the scale factor, i.e., $\Omega(a) = \alphaMtoday \, a$, see \sect{\ref{sec:eftcamb}}.
The equation of state parameter $w_\de$ is fixed to $-1$,
and therefore, $\Omega_0 = 0$ will correspond to the $\Lambda$CDM model.
Here \textit{Planck} means \planckTT. Adding CMB lensing 
to the data sets does not change the results significantly;
high-$\ell$ polarization tightens the constraints by a few percent,
as shown in Tab.~\ref{tbl:Lom}.}
\label{fig:linearEFT}
\end{center}
\end{figure}

Integrating \eq{\ref{eq:alphaMomega}} we obtain
\begin{equation}
\Omega(a) = \exp\left\{\frac{\alphaMtoday}{\beta} a^\beta \right\} - 1,
\end{equation}
which coincides with the built-in exponential model of {\tt EFTCAMB} for
$\Omega_0 = \alphaMtoday/\beta$.
The marginalized posterior distributions for the two parameters $\Omega_0$
and $\beta$ are plotted in \fig{\ref{fig:EXPom}} for different combinations
of data.
For $\alphaMtoday = 0$ we recover \LCDM.
For small values of $\Omega_0$ and for $\beta=1$, the exponential reduces to
the built-in linear evolution in {\tt EFTCAMB},
\begin{equation}
\Omega(a) = \Omega_0 \, a \, .
\end{equation}
The results of the MCMC analysis are shown in Table~\ref{tbl:Lom}. 
For both the exponential and the linear model we use a
flat prior $\Omega_0\in [0,1]$. For the scaling exponent $\beta$ of the
exponential model we use a flat prior $\beta\in (0,3]$.
For $\beta \rightarrow 0$ the MG parameter $\alpha_M$ remains constant and
does not go to zero in the early Universe, while for $\beta=3$ the scaling
would correspond to M functions in the action (\ref{Eqn:EFTLag}) which are of the same order as the
relative energy density between DE and the dark matter background, similar
to the suggestion in \cite{Bellini:2014fua}.
An important feature visible in Fig.\ \ref{fig:EXPom} is the sharp cutoff at $\beta\approx1.5$.
This cutoff is due to ``viability conditions'' that are enforced by {\tt EFTCAMB} and that reject
models due to a set of theoretical criteria (see \cite{2014arXiv1405.3590H} for a full list of theoretical priors
implemented in {\tt EFTCAMB}). Disabling some of these conditions allows to extend the
acceptable model space to larger $\beta$, and we find that the constraints on $\alpha_{M0}$ continue to weaken as $\beta$ grows further,
extending Fig.\ \ref{fig:EXPom} in the obvious way. We prefer however to use here
the current public {\tt EFTCAMB} version without modifications. A better understanding of whether all stability conditions implemented in the code are really necessary or exclude a larger region than necessary in parameter space will have to be addressed in the future.
The posterior distribution of the linear evolution for $\Omega$ is shown in
Fig.~\ref{fig:linearEFT} and is compatible with \LCDM.
Finally, it is interesting to note that in both the exponential and the linear expansion, the inclusion of WL data set weakens constraints with respect to \planckTT\ alone. This is due to the fact that in these EFT theories, WL and \planckTT\ are in tension with each other, WL preferring higher values of the expansion rate with respect to \Planck.

\begin{table*}
\tiny
\nointerlineskip
\setbox\tablebox=\vbox{
\newdimen\digitwidth % These five lines change what an asterisk
\setbox0=\hbox{\rm 0} % means to TeX. Instead of meaning
\digitwidth=\wd0 % "print an '*' here", it now means "leave
\catcode`*=\active % as much blank space as a single number
\def*{\kern\digitwidth} % takes up".
\newdimen\signwidth % These five lines change the meaning of an
\setbox0=\hbox{{\rm +}} % exclamation mark in the same way, so that it
\signwidth=\wd0 % leaves as much space as a plus or minus sign.
\catcode`!=\active % These definitions will disappear at the end of
\def!{\kern\signwidth} % the \vbox.
\newdimen\pointwidth % These five lines change the meaning of a
\setbox0=\hbox{\rm .} % question mark in the same way, so that it
\pointwidth=\wd0 % leaves as much space as a period.
\catcode`?=\active % These definitions will disappear at the end of
\def?{\kern\pointwidth} % the \vbox.
\halign{\hbox to 1.20in{#\leaderfil}\tabskip=1.0em&
\hfil#\hfil\tabskip=1.0em&
\hfil#\hfil\tabskip=1.0em&
\hfil#\hfil\tabskip=1.0em&
\hfil#\hfil\tabskip=1.0em&
\hfil#\hfil\tabskip=1.0em&
\hfil#\hfil\tabskip=0.0em\cr
\noalign{\doubleline}
\omit Parameter\hfil& \shortTT+BSH&\shortTT+WL& \shortTT+BAO/RSD& \shortTT+BAO/RSD+WL& \TTTEEE& \TTTEEE+BSH\cr
\noalign{\vskip 3pt\hrule\vskip 4pt}
Linear EFT & & & & & & \cr
$\alphaMtoday                 $& $<0.052 (95\,\% \mbox{CL})$&$<0.072 (95\,\% \mbox{CL})$&$<0.057(95\,\% \mbox{CL})$&$<0.074(95\,\% \mbox{CL})$&$<0.050(95\,\% \mbox{CL})$&$<0.043 (95\,\% \mbox{CL})$\cr
$H_0                          $&$67.69 \pm 0.55$&$67.75 \pm 0.95$&$67.63 \pm 0.63$&$67.89 \pm 0.62 $&$67.17 \pm 0.66 $&$67.60 \pm 0.48$\cr
$\sigma_8                     $&$*0.826\pm0.015$&$*0.818\pm0.014$&$*0.822\pm0.014$&$*0.814\pm 0.014$&$*0.830\pm 0.013$&$*0.830 \pm 0.014$\cr
\noalign{\vskip 3pt\hrule\vskip 4pt}
Exponential EFT & & & & & & \cr
$\alphaMtoday           $& $<0.063 (95\,\% \mbox{CL})$&$<0.092 (95\,\% \mbox{CL})$&$<0.066(95\,\% \mbox{CL})$&$<0.097(95\,\% \mbox{CL})$&
$<0.054(95\,\% \mbox{CL})$&$<0.062(95\,\% \mbox{CL})$\cr
$\beta                 $& $0.87^{+0.57}_{-0.27}  $&$0.91^{+0.54}_{-0.26}  $&$0.88^{+0.56}_{-0.28}  $&$0.92^{+0.53}_{-0.25}$&$0.90^{+0.55}_{-0.26}$&$0.92^{+0.53}_{-0.24}$\cr
\noalign{\vskip 2pt}
$H_0                          $&$67.70\pm 0.56 $&$67.78\pm 0.96 $&$67.60\pm 0.62 $&$67.87\pm 0.63 $&$67.15\pm 0.65 $&$67.58\pm 0.46$\cr
$\sigma_8                     $&$*0.826\pm0.015$&$*0.817\pm0.014$&$*0.821\pm0.014$&$*0.814\pm0.014$&$*0.830\pm0.013$&$*0.830 \pm 0.013$\cr
\noalign{\vskip 3pt\hrule\vskip 4pt}
}}
\endtable
\caption{Marginalized mean values and $68\,\%$ CL intervals for the EFT parameters, both in the linear model, $\alphaMtoday$, 
and in the exponential one, $\{\alpha_{{\rm M}0}, \beta \}$ (see \sect{\ref{sec:eftcamb}}). Adding CMB lensing does not improve the constraints,
while small-scale polarization can more strongly constraint $\alphaMtoday$.}
    \label{tbl:Lom}
\end{table*}

\subsubsection{Modified gravity and the gravitational potentials}
\label{sec:genpar}

The second approach used in this paper to address MG is more phenomenological
and, as described in \sect{\ref{subsubsec:mgcamb}}, starts from directly
parameterizing the functions of the gravitational potentials listed in
Eqs.~(\ref{eq:Qdef})--(\ref{eq:etadef}). Any choice of two of those functions
will fully parameterize the deviations of the perturbations from a smooth
DE model and describe the cosmological observables of an MG model.

In \cite{Simpson:2012ra} the amplitude of the deviation with respect to \LCDM\
was parameterized similarly to the DE-related case that we will define 
as case \ref{casederelated} below, but using $\mu(a)$ and $\Sigma(a)$ instead of
$\mu(a,k)$ and $\eta(a,k)$\footnote{The parameterization of $\mu$ and
$\Sigma$ in \cite{Simpson:2012ra} uses $\Omega_{\rm{DE}}(a)/\Omega_{\rm{DE}}$
instead of $\Omega_{\rm{DE}}(a)$; their $\mu_0$ and $\Sigma_0$ correspond to our
$\mu_0-1$ and $\Sigma_0-1$ respectively.}. They found the constraints
$\mu_0-1=0.05\pm0.25$ and $\Sigma_0-1=0.00\pm0.14$ using RSD data from the
WiggleZ Dark Energy Survey \citep{Blake:11} and 6dF Galaxy Survey (6dFGS)
\citep{Beutler:2012px}, together with CFHTLenS WL. 
\cite{Baker:2014zva} provided forecasts on $\mu_0-1$ and
$\Sigma_0-1$ for a future experiment that combines galaxy clustering and
tomographic weak lensing measurements. 
The amplitude of departures from the standard values was parameterized as in
\cite{Simpson:2012ra}, but a possible scale dependence was introduced.
In \cite{Zhao:2010dz}, the authors constrained $\mu_0$ and $\eta_0$ and derived
from those the limits on $\Sigma_0$, using WMAP-5 data along with CFHTLenS and
ISW data. Together with a principal component analysis, they also constrained
$\mu$ and $\eta$ assuming a time evolution of the two functions, introducing a
transition redshift $z_{\rm{s}}$ where the functions move smoothly from an
early time value to a late time one; 
they obtained $\mu_0=1.1^{+0.62}_{-0.34},\ \eta_0=0.98^{+0.73}_{-1.0}$ for
$z_{\rm s}=1$ and $\mu_0=0.87\pm0.12,\ \eta_0=1.3\pm0.35$ for $z_{\rm s}=2$.
A similar parametrization was also used in \cite{2010PhRvD..81l3508D} in terms of $\mu_0$ and $\varpi$ (equivalent to $\mu_0-1$ and $\eta_0-1$ in our convention) 
using WMAP5, Union2, COSMOS and CFHTLenS data, both binning these functions in redshift
and assuming a time evolution (different from the one we will assume in the following), obtaining $-0.83<\mu_0<2.1$ and $-1.6<\varpi<2.7$ at 95$\%$ confidence level for their present values.
In \cite{Macaulay:2013swa} the authors instead parameterized $\Psi/\Phi$
(the inverse of $\eta$) as $(1-\zeta)$ and use RSD data from 6dFGS, BOSS, LRG,
WiggleZ and VIPERS galaxy redshift surveys to constrain departures from \LCDM;
they did not assume a functional form for the time evolution of $\zeta$, but
rather constrained its value at two different redshifts ($z=0$ and $z=1$),
finding a $2\,\sigma$ tension with the \LCDM\ limit ($\zeta=0$) at $z=1$.

In this paper, we choose the pair of functions $\mu(a,k)$ (related to the
Poisson equation for $\Psi$) and $\eta(a,k)$ (related to the gravitational slip), as defined in Eqs.~(\ref{eq:mudef}) and (\ref{eq:etadef}), 
since these are the functions directly implemented in the publicly
available code {\tt MGCAMB}\footnote{Available at
\url{http://www.sfu.ca/\textasciitilde aha25/MGCAMB.html} (Feb. 2014 version),
see appendix~A of \citep{Zhao:2008bn} for a detailed description of the
implementation.} \citep{Zhao:2008bn,Hojjati:2011ix} integrated in the latest version of \COSMOMC.

\begin{table*}
\footnotesize 
\nointerlineskip
\setbox\tablebox=\vbox{
\newdimen\digitwidth % These five lines change what an asterisk
\setbox0=\hbox{\rm 0} % means to TeX. Instead of meaning
\digitwidth=\wd0 % "print an '*' here", it now means "leave
\catcode`*=\active % as much blank space as a single number
\def*{\kern\digitwidth} % takes up".
\newdimen\signwidth % These five lines change the meaning of an
\setbox0=\hbox{{\rm +}} % exclamation mark in the same way, so that it
\signwidth=\wd0 % leaves as much space as a plus or minus sign.
\catcode`!=\active % These definitions will disappear at the end of
\def!{\kern\signwidth} % the \vbox.
\newdimen\pointwidth % These five lines change the meaning of a
\setbox0=\hbox{\rm .} % question mark in the same way, so that it
\pointwidth=\wd0 % leaves as much space as a period.
\catcode`?=\active % These definitions will disappear at the end of
\def?{\kern\pointwidth} % the \vbox.
\halign{\hbox to 0.9in{#\leaderfil}\tabskip=1.0em&
\hfil#\hfil\tabskip=1.0em&
\hfil#\hfil\tabskip=1.0em&
\hfil#\hfil\tabskip=1.0em&
\hfil#\hfil\tabskip=1.0em&
\hfil#\hfil\tabskip=1.0em&
\hfil#\hfil\tabskip=0.0em\cr
\noalign{\doubleline}
\omit Parameter\hfil& \planckTT& \planckTT& \planckTT& \planckTT& \planckTT& \planckall\cr
\omit& \omit& +BSH& +WL& +BAO/RSD& +WL+BAO/RSD& +BSH\cr
\noalign{\vskip 3pt\hrule\vskip 4pt} 
$E_{11}$    & $0.099^{+0.34}_{-0.73}$& $0.06^{+0.32}_{-0.69}$  & $-0.20^{+0.19}_{-0.47}$   & $-0.24^{+0.19}_{-0.33}$& $-0.30^{+0.18}_{-0.30}$ & $0.08^{+0.33}_{-0.69}$\cr
$E_{22}$    & $0.99\pm 1.3$          & $1.03\pm 1.3$           & $1.92^{+1.4}_{-0.96}$     & $1.77\pm 0.88$         & $2.07\pm 0.85$          & $0.9\pm 1.2$\cr
$\mu_0-1$   & $0.07^{+0.24}_{-0.51}$ & $0.04^{+0.22}_{-0.48}$  & $-0.14^{+0.13}_{-0.34}$   & $-0.17^{+0.14}_{-0.23}$& $-0.21^{+0.12}_{-0.21}$ & $0.06^{+0.23}_{-0.48}$\cr
$\eta_0-1$  & $0.70\pm 0.94$         & $0.72\pm 0.90$          & $1.36^{+1.0}_{-0.69}$     & $1.23\pm 0.62$         & $1.45\pm 0.60$          & $0.60\pm 0.86$\cr
$\Sigma_0-1$& $0.28\pm 0.15$         & $0.27\pm 0.14$          & $0.34^{+0.17}_{-0.14}$    & $0.29\pm 0.13$         & $0.31\pm 0.13$          & $0.23\pm 0.13$\cr
$\tau$      & $0.065\pm 0.021$       & $0.063\pm 0.020$        & $0.061^{+0.020}_{-0.022}$ & $0.062\pm 0.019$       & $0.057\pm 0.019$        & $0.060\pm 0.019$\cr
$H_0$ \rm{(km/s/Mpc)}       & $68.5\pm 1.1$          & $68.17\pm 0.58$         & $69.2\pm 1.1$             & $68.26\pm 0.69$        & $68.55\pm 0.66$         & $67.90\pm 0.48$\cr
$\sigma_8$  & $0.817^{+0.034}_{-0.055}$ & $0.816^{+0.031}_{-0.051}$ & $0.786^{+0.021}_{-0.037}$ & $0.792^{+0.021}_{-0.025}$ & $0.781^{+0.019}_{-0.023}$ & $0.816^{+0.031}_{-0.051}$ \cr
\noalign{\vskip 3pt\hrule\vskip 4pt}
}}
\endtable
\caption{Marginalized mean values and 68\,\% C.L. errors on cosmological
parameters and the parameterizations of Eqs.~(\ref{eqMG:DErelated1}) and
(\ref{eqMG:DErelated2}) in the DE-related case (see \sect{\ref{sec:genpar}}), for
the scale-independent case.}
\label{tab:MGDEres}
\end{table*}

\begin{table*}
\footnotesize 
\nointerlineskip
\setbox\tablebox=\vbox{
\newdimen\digitwidth % These five lines change what an asterisk
\setbox0=\hbox{\rm 0} % means to TeX. Instead of meaning
\digitwidth=\wd0 % "print an '*' here", it now means "leave
\catcode`*=\active % as much blank space as a single number
\def*{\kern\digitwidth} % takes up".
\newdimen\signwidth % These five lines change the meaning of an
\setbox0=\hbox{{\rm +}} % exclamation mark in the same way, so that it
\signwidth=\wd0 % leaves as much space as a plus or minus sign.
\catcode`!=\active % These definitions will disappear at the end of
\def!{\kern\signwidth} % the \vbox.
\newdimen\pointwidth % These five lines change the meaning of a
\setbox0=\hbox{\rm .} % question mark in the same way, so that it
\pointwidth=\wd0 % leaves as much space as a period.
\catcode`?=\active % These definitions will disappear at the end of
\def?{\kern\pointwidth} % the \vbox.
\halign{\hbox to 0.9in{#\leaderfil}\tabskip=1.0em&
\hfil#\hfil\tabskip=1.0em&
\hfil#\hfil\tabskip=1.0em&
\hfil#\hfil\tabskip=1.0em&
\hfil#\hfil\tabskip=1.0em&
\hfil#\hfil\tabskip=1.0em\cr
\noalign{\doubleline}
\omit Max. degeneracy\hfil& \planckTT& \planckTT& \planckTT& \planckTT& \planckTT \cr
\omit& \omit& +BSH& +WL& +BAO/RSD& +WL+BAO/RSD \cr
\noalign{\vskip 3pt\hrule\vskip 4pt} 
DE-related                  & $0.84^{+0.30}_{-0.40}\,\,\, (2.1 \sigma)$ & $0.80^{+0.28}_{-0.39}\,\,\, (2.1 \sigma)$ & $1.08^{+0.35}_{-0.42}\,\,\, (2.6 \sigma)$ 
                            & $0.90^{+0.33}_{-0.37}\,\,\, (2.4 \sigma)$ & $1.03\pm 0.34\,\,\, (3.0\sigma)$  \cr
\,\,\,\,\,\,\,+ CMB lensing & $0.42^{+0.18}_{-0.34}\,\,\, (1.2 \sigma)$ & $0.38^{+0.18}_{-0.28}\,\,\, (1.4 \sigma)$ & $0.58^{+0.24}_{-0.37}\,\,\, (1.6 \sigma)$ 
                            & $0.40^{+0.18}_{-0.28}\,\,\, (1.4 \sigma)$ & $0.51^{+0.21}_{-0.30}\,\,\,\,\,\,\,\,\,\, (1.7 \sigma)$  \cr
\noalign{\vskip 3pt\hrule\vskip 4pt}
Time-related                & $0.67^{+0.26}_{-0.66}\,\,\, (1.0 \sigma)$ & $0.69^{+0.25}_{-0.67}\,\,\, (1.0 \sigma)$ & $1.12^{+0.40}_{-0.64}\,\,\, (1.8 \sigma)$ 
                            & $0.55^{+0.25}_{-0.32}\,\,\, (1.7 \sigma)$ & $0.70^{+0.27}_{-0.33}\,\,\,\,\,\,\,\,\,\, (2.1 \sigma)$  \cr
\noalign{\vskip 3pt\hrule\vskip 4pt}
}}
\endtable
\caption{Marginalized mean values and 68\,\% C.L. errors on the present day value of the function 
$2[\mu(z,k)-1]+[\eta(z,k)-1]$, 
which corresponds to the (approximate) maximum degeneracy line identified within the 2 dimensional posterior distributions. This function gives a quick idea of the maximum possible 
tension found for each data set combination in these classes of models, for the scale-independent case.
The upper part of the table refers to the DE-related parametrisation, with and without CMB lensing, 
while the lower part refers to the time-related one (see \sect{\ref{sec:genpar}}). For convenience, we write explicitly in brackets for each case the tension in units of 
$\sigma$ with respect to the standard \lcdm\ zero value. The DE-related case is more in tension than the time-related parameterization, with a maximum tension that ranges between 
2.1 $\sigma$ and $3 \sigma$, depending on the data sets. When CMB lensing is included, also the DE-related parameterization becomes compatible with \LCDM, with a maximum possible 
`tension' of at most 1.7 $\sigma$ when WL and BAO/RSD are included.}
\label{tab:maxdeg}
\end{table*}

Other functional
choices can be easily derived from them \citep{Baker:2014zva}.
We then parameterize $\mu$ and $\eta$ as follows. 
Since the \Planck\ CMB data span three orders of magnitude in $\ell$,
it seems sensible to allow for two scales to be present:
\begin{eqnarray}\label{eq:mgpar}
\mu(a,k) &=& 1 + f_1(a) \frac{1 + c_1 (\lambda H/k)^2}{1+(\lambda H/k)^2};
 \label{eqMG:DErelated1} \\
\eta(a,k) &=& 1 + f_2(a) \frac{1 + c_2 (\lambda H/k)^2}{1+(\lambda H/k)^2}.
 \label{eqMG:DErelated2}
\end{eqnarray}
For large length scales (small $k$), the two functions reduce to
$\mu \rightarrow 1+f_1(a) c_1$ and $\eta \rightarrow 1+f_2(a) c_2$;
for small length scales (large $k$), one has $\mu \rightarrow 1+f_1(a)$
and $\eta \rightarrow 1+f_2(a)$. In other words, we implement scale dependence
in a minimal way, allowing $\mu$ and $\eta$ to go to two different limits for
small and large scales.  Here the $f_{\rm{i}}$ are functions of time only,
while the $c_{\rm{i}}$ and $\lambda$ parameters are constants.  The
$c_{\rm{i}}$ give us information on the scale dependence of $\mu$ and $\eta$,
but the $f_{\rm{i}}$ measure the amplitude of the deviation from standard
GR, corresponding to $\mu = \eta = 1$.

We choose to parameterize the time dependence of the $f_{\rm{i}}(a)$ functions
as
\begin{enumerate} \label{def_paramMG}
\item coefficients related to the DE density, $f_{\rm{i}}(a) = E_{\rm{ii}}
 \Omega_{\rm{DE}}(a)$, \label{casederelated}
\item time-related evolution,
 $f_{\rm{i}}(a) = E_{\rm{i1}} + E_{\rm{i2}} (1-a)$.\label{casetimerelated}
\end{enumerate}
The first choice is motivated by the expectation that the contribution of MG
to clustering and to the anisotropic stress is proportional to its effective
energy density, as is the case for matter and relativistic particles. The
second parameterization provides a complementary approach to the first:
$E_{\rm{i1}}$ describes the MG contribution at late times, while
$E_{\rm{i2}}$ is relevant at early times.
Therefore the adoption of the time-related evolution allows, in principle, for
deviations from the standard behaviour also at high redshift, while the
parameterization connected to the DE density leads by definition to
$(\mu,\eta)\rightarrow1$ at high redshift, since the redshift evolution is
tied to that of $\Omega_{\rm{DE}}(z)$.

For case~1 (referred to as ``DE-related'' parameterization) we then have five
free parameters, $E_{11}$, $c_1$, $E_{22}$, $c_2$, and $\lambda$, while for
case~2 (the ``time-related'' parameterization) we have two additional parameters,
$E_{12}$ and $E_{21}$.
The choice above looks very similar to the BZ parameterization
\citep{Bertschinger:2008zb} for the quasi-static limit of $f(R)$ and
scalar-tensor theories. However, we emphasise that
Eqs.~(\ref{eqMG:DErelated1})--(\ref{eqMG:DErelated2}) should not be seen as a
quasi-static limit of any specific theory, but rather as a (minimal) way to
allow for (arbitrary) scale dependence, since the data cover a sufficiently
wide range of scales. Analogously to the EFT approach discussed in the
previous section, we set the background evolution to be the same as in \lcdm,
so that $w=-1$. In this way the additional parameters purely probe the
perturbations.

The effect of the $E_{\rm{ii}}$ parameters on the CMB temperature
and lensing potential power spectra has been shown in \Fig{\ref{fig:MGcls}}
for the ``DE-related'' choice. In the temperature spectrum the amplitude of the ISW effect is modified; the lensing potential changes more than the
temperature spectrum for the same amplitude of the $E_{\rm{ii}}$ parameter. 

We ran Monte Carlo simulations to compare the theoretical predictions with
different combinations of the data for both cases \ref{casederelated} and \ref{casetimerelated}.  For both
choices we tested whether scale dependence plays a role (via the parameters
$c$ and $\lambda$) with respect to the scale-independent case in which we fix
$c_1=c_2=1$.
Results show that a scale dependence of $\mu$ and $\eta$ does not lead to a
significantly smaller $\chi^2$ with respect to the scale-independent case,
both for the DE-related and time-related parameterizations.  Therefore there is no
gain in adding $c_{\rm{i}}$ and $\lambda$ as extra degrees of freedom.  For this reason,
in the following we will mainly show results obtained for the scale-independent
parameterization.

\begin{figure*}[htb!]
\begin{center}
\hspace*{-1cm}
\begin{tabular}{cc}
\includegraphics[width=.45\textwidth]{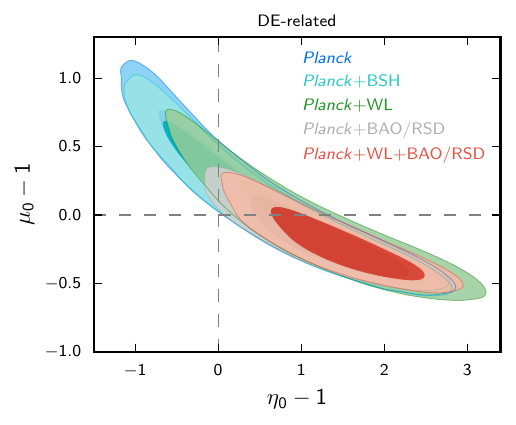}&
\includegraphics[width=.45\textwidth]{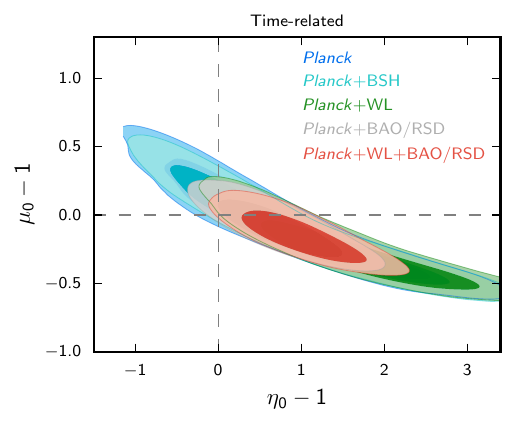}
\end{tabular}
\caption{68\,\% and 95\,\% contour plots for the two parameters
$\{\mu_0-1,\eta_0-1\}$ obtained by evaluating Eqs.~(\ref{eqMG:DErelated1})
and (\ref{eqMG:DErelated2}) at the present time when no scale dependence is
considered (see \sect{\ref{sec:genpar}}). We consider both the DE-related (left panel) and time-related
evolution cases (right panel).  Results are shown for the
scale-independent case ($c_1 = c_2 = 1$).
In the labels, \Planck\ stands for \planckTT.}
\label{fig:mueta}
\end{center}
\end{figure*}

\Tbl{\ref{tab:MGDEres}} shows results for the DE-related case for
different combinations of the data. Adding the BSH data sets to the \planckTT\
data does not significantly increase the constraining power on MG parameters;
\Planck\ polarization also has little impact. On the contrary, the addition of
RSD data tightens the constraints significantly. The WL contours, including the
ultraconservative cut that removes dependence on nonlinear physics, result in
weaker constraints. In the table, $\mu_0-1$ and $\eta_0-1$ are obtained by
reconstructing Eqs.~(\ref{eqMG:DErelated1}) and (\ref{eqMG:DErelated2})
from $E_{11}$ and $E_{22}$ at the present time.
In addition, the present value of the $\Sigma$ parameter, defined in
\eq{\ref{eq:sigmadef}}, can be obtained from $\mu$ and $\eta$ as
$\Sigma=(\mu/2)(1+\eta)$ using Eqs.~(\ref{eq:mudef}) and (\ref{eq:etadef}).

\begin{figure}[htb!]
\begin{center}
\hspace*{-1cm}
\begin{tabular}{cc}
\includegraphics[width=.45\textwidth]{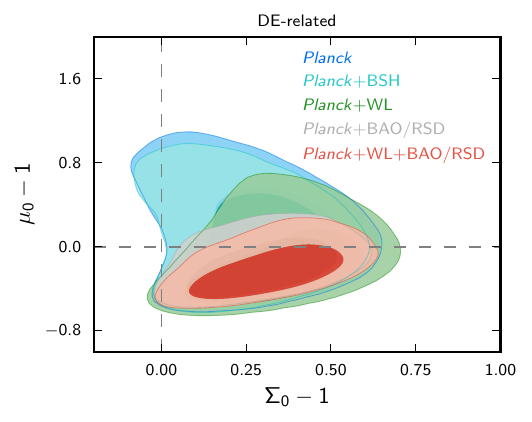}&
\end{tabular}
\caption{Marginalized posterior distributions for 68\% and 95\% C.L. for the two parameters
$\{\mu_0-1,\Sigma_0-1\}$ obtained by evaluating Eqs.~(\ref{eqMG:DErelated1})
and (\ref{eqMG:DErelated2}) at the present time in the DE-related parametrization when no scale dependence is
considered (see \sect{\ref{sec:genpar}}). $\Sigma$ is obtained as $\Sigma=(\mu/2)(1+\eta)$. The time-related evolution would give similar contours. In the labels,
\Planck\ stands for \planckTT.}
\label{fig:musigma}
\end{center}
\end{figure}

\begin{figure*}[htb!]
\begin{center}
\hspace*{-1cm}
\begin{tabular}{cc}
\includegraphics[width=.45\textwidth]{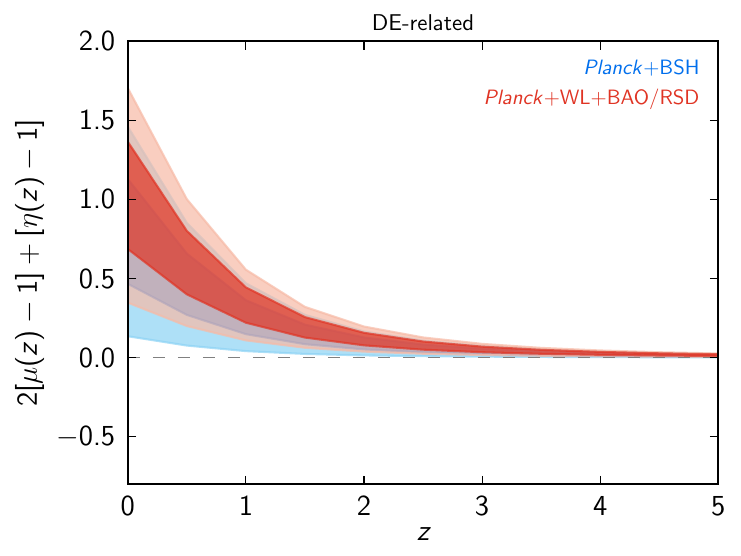}&
\includegraphics[width=.45\textwidth]{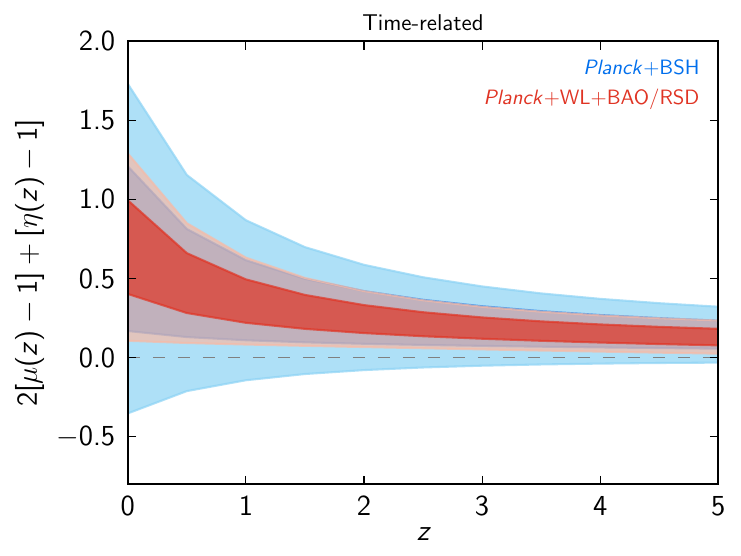}\\
\end{tabular}
\caption{Redshift dependence of the function $2[\mu(z,k)-1]+[\eta(z,k)-1]$, defined in \eqs{\ref{eqMG:DErelated1},\ref{eqMG:DErelated2}}, 
which corresponds to the maximum degeneracy line identified within the 2 dimensional posterior distributions. This combination shows the 
strongest allowed tension with \LCDM. The left panel refers to the DE-related case while the right panel refers to the time-related evolution 
(see \sect{\ref{sec:genpar}}). In both panels, no scale dependence is considered. The coloured areas show the regions containing 68\,\% and 
95\,\% of the models. In the labels, \Planck\ stands for \planckTT.}
\label{fig:muetarecon}
\end{center}
\end{figure*}

\begin{figure*}[htb!]
\begin{center}
\hspace*{-1cm}
\begin{tabular}{cc}
\includegraphics[width=.45\textwidth]{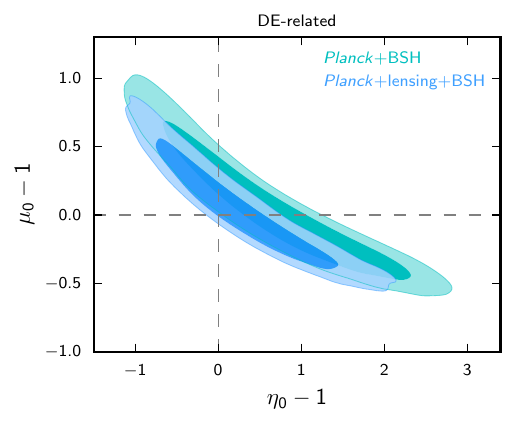}&
\includegraphics[width=.45\textwidth]{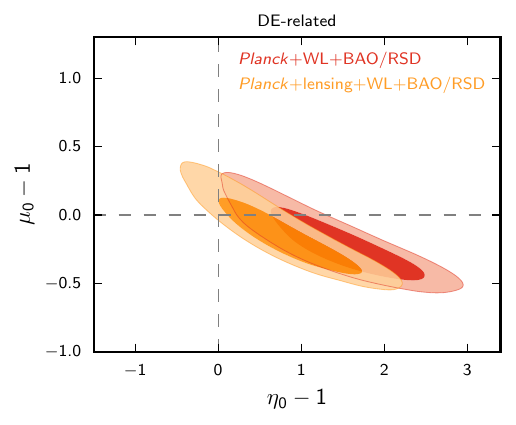}\\
\end{tabular}
\caption{68\,\% and 95\,\% marginalised posterior distributions for the two parameters
$\{\mu_0-1,\eta_0-1\}$ obtained by evaluating Eqs.~(\ref{eqMG:DErelated1})
and (\ref{eqMG:DErelated2}) at the present time when no scale dependence is
considered (see \sect{\ref{sec:genpar}}). Here we show the effect of CMB lensing, which shifts the contours towards \LCDM. In the labels,
\Planck\ stands for \planckTT.}
\label{fig:lenscomp}
\end{center}
\end{figure*}

Some tension appears, in particular, when plotting the marginalized posterior distributions in the planes ($\mu_0-1$, $\eta_0-1$)
and ($\mu_0-1$, $\Sigma_0-1$), as shown in Figs.~\ref{fig:mueta} and
\ref{fig:musigma}. Here the constraints on the two parameters that
describe the perturbations in MG are simultaneously taken into account. 
In Fig.\ \ref{fig:mueta}, left and right panels refer to the DE-related and time-related parameterizations
defined in \ref{def_paramMG}, respectively, while the dashed lines indicate the
values predicted in \lcdm. Interestingly, results appear similar in both parameterizations.
In the DE-related case (left panel), the \LCDM\ point lies at the border of the $2 \sigma$ contour, already when considering \planckTT\ alone. 
More precisely, when looking at the goodness of fit, with respect to the standard \lcdm\ assumption, the MG scenario (which includes two extra parameters $E_{11}$ and $E_{22}$) leads to an
improvement of $\Delta\chi^2=-6.3$ when using \planckTT\ (similarly divided between lowP and TT)
and of $\Delta\chi^2=-6.4$ when including BSH (with a $\Delta
\chi^2_{\rm{CMB}} \sim - 5.6$ equally divided between TT and lowP).
When \Planck\ data (TT+lowP) are combined also with WL data, the tension increases to $\Delta \chi^2 = -10.6$ (with the CMB still contributing about the same amount, $\Delta \chi^2_{\rm CMB} = -6.0$). 
When considering \planckTT+BAO/RSD, $\Delta \chi^2 = -8.1$ with respect to \lcdm\ while, when combining both WL and BAO/RSD, the tension is maximal, with $\Delta\chi^2=-10.8$ and $\chi^2_{\rm CMB} = -6.9$.
There is instead less tension for the time-related parameterization, as is visible in the right panel of
\Fig{\ref{fig:mueta}}.

Once the behaviour of the coefficients in the two parameterizations is known,
we can use Eq.~(\ref{eq:mgpar}) to reconstruct the evolution of $\mu$ and
$\eta$ with scale factor (or redshift, equivalently). In 
\Fig{\ref{fig:muetarecon}} we choose to show the linear combination 
$2[\mu(z,k)-1]+[\eta(z,k)-1]$, which corresponds approximately to the maximum degeneracy line in the 2 dimensional $\mu - 1, \eta - 1$ parameter space, which allows to better
visualize the joint constraints on $\mu$ and $\eta$ and their maximal allowed departure from \lcdm. As expected, the DE-related dependence forces the combination to be compatible with \LCDM\ in the past, when the DE density is negligible; the time-related parameterization, instead allows for a larger variation in the past.

The tension can be understood by noticing that the best fit power spectrum corresponds to a value of $\mu$ and $\eta$ ($E_{11}=-0.3$, $E_{22}=2.2$ for \planckTT) 
close to the thick  long dashed line shown in \Fig{\ref{fig:MGcls}} for demonstration. This model leads to less power in the CMB at large scales and a higher lensing potential, which is slightly preferred by the data points with respect to \LCDM. This explains also why the MG parameters are somewhat degenerate with the lensing amplitude $A_{\rm L}$ (which is an `unphysical' parameter redefining the lensing amplitude that affects the CMB power spectrum). As discussed in \cite{planck2014-a15} (see for ex. \sect{5.1.2}), \LCDM\ would lead to a value of $A_L$ \citep{calabrese_etal_2008} somewhat larger than 1. When varying it in MG, we find a mean value of $A_{\mathrm{L}}=1.116^{+0.095}_{-0.13}$ which is compatible with $A_L = 1$ at 1 $\sigma$. The price to pay is the tension with \LCDM\ in MG parameter space, which compensates the need for a higher $A_L$ that one would have in \LCDM. The CMB lensing likelihood extracted from the 4-
point function of the \Planck\ maps \citep{planck2014-a17} on the other hand does not prefer a higher lensing potential and agrees well with \LCDM . For this reason the tension is reduced when we add CMB lensing, as shown in \fig{\ref{fig:lenscomp}.}
We also note that constraints for this class of model are sensitive
to the estimation of the optical depth $\tau$. Smaller values of
$\tau$ tend to shift the results further away from \LCDM.

In order to have a quick overall estimate of the tension for all cases discussed above, we then show in \tbl{\ref{tab:maxdeg}} the marginalized mean and $68\%$ CL errors for the linear combination $2[\mu(z,k)-1]+[\eta(z,k)-1]$ that . In the table, we indicate in brackets, for convenience, the `tension' with \LCDM\ for each case. This is the maximum allowed tension, since it is calculated along the maximum degeneracy direction. The DE-related parameterization is more in tension with \LCDM\ than the time-related one. The maximum tension reaches 3 $\sigma$ when including WL and BAO/RSD, being therefore mainly driven by external data sets. The inclusion of CMB lensing shifts results towards \LCDM, as discussed.

\begin{figure}[hb!]
\begin{center}
\hspace*{-1cm}
\begin{tabular}{cc}
\includegraphics[width=.45\textwidth]{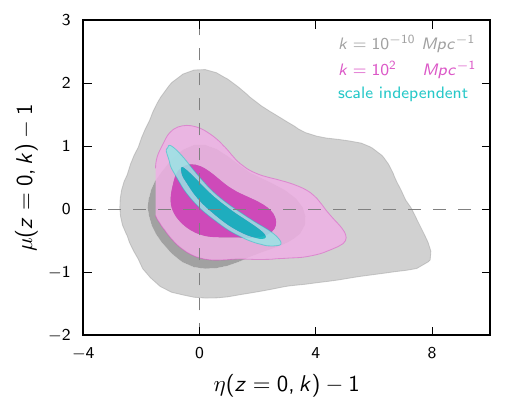}&
\end{tabular}
\caption{68\,\% and 95\,\% contour plots for the two parameters
$\{\mu_0(k)-1,\eta_0(k)-1\}$ obtained by evaluating
Eqs.~(\ref{eqMG:DErelated1}) and (\ref{eqMG:DErelated2}) at the present time
for the DE-related parameterization (see \sect{\ref{sec:genpar}}). We consider both the scale-independent
and scale-dependent cases, choosing $k$ values of $10^{-10}{\rm Mpc}^{-1}$
and $10^{2}{\rm Mpc}^{-1}$.}
\label{fig:mueta_scale}
\end{center}
\end{figure}

Finally, in general, $\mu$ and $\eta$ depend not only on redshift but also on scale,
via the parameters ($c_i$, $\lambda$). When marginalizing over them,
constraints become weaker, as expected. The comparison with the
scale-independent case is shown in \Fig{\ref{fig:mueta_scale}} for
\planckTT+BSH and different values of $k$. 
When allowing for scale dependence, the tension with \LCDM\ is washed out by
the weakening of the constraints and the goodness of fit does not improve with respect to the scale independent case.

\subsection{Further examples of particular models} 
\label{sec:examplemodels}
Quite generally, DE and MG theories deal with at least one extra degree of freedom
that can usually be associated with a scalar field. For `standard' DE theories
the scalar field couples minimally to gravity, while in MG theories the
field  can be seen as the
mediator of a fifth force in addition to standard interactions. This happens
in scalar-tensor theories (including $f(R)$ cosmologies), massive gravity,
and all coupled DE models, both when matter is involved or when neutrino
evolution is affected.  Interactions and fifth forces are therefore a common
characteristic of many proposed models, the difference being whether the
interaction is universal (i.e., affecting all species with the same coupling,
as in scalar-tensor theories) or is different for each species (as in coupled
DE, \citealt{Wetterich:1994bg, amendola_2000} or growing neutrino models,
\citealt{Fardon:2003eh, Amendola_etal_2008}). 
In the following we will test well known examples of particular models within
all these classes.

\subsubsection{Minimally coupled DE: sound speed and k-essence} \label{sec:kessence}
In minimally coupled quintessence models, the sound speed is $c_{\rm{s}}^2 = 1$
and DE does not contribute significantly to clustering. However, in so-called
``k-essence'' models, the kinetic term in the action is generalised
to an arbitrary function of $(\nabla\phi)^2$~\citep{ArmendarizPicon:2000dh}: the sound speed can then be different from the speed of light
and if $c_{\rm{s}} \ll 1$, the DE perturbations can become non-negligible
on sub-horizon scales and impact structure formation. To test this scenario we
have performed a series of analyses where we allow for a constant equation of
state parameter $w$ and a constant speed of sound $c_{\rm{s}}^2$ (with a
uniform prior in $\log c_{\rm{s}}$). We find that the limits on $w$ do not
change from the quintessence case and that there is no significant constraint
on the DE speed of sound using current data. This can be understood as follows:
on scales larger than the sound horizon and for $w$ close to $-1$, DE
perturbations are related to dark matter perturbations through
$\Delta_{\rm DE} \simeq (1+w)\Delta_{\rm{m}}/4$ and inside the sound horizon
they stop growing because of pressure support
\citep[see e.g.,][]{2009JCAP...02..018C, Sapone:2009mb}. In addition, at early times the DE density
is much smaller than the matter density, with $\rho_{\rm DE}/\rho_{\rm{m}} =
[(1-\Omega_{\rm{m}})/\Omega_{\rm{m}}] a^{-3 w}$. Since the relative DE
contribution to the perturbation variable $Q(a,k)$ defined in
\eq{\ref{eq:Qdef}} scales like
$\rho_{\rm DE}\Delta_{\rm DE}/(\rho_{\rm{m}} \Delta_{\rm{m}})$,
in k-essence type models the impact of the DE perturbations on the total
clustering is small when $1+w \approx 0$. For the DE perturbations in
k-essence to be detectable, the sound speed would have had to be very small,
and $|1+w|$ relatively large.

\subsubsection{Massive gravity and generalized scalar field models} \label{sec:EOS}
We now give two examples of subclasses of Horndeski models, written in terms of an alternative pair of DE perturbation functions (with respect to $\mu$ and $\eta$ used before, for example), given by the anisotropic stress $\sigma$ and the
entropy perturbation $\Gamma$:
\beq
w\Gamma={\delta p\over\rho}-{dp\over d\rho}\delta\,.
\eeq
When $\Gamma=0$ the perturbations are adiabatic, that is
$\delta p={dp\over d\rho}\delta \rho$.

For this purpose, it is convenient to adopt the `equation of state' approach described in \cite{2015JCAP...04..048B, 2015JCAP...02..037S}.
%Battye:2012eu
The gauge-invariant quantities $\Gamma$ and $\sigma$
can be specified in terms of the other perturbation variables, namely $\delta \rho$, $\theta$,
$h$ and $\eta$ in the scalar sector, and their
derivatives. 

We then show results for two limiting cases in this formalism, corresponding to Lorentz-violating massive gravity (LVMG) for which ($\sigma \neq 0, \Gamma = 0$)
 and generalized scalar field models (GSF) in which the anisotropic stress is zero ($\sigma = 0, \Gamma \neq 0$).

\paragraph{Lorentz-violating massive gravity (LVMG)}
If the Lagrangian is
${\cal L}\equiv {\cal L}(g_{\mu\nu})$ (i.e. only written in terms of metric perturbations, as in the EFT action) and one imposes time translation
invariance (but not spatial translational invariance), one finds that this
corresponds to an extra degree of freedom, $\xi^i$, that has a physical
interpretation as an elastic medium, or as Lorentz-violating massive
gravity~\citep{2004JHEP...10..076D, 2008PhyU...51..759R, Battye:2013er}. In this case, the scalar equations are characterized by $\Gamma=0$ (the model
is adiabatic) and a non-vanishing anisotropic stress:
\beq
\sigma=(w-c^2_{\rm s})\left[\frac{\delta}{1+w}-3\eta\right]\,,
\eeq
including one degree of freedom, the sound speed $c^2_{\rm s}$, which can be
related equivalently to the shear modulus of the elastic medium or the
Lorentz violating mass. Tensor (gravitational wave) equations will also include a mass term. 
The low sound speed may lead to clustering of the DE fluid, which allows the data to place
constraints on $c^2_{\rm s}$. But as $w$ approaches $-1$, the DE perturbations are suppressed
and the limits on the sound speed weaken. We can take this degeneracy between $1+w$ and $c^2_{\rm s}$
into account by using the combination $\lambda_{\rm c} = |1+w|^{\alpha} \log_{10} c_{\rm s}^2$ in the MCMC analysis, where $\alpha=0.35$ was chosen to decorrelate $w$ and $\lambda_{\rm c}$. With this, we find  \Planck\ TT+lowP+lensing gives lower limit of $\lambda_{\rm c}>-1.6$ at $2\sigma$ and a tighter one when including BAO/RSD and WL, with $\lambda_{\rm c}>-1.3$ at $2\sigma$. 
For any $w\neq -1$ these limits can be translated into limits on $\log_{10} c_{\rm s}^2$ by computing $\lambda_c/|1+w|^{\alpha}$.
The \LCDM\ limit is however fully compatible with the data, i.e.\ there is no detection of any
deviation from $w=-1$ (and in this limit $c^2_{\rm s}$ is unconstrained).

\paragraph{Generalized scalar field models (GSF)}
One can allow for generalized
scalar fields by considering a Lagrangian
${\cal L}\equiv {\cal L}(\phi,\partial_\mu\phi,\partial_\mu\partial_\nu\phi,
g_{\mu\nu},\partial_\alpha g_{\mu\nu})$, in which the dependence on the scalar fields is made explicit, imposing full reparameterization
invariance ($x^{\mu}\rightarrow x^{\nu}+\xi^{\mu}$), allowing for only linear
couplings in $\partial_\alpha g_{\mu\nu}$ and second-order field equations.
In this case the anisotropic stresses are zero and 
\begin{eqnarray}
 && w\Gamma = (\alpha-w)   \left \{ \delta-3\beta_1{\cal H}(1+w)\theta-
  {3\beta_2{\cal H}(1+w)\over 2k^2-6(\dot{\cal H}-{\cal H}^2)}\dot h  \right.
 \nonumber \\  
&&  \label{Eqn:EFTLag2} \quad + \left.  {3(1-\beta_2-\beta_1){\cal H}(1+w)
 \over 6\ddot{\cal H}-18{\cal H}\dot{\cal H}+6{\cal H}^3+2k^2{\cal H}}
 \ddot h  \right\}\,.
\end{eqnarray} 
This has three extra parameters ($\alpha,\beta_1,\beta_2$), in addition to $w$.
If $\beta_1=1$ and $\beta_2=0$ this becomes the generalized k-essence model.
An example of this class of models is
``kinetic gravity braiding''~\citep{Deffayet:2010qz} and similar to the non-minimally coupled k-essence discussed via EFT in \sect{\ref{sec:eftcamb}}. 
The $\alpha$ parameter in \eq{\ref{Eqn:EFTLag2}} can be now interpreted as a sound speed, 
unconstrained as in results above. 
There are however two additional degrees of freedom, $\beta_1$ and $\beta_2$. 
RSD data are able to constrain them, with the addition of \Planck\ lensing and WL making only a minor change to the
joint constraints. As in the LVMG case, 
we use a new basis  $\gamma_i = |1+w|^{\alpha_i} \beta_i$ in the MCMC analysis, where $\alpha_1=0.2$ and $\alpha_2=1$ were chosen to decorrelate  $w$ and $\gamma_{i}$. The resulting $2\sigma$ upper limits are $\gamma_1<0.67$ and $\gamma_2<0.61$ (for $w>-1$), $\gamma_2<2.4$ (for $w<-1)$  for  the combination of \Planck\ TT+lowP+lensing+BAO/RSD+WL. As for the LVMG case, there is no detection of a deviation from \LCDM\, and for $w=-1$ there are no constraints on $\beta_1$ and $\beta_2$.

\subsubsection{Universal couplings: $f(R)$ cosmologies} \label{sec:fr}
A well-investigated class of MG models is constituted by the $f(R)$ theories
that modify the Einstein-Hilbert action by substituting the Ricci scalar with
a more general function of itself:
\begin{equation}
S=\frac{1}{2\kappa^2}\int d^4x\sqrt{-g}(R+f(R)) + \int d^4x
 \mathcal{L}_{\rm M}(\chi_i, g_{\mu \nu}),
\end{equation}
where $\kappa^2=8 \pi G$. $f(R)$ cosmologies can be
mapped to a subclass of scalar-tensor theories, where the coupling of the
scalar field to the matter fields is universal. 

For a fixed background, the Friedmann equation provides a second-order
differential equation for $f(R(a))$ \citep[see e.g.,][]{song,levon}.
One of the initial conditions is usually set by requiring
\begin{equation}
\lim_{R \to \infty} \frac{f(R)}{R} = 0,
\end{equation}
and the other initial (or boundary condition), usually called $B_0$,
is the present value of
\begin{equation}
\label{eqn:b0}
B(z) = \frac{f_{RR}}{1+f_R}\frac{\hub \dot{R}}{\dot{\hub}-\hub^2} . 
\end{equation}
Here, $f_R$ and $f_{RR}$ are the first and second derivatives of $f(R)$,
and a dot means a derivative with respect to conformal time.
Higher values of $B_0$ suppress power at large scales in the CMB power
spectrum, due to a change in the ISW effect. This also changes the CMB lensing
potential, resulting in slightly smoother peaks at higher $\ell$s
\citep{2007PhRvD..76f3517S, 2008PhRvD..78d3002S, 2008PhRvD..78b4015B,
Marchini:2013lpp}.

It is possible to restrict \EFTCAMB\ to describe $f(R)$-cosmologies. Given an
evolution history for the scale factor and the value of $B_0$, \EFTCAMB\
effectively solves the Friedmann equation for $f(R)$. It then uses this
function at the perturbation level to evolve the metric potentials and matter
fields. The merit of \EFTCAMB\ over the other available similar codes is that
it checks the model against some stability criteria and does not assume the
quasi-static regime, where the scales of interest are still linear but smaller
than the horizon and the time derivatives are ignored.

\begin{figure}[htb!]
\begin{center}
\hspace*{-1cm}
\begin{tabular}{cc}
\includegraphics[width=.45\textwidth]{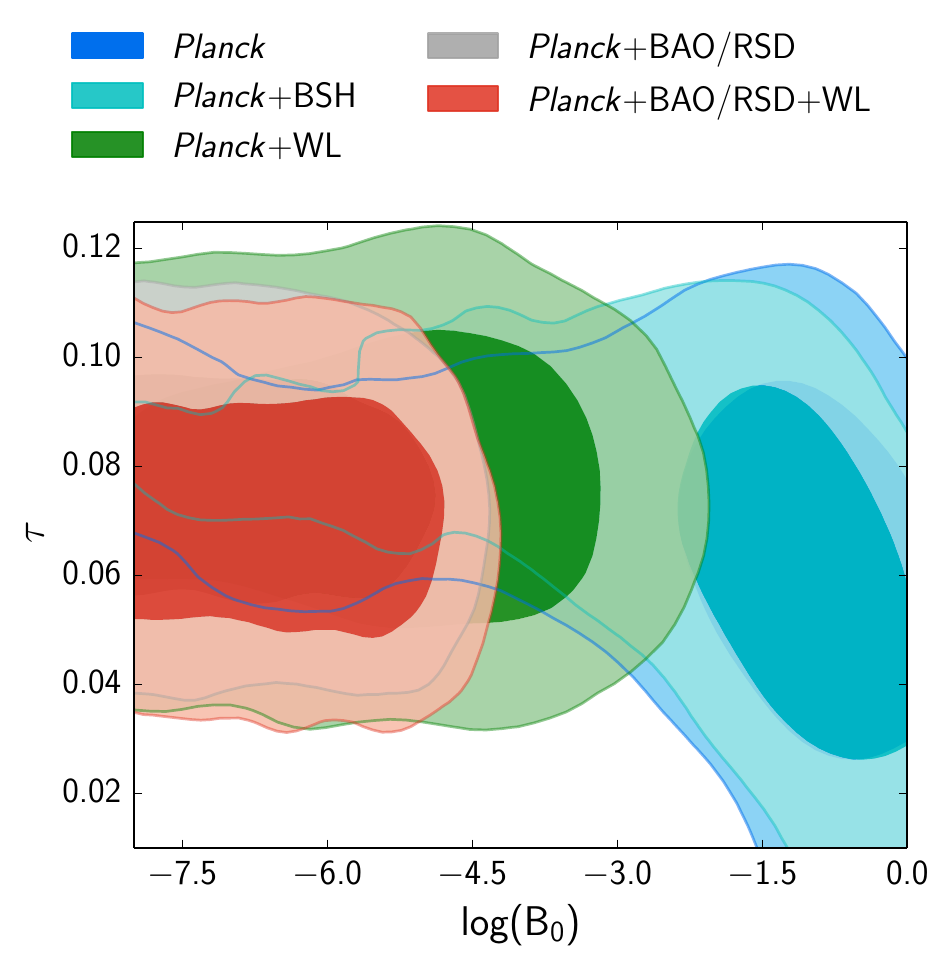}
 \end{tabular}
\caption{$68\,\%$ and $95\,\%$ contour plots for the two parameters,
$\{\rm{Log}_{10}(B_0),\tau\}$ (see \sect{\ref{sec:fr}}).
There is a degeneracy between the two parameters for \planckTT+BSH. Adding
lensing will break the degeneracy between the two. Here \textit{Planck}
indicates \planckTT.}
\label{fig:logB0-tau}
\end{center}
\end{figure}

\begin{figure}[htb!]
\begin{center}
\hspace*{-1cm}
\begin{tabular}{cc}
\includegraphics[width=.45\textwidth]{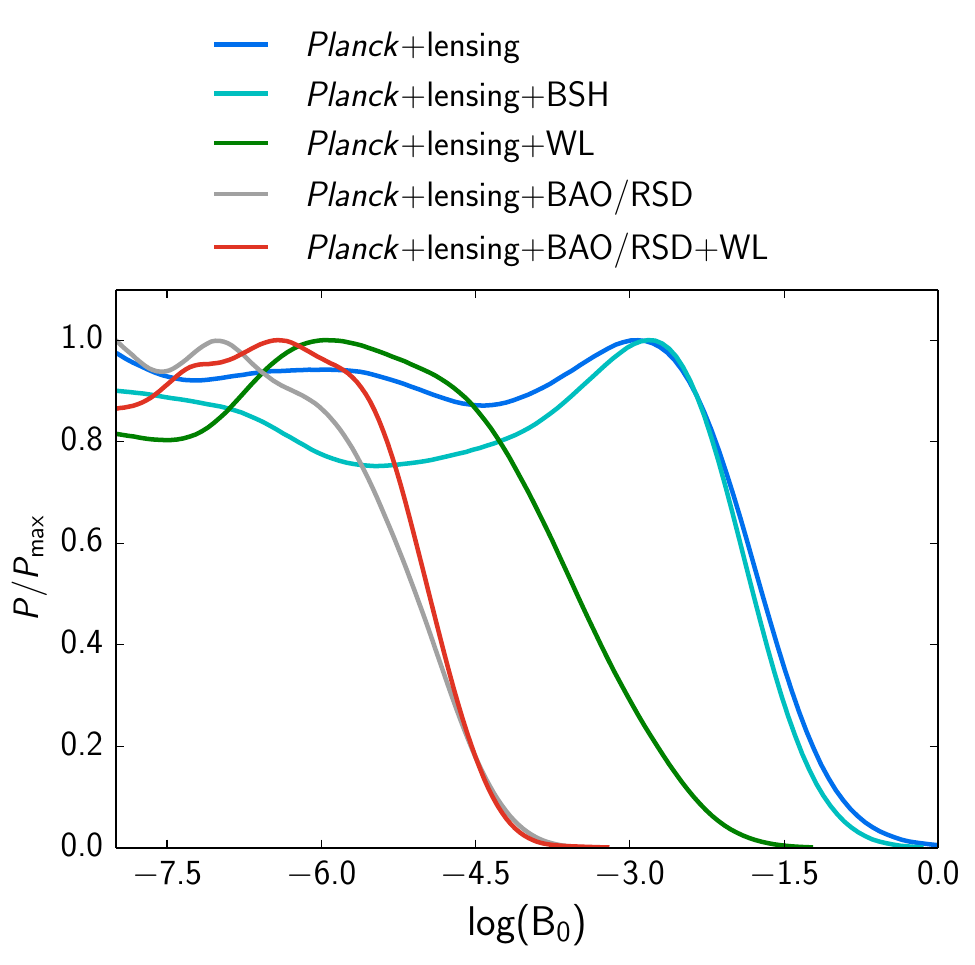}
 \end{tabular}
\caption{Likelihood plots of the $f(R)$ theory parameter, $B_0$ (see \sect{\ref{sec:fr}}).
CMB lensing breaks the degeneracy between
$B_0$ and the optical depth, $\tau$, resulting in lower upper bounds.}
\label{fig:logB0}
\end{center}
\end{figure}

As shown in Fig.~\ref{fig:logB0-tau}, there is a degeneracy between the
optical depth, $\tau$, and the $f(R)$ parameter, $B_0$. 
Adding any structure formation probe, such as WL, RSD or CMB lensing,
breaks the degeneracy.
Figure~\ref{fig:logB0} shows the likelihood of the $B_0$ parameter using
\EFTCAMB, where a $\Lambda$CDM background evolution is assumed, i.e.,
$w_{\rm DE} = -1$. 

As the different data sets provide constraints
on $B_0$ that vary by more than four orders of magnitude, we show plots for
$\log_{10} B_0$; to make these figures we use
a uniform prior in $\log_{10} B_0$ to avoid distorting the posterior due to
prior effects. However, 
for the limits quoted in the tables
we use $B_0$ (without $\log$) as the fundamental quantity and quote
$95\,\%$ limits based on $B_0$. In this way the
upper limit on $B_0$ is effectively given by the location of the drop in
probability visible in the figures, but not influenced by
the choice of a lower limit of $\log_{10} B_0$. Overall this appears to be
the best compromise to present the constraints on the $B_0$ parameter. 
In the plots, the GR value ($B_0 = 0$) is reached by a plateau
stretching towards minus infinity. 

\begin{table*}
\tiny
\nointerlineskip
\setbox\tablebox=\vbox{

\newdimen\digitwidth % These five lines change what an asterisk
\setbox0=\hbox{\rm 0} % means to TeX. Instead of meaning
\digitwidth=\wd0 % "print an '*' here", it now means "leave
\catcode`*=\active % as much blank space as a single number
\def*{\kern\digitwidth} % takes up".

\newdimen\signwidth % These five lines change the meaning of an
\setbox0=\hbox{{\rm +}} % exclamation mark in the same way, so that it
\signwidth=\wd0 % leaves as much space as a plus or minus sign.
\catcode`!=\active % These definitions will disappear at the end of
\def!{\kern\signwidth} % the \vbox.

\newdimen\pointwidth % These five lines change the meaning of a
\setbox0=\hbox{\rm .} % question mark in the same way, so that it
\pointwidth=\wd0 % leaves as much space as a period.
\catcode`?=\active % These definitions will disappear at the end of
\def?{\kern\pointwidth} % the \vbox.

\halign{\hbox to 1.0in{#\leaderfil}\tabskip=1.0em&
\hfil#\hfil\tabskip=1.0em&
\hfil#\hfil\tabskip=1.0em&
\hfil#\hfil\tabskip=1.0em&
\hfil#\hfil\tabskip=1.0em&
\hfil#\hfil\tabskip=0.0em\cr
\noalign{\doubleline}
\omit $f(R)$ models \hfil& \planckTT&\planckTT& \planckTT& \planckTT& \planckTT\cr
\omit& & +BSH& +WL& +BAO/RSD& +WL+BAO/RSD\cr
\noalign{\vskip 3pt\hrule\vskip 4pt} 
$B_0$& $<0.79$ (95\,\% CL)& $<0.69$ (95\,\% CL)& $<0.10$ (95\,\% CL)& $<0.90 \times10^{-4}$ (95\,\% CL) & $< 0.86 \times10^{-4}$  (95\,\% CL)\cr
$ B_0\,\,\,\,\,(+ \lensing)$& $<0.12$ (95\,\% CL)& $<0.07$ (95\,\% CL)& $<0.04$ (95\,\% CL)& $<0.97 \times{10^{-4}}$ (95\,\% CL)& $< 0.79 \times10^{-4}$(95\,\% CL) \cr
\noalign{\doubleline}
}}
\endtable
\caption{$95\,\%$ CL intervals for the $f(R)$ parameter,
$B_0$ (see \sect{\ref{sec:fr}}). While the plots are produced for $\log_{10} B_0$, the numbers 
in this table are produced via an analysis on $B_0$ since
the GR best fit value ($B_0 = 0$) lies out of the bounds in a $\log_{10} B_0$
analysis and its estimate would be prior dependent.
}
    \label{tbl:eftB0}
\end{table*} 

Finally, we note that $f(R)$ models can be studied also with the \MGCAMB\
parametrization, assuming the quasi-static limit. We find that for the
allowed range of the $B_0$ parameter, the results with and without the
quasi-static approximation are the same within the uncertainties.
The $95\,\%$ confidence intervals are reported in Table~\ref{tbl:eftB0}. 
These values show an improvement over the WMAP analysis made
with \MGCAMB\, ($B_0 < 1$ ($95\, \%$ C.L.) in \cite{song}) and are similar to the limits
obtained in \cite{Marchini:2013oya, 2013PhRvD..88l3514H} (see also \cite{2014JCAP...03..046D} where data from WiggleZ were used, \cite{2015PhRvD..92d4009C} where they considered galaxy clusters).

\subsubsection{Non-universal couplings: coupled Dark Energy} \label{sec:cq}
Universal couplings discussed in the previous subsection generally require
screening mechanisms to protect baryonic interactions in high density
environments, where local measurements are tightly constraining (see e.g.~\citealt{Khoury:2010xi} and \cite{2014arXiv1407.6044V} for astrophysical constraints).
An alternative way to protect baryons is to allow for non-universal couplings,
in which different species can interact with different strengths: baryons are
assumed to be minimally coupled to gravity while other species (e.g.,
dark matter or neutrinos) may feel a ``fifth force,'' with a range at
cosmological scales. 

A fifth force between dark matter particles, mediated by the DE scalar field,
is the key ingredient for the coupled DE scenario \cite{amendola_2000}.
In the Einstein frame, the interaction is described by the
Lagrangian
\be \label{L_phi} {\cal L} =
 -\frac{1}{2}\partial^\mu \phi \partial_\mu \phi - V(\phi) -
 m(\phi)\bar{\psi}\psi + {\cal L}_{\rm kin}[\psi] \,,
\ee
in which the mass of
matter fields $\psi$ is not a constant (as in the standard cosmological model),
but rather a function of the DE scalar field $\phi$.
A coupling between matter and DE can be
reformulated in terms of scalar-tensor theories or $f(R)$ mo\-dels
\citep{Wetterich:1994bg, pettorino_baccigalupi_2008, 2014arXiv1402.5031W} via a Weyl scaling from the Einstein frame (where matter is coupled and gravity is standard) to the Jordan frame (where the gravitational coupling to the Ricci scalar is modified and matter is uncoupled).
This is exactly true when the contribution of baryons is neglected. 

Dark matter (indicated with the subscript c) and DE densities are then not conserved separately, but
coupled to each other: 
 \bea
 \label{rho_conserv_eq_phi} \rho_{\phi}' &=& -3 {\cal{H}}
 \rho_{\phi} (1 + w_\phi) + {\beta} \rho_{\rm{c}} \phi'  \,, \\
 \label{rho_conserv_eq_c} \rho_{\rm{c}}' &=& -3 {\cal{H}} \rho_{\rm{c}}
 - {\beta} \rho_{\rm{c}} \phi'  \, .
 \nonumber
\eea
Here each component is treated as a fluid with stress energy tensor
 ${T^\nu}_{(\alpha)\mu} = (\rho_\alpha + p_\alpha) u_\mu u^\nu + p_\alpha
 \delta^\nu_\mu$, where $u_\mu = (-a, 0, 0, 0)$ is the fluid 4-velocity and
 $w_\alpha \equiv p_\alpha/\rho_\alpha$ is the equation of state.
Primes denote derivatives with respect to conformal time and $\beta$ is
assumed, for simplicity, to be a constant.
This choice corresponds to a Lagrangian in which dark matter fields have an
exponential mass dependence $m(\phi) = m_0 \exp^{-\beta {\phi}}$
(originally motivated by Weyl scaling scalar-tensor theories),
where $m_0$ is a constant. 
The DE scalar field (expressed in units of the reduced Planck mass
$M = (8 \pi G_N)^{-1/2}$) evolves according to the
Klein-Gordon equation, which now includes an extra term that depends on the
density of cold dark matter:
\be
 \label{kg} \phi'' + 2{\cal H} \phi' + a^2 \frac{dV}{d \phi} = a^2 \beta
 \rho_{\rm{c}} \, .
\ee
Following \cite{pettorino_baccigalupi_2008}, we choose an inverse power-law
potential defined as $ V = V_0 \phi^{-\alpha}$, with $\alpha$ and $V_0$ being
constants. The amplitude $V_0$ is fixed thanks to an iterative routine, as
implemented by \cite{amendola_etal_2012, pettorino_etal_2012}. 
To a first approximation $\alpha$ only affects late-time cosmology.
For numerical reasons, the iterative routine finds the initial value of the
scalar field, in the range $\alpha \ge 0.03$, which is close to the
$\Lambda$CDM value $\alpha = 0$ and extends the range of validity with
respect to past attempts; the equation of state $w$ is approximately related
to $\alpha$ via the expression \citep{amendola_etal_2012}: $w = -2/(\alpha + 2)$ so that a value of
$\alpha = 0.03$ corresponds approximately to $w(\alpha = 0.03) = -0.99$.
The equation of state $w \equiv p/\rho$ is not an independent parameter within
coupled DE theories, being degenerate with the flatness of the potential.
Dark matter particles then feel a fifth force with an effective gravitational
constant $G_{\rm eff}$ that is stronger than the Newtonian
one by a factor of $\beta^2$, i.e.
\begin{equation}
G_{\rm eff} = G(1+2 \beta^2) \,, 
\end{equation}
so that a value of $\beta = 0 $ recovers the standard gravitational
interaction. The coupling affects the dynamics of the gravitational potential
(and therefore the late ISW effect), hence the shape and amplitude of
perturbation growth, and shifts the position of the acoustic peaks to larger
multipoles, due to an increase in the distance to the last-scattering surface;
furthermore, it reduces the ratio of baryons to dark matter at early times
with respect to its present value, since coupled dark matter dilutes faster
than in an uncoupled model. The strength of the coupling is known to be
degenerate with  a combination of $\Omega_{\rm c}$, $n_{\rm s}$ and $H(z)$ \citep{amendola_quercellini_2003, pettorino_baccigalupi_2008, Bean:2008ac,
amendola_etal_2012}.
Several analyses have previously been carried out, with hints of coupling
different from zero, e.g., by \cite{pettorino_2013}, who
found $\beta = 0.036 \pm 0.016$  (using \Planck\ 2013 + WMAP polarization
+ BAO) different from zero at 2.2$\,\sigma$ (the significance increasing to
3.6$\,\sigma$ when data from HST were included).

\begin{figure}[htb!]
\begin{center}
\hspace*{-1cm}
\begin{tabular}{cc}
\includegraphics[width=0.42\textwidth]{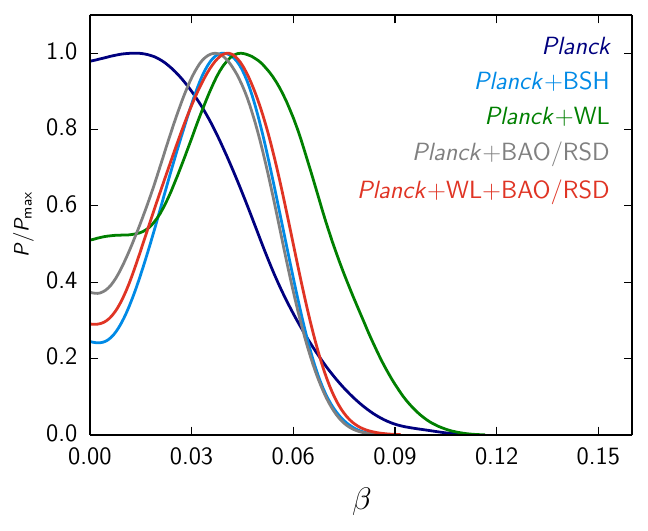}
 \end{tabular}
\caption{Marginalized posterior distribution for the coupling $\beta$ (see \sect{\ref{sec:cq}}). The
value corresponding to standard gravity is zero. Results and goodness of fit
are discussed in the text.}
\label{fig:cq1D}
\end{center}
\end{figure}

The marginalized posterior distribution, using \Planck\ 2015 data, for the coupling parameter
$\beta$ is shown in \fig{\ref{fig:cq1D}}, while the corresponding mean values
are shown in Table~\ref{tab:cq}. \planckTTonly\ data alone gives constraints
compatible with zero coupling and the slope of the potential is consistent
with a cosmological constant value of $\alpha = 0$ at 1.3$\,\sigma$. When other
data sets are added, however, both the value of the coupling and the slope of
the potential are pushed to non-zero values, i.e., further from \LCDM.
In particular, \Planck+BSH gives a value which is $\sim 2.5\,\sigma$ in
tension with \LCDM, while, separately, \Planck+WL+BAO/RSD gives a value of the
coupling $\beta$ compatible with the one from \Planck+BSH and about
2.3$\,\sigma$ away from \LCDM. 

\begin{figure*}[htb!]
\begin{center}
\hspace*{-1cm}
\begin{tabular}{cc}
\includegraphics[width=1\textwidth]{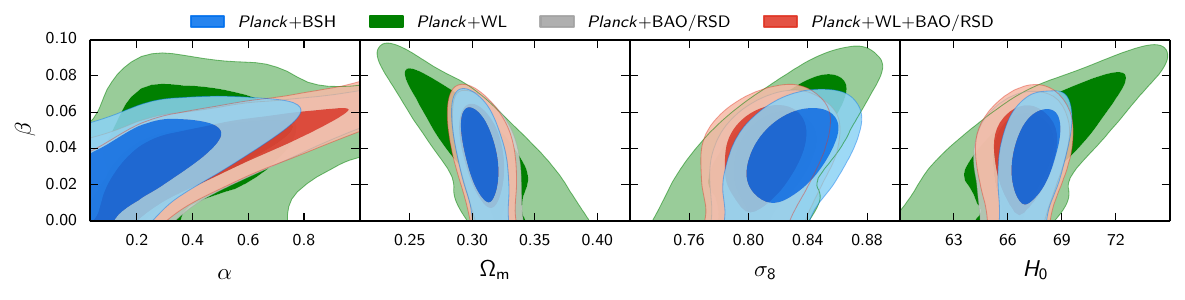}
 \end{tabular}
\caption{Marginalized posterior distribution for coupled DE and different
combinations of the data sets (see \sect{\ref{sec:cq}}). Here \Planck\ refers to \planckTT. We show the
degeneracy of the coupling $\beta$ with $\alpha$, $\Omega_{\rm m}$, $\sigma_8$ and $H_0$.}
\label{fig:cq2D_H0}
\end{center}
\end{figure*}

When comparing with \LCDM, however, the goodness of fit does not improve, despite the additional parameters.
Only the $\chi^2_{BAO/RSD}$ improves by
$\approx 1$ in CDE with respect to \LCDM, the difference not being significant
enough to justify the additional parameters. The fact that the marginalized
likelihood does not improve, despite the apparent 2$\,\sigma$ tension, may hint
at some dependence on priors: for example, the first panel in \Fig\ref{fig:cq2D_H0} shows that there is some degeneracy between the coupling $\beta$ and the potential slope $\alpha$; while contours are almost compatible with \LCDM\ in the 2 dimensional plot, the marginalization over $\alpha$ takes more contributions from higher values of $\beta$, due to the degeneracy, and seems to give a slight more significant peak in the one dimensional posterior distribution shown in \Fig \ref{fig:cq1D}. Whether other priors also contribute to the peak remains to be understood. In any case, the goodness of fit does not point towards a preference for non-zero coupling.
Degeneracy between the coupling and other cosmological parameters is shown in
the other panels of the same figure, with results compatible with those
discussed in \cite{amendola_etal_2012} and \cite{pettorino_2013}. Looking at the conservation equations
(i.e., Eqs.~\ref{rho_conserv_eq_phi} and \ref{rho_conserv_eq_c}), larger
positive values of $\beta$ correspond to a larger transfer of energy from
dark matter to DE (effectively adding more DE in the recent past, with roughly
$\Omega_\phi \propto \beta^2$ for an inverse power-law potential) and therefore
lead to a smaller $\Omega_{\rm m}$ today; as a consequence, the distance to
the last-scattering surface and the expansion rate are modified,
with $H'/H = -3/2(1+w_{\rm eff})$, where $w_{\rm eff}$ is the effective
equation of state given by the ratio of the total pressure over total
(weighted) energy density of the coupled fluid; a larger coupling prefers
larger $H_0$ and higher $\sigma_8$. 

The addition of polarization tightens the bounds on the coupling, increasing
the tension with \LCDM, reaching $2.8\,\sigma$ and $2.7\,\sigma$ for \Planck+BSH and
\Planck+WL+BAO/RSD, respecti\-vely. Also in this case the overall $\chi^2$ does not improve
between coupled DE and \LCDM. 

\begin{table*}
\tiny
\nointerlineskip
\setbox\tablebox=\vbox{
\newdimen\digitwidth % These five lines change what an asterisk
\setbox0=\hbox{\rm 0} % means to TeX. Instead of meaning
\digitwidth=\wd0 % "print an '*' here", it now means "leave
\catcode`*=\active % as much blank space as a single number
\def*{\kern\digitwidth} % takes up".
\newdimen\signwidth % These five lines change the meaning of an
\setbox0=\hbox{{\rm +}} % exclamation mark in the same way, so that it
\signwidth=\wd0 % leaves as much space as a plus or minus sign.
\catcode`!=\active % These definitions will disappear at the end of
\def!{\kern\signwidth} % the \vbox.
\newdimen\pointwidth % These five lines change the meaning of a
\setbox0=\hbox{\rm .} % question mark in the same way, so that it
\pointwidth=\wd0 % leaves as much space as a period.
\catcode`?=\active % These definitions will disappear at the end of
\def?{\kern\pointwidth} % the \vbox.
\halign{\hbox to 1.0in{#\leaderfil}\tabskip=0.em&
\hfil#\hfil\tabskip=0.7em&
\hfil#\hfil\tabskip=0.7em&
\hfil#\hfil\tabskip=0.7em&
\hfil#\hfil\tabskip=0.7em&
\hfil#\hfil\tabskip=0.0em\cr
\noalign{\doubleline}
\omit CDE models\hfil& \planckTT& \planckTT& \planckTT& \planckTT& \planckTT\cr
\omit & & +BSH&+WL& +BAO/RSD& +WL+BAO/RSD\cr
\noalign{\vskip 3pt\hrule\vskip 4pt} 
$\beta$& $<0.066$ ($95\,\%$)& $0.037^{+0.018}_{-0.015}$& $0.043^{+0.026}_{-0.022}$& $0.034^{+0.019}_{-0.016}$& $0.037^{+0.020}_{-0.016}$\cr
$\alpha$& $0.43^{+0.15}_{-0.33}$& $0.29^{+0.077}_{-0.26}$& $0.44^{+0.18}_{-0.29}$& $0.40^{+0.15}_{-0.29}$& $0.45^{+0.17}_{-0.33}$\cr
$H_{0}$ \rm{(km/s/Mpc)}& $65.4^{+3.2}_{-2.6}$& $67.47^{+0.88}_{-0.79}$& $67.6\pm 2.8$& $66.7\pm 1.1$& $66.9\pm 1.1$\cr
$\sigma_8  $ & $0.812^{+0.031}_{-0.026}  $& $0.829\pm 0.018$& $0.819^{+0.031}_{-0.026} $& $0.817\pm 0.017$& $0.810\pm 0.017$\cr
\noalign{\vskip 6pt\hrule\vskip 4pt} 
& \TTTEEE+\lowTEB& \TTTEEE+\lowTEB& \TTTEEE+\lowTEB& \TTTEEE+\lowTEB& \TTTEEE+\lowTEB\cr
& & +BSH& +WL& +BAO/RSD& +WL+BAO/RSD\cr
\noalign{\vskip 3pt \vskip 4pt} 
$\beta$& $< 0.062$ ($95\,\%$)& $0.036^{+0.016}_{-0.013}$& $0.036^{+0.019}_{-0.026}$& $0.034^{+0.018}_{-0.015}$& $0.038^{+0.018}_{-0.014}$\cr
$\alpha$& $0.42^{+0.14}_{-0.32}$& $<0.58$ ($95\,\%$)& $0.42^{+0.13}_{-0.33}$& $0.37^{+0.13}_{-0.28}$& $0.42^{+0.16}_{-0.31}$\cr
$H_{0}$ \rm{(km/s/Mpc)}& $65.4^{+2.8}_{-2.2}$& $67.39^{+0.85}_{-0.75}$& $66.5^{+2.8}_{-2.4}$& $66.7^{+1.1}_{-0.95}$& $66.8^{+1.2}_{-0.94}$\cr
$\sigma_8  $ & $0.814^{+0.030}_{-0.025}$& $0.832\pm 0.016$& $0.817^{+0.031}_{-0.025}$& $0.823\pm 0.015$& $0.818\pm 0.015$\cr
\noalign{\vskip 3pt\hrule\vskip 4pt} 
}}
\endtable
\caption{Marginalized mean values and $68\,\%$ C.L. intervals for coupled DE (see \sect{\ref{sec:cq}}).
 \Planck\ here refers to
\planckTT.
Results and goodness of fits are discussed in the text. CMB lensing
does not improve the constraints
significantly.}
    \label{tab:cq}
\end{table*} 

\section{Conclusions}\label{sec:conclusions}
The quest for Dark Energy and Modified Gravity is far from over.
A variety of different theoretical scenarios have been proposed in literature and need
to be carefully compared with the data. This effort is still in its early
stages, given the variety of theories and parameterizations that have been
suggested, together with a lack of well tested numerical codes that allow
us to make detailed predictions for the desired range of parameters.
In this paper, we have provided a systematic analysis covering a gene\-ral
survey of a variety of theoretical models, including the use of different
numerical codes and observational data sets.
Even though most of the weight in the \Planck\ data lies at high redshift,
\Planck\ can still provide tight constraints on DE and MG,
especially when used in combination with other probes.
Our focus has been on the scales where linear theory is applicable, 
since these are the most theoretically robust. 
Overall, the constraints that we find are consistent with the simplest
scenario, $\Lambda$CDM, with
constraints on DE models (including minimally-coupled scalar
field models or evolving equation of state models) and MG
models (including effective field theory, phenomenological parameterizations, $f(R)$ and 
coupled DE models) that are signifi\-cantly improved with respect to past analyses.
We discuss here our main results, drawing our conclusions for each of them and summarizing the story-line we have followed in this paper to discuss DE and MG.

Our journey started from distinguishing background and perturbation parameterizations. In the first case, the background is modified (which in turn affects the perturbations), leading to the following results.
\begin{enumerate}
\item The equation of state $w(z)$ as a function of redshift has been tested
for a variety of parameterizations.
\begin{enumerate} 
\item In ($w_0, w_{\rm{a}}$), \planckTT +BSH is compatible with \lcdm, as well as BAO/RSD. When adding WL to \planckTT, both WL and CMB prefer the ($w_0, w_{\rm{a}}$) model with respect to \LCDM\ at about $2 \sigma$, although with a preference for high values of $H_0$ (third panel of \Fig{\ref{fig:w1D}}) that are excluded when including BSH.
\item We have reconstructed the equation of state in redshift, testing a Taylor expansion up to the third order in the scale factor and by doing a Principal Component Analysis of $w(z)$ in different redshifts bins. In addition, we have tested an alternative parametrization, that allows to have a varying $w(z)$ that depends on one parameter only. All tests on time varying $w(z)$ are compatible with \LCDM\ for all data sets tested.
\end{enumerate}
\item 
`Background' Dark Energy models are generally of quintessence type where a scalar field
rolls down a potential. 
We have shown via the ($\epsilon_{\rm{s}}, \epsilon_{\rm{\infty}}$) parameterization, related respectively to late and early time evolution, that 
the quintessence/phantom potential at low redshift must be
relatively flat: $d \ln V/d \phi < 0.35/M_{\rm P}$ for quintessence;
and $d\ln V /d\phi < 0.68/M_{\rm P}$ for phantom models.
A zero slope ($\Lambda$CDM) remains consistent with the data and
compared to previous studies, the uncertainty has been reduced by about
10\,\%. We have produced a new plot (\Fig{\ref{fig:epsinf}}) that helps to visualize minimally coupled scalar field models, similarly to analogous plots often used in inflationary theories.
\item 
Information on DE, complementary to ($w_0, w_{\rm{a}}$), comes from
asking whether there can be any DE at early times. First, we have obtained constraints on early DE parametrizations, assuming a constant DE relative density at all epochs until it matches \LCDM\ in recent times: we have improved previous constraints by a factor $\sim 3-4$, leading to $\Omega_{\rm{e}} < 0.0071$ at $95\%$ C.L. from \planckTT+\lensing+BSH and $\Omega_{\rm{e}} < 0.0036$ for \planckall +BSH.
In addition, we have also asked how much such tight constraints are weakened when the fraction of early DE is only present in a limited range in redshift and presented a plot of $\Omega_{\rm{e}}(z)$ as a function of 
$z_{\rm{e}}$, the redshift starting from which a fraction $\Omega_{\rm{e}}$ is present. Also in this case constraints are very tight, with $\Omega_{\rm{e}} \la 2\,\% \,\,(95\,\%$ C.L.) even for $z_{\rm{e}}$ as late as $\approx 50$.
\end{enumerate}
The background is then forced to be very close to \LCDM, unless the tight constraints on early DE can somehow be evaded in a realistic model by counter balancing effects. 

In the second part of the paper, we then moved on to understanding what \planckonly\ can say when the evolution of the perturbations is modified independently of the background, as is the case
in most MG models. For that, we followed two complementary approaches: one (top-down) that starts from a very general EFT action for DE (\sect{\ref{EFT}}); the other (bottom-up) that starts from parameterizing directly observables (\sect{\ref{sec:genpar}}). In both cases we have assumed that the background is exactly \LCDM, in order to disentangle the effect of perturbations. We summarize here our results.
\begin{enumerate}
\item Starting from EFT theories for DE, which include (almost) all universally coupled models in MG via nine generic fun\-ctions of time, we have discussed how to restrict them to Horndeski theories, described in terms of five free functions of time. Using the publicly available code \EFTCAMB, we have then varied three of these functions (in the limits allowed by the code) which correspond to a non-minimally coupled K-essence model (i.e. $\alphaB$, $\alphaM$, $\alphaK$ are varying functions of the scale factor). We have found limits on the present value ${\alphaM}_0 < 0.052$ at $95 \%$ C.L. (in the linear EFT approximation), in agreement with \LCDM. Constraints depend on the stabi\-lity routines included in the code, which will need to be further tested in the future, together with allowing for a larger set of choices for the Horndeski functions, not available in the present version of the numerical code.
\item When starting from observables, two functions of time and scale are required to describe perturbations completely, in any model. Among the choices available (summarized in \sect{\ref{subsubsec:mgcamb}}), we choose $\mu(a, k)$ and $\eta(a,k)$ (other observables can be derived from them). In principle, constraints on these functions are dependent on the chosen parameterization, which needs to be fixed. We have tested two different time dependent parameterizations (DE-related and time-related) and both lead to similar results, although the first is slightly more in tension than the other with \LCDM\ (\Fig{\ref{fig:mueta}}).  
In this framework, \lcdm\ lies at the $2\sigma$ limit when \planckTT +BSH is considered, the tension increa\-sing to about $3 \sigma$ 
when adding WL and BAO/RSD to \planckTT.
As discussed in the text, the mild tension with \planckTT\ is related to lower power in the TT spectrum and a larger lensing potential in the MG model, with respect to \LCDM. 
The inclusion of CMB lensing shifts all contours back to \LCDM. We have reconstructed the two observables in redshift for both parameterizations, along the maximum degeneracy line (\Fig{\ref{fig:muetarecon}}). When scale dependence is also included, constraints become much weaker and the goodness of fit does not improve, indicating that the data do not seem to need the addition of additional scale dependent parameters.
\end{enumerate}
The last part of the paper discusses a selection of particular MG models of interest in literature. 
\begin{enumerate}
\item We first commented on the simple case of a minimally coupled scalar field in which not only the equation of state is allowed to vary but the sound speed of the DE fluid is not forced to be 1, as it would be in the case of quintessence. Such a scenario corresponds to k-essence type models. As expected, given that the equation of state is very close to the \LCDM\ value, the total impact of DE perturbations on the clustering is small.
\item We adopt an alternative way to parameterize observables (the equation of state approach) in terms of gauge invariant quantities $\Gamma$ and $\sigma$. We have used this approach to investigate Lorentz-breaking massive gravity and generalised scalar fields models, updating previous bounds.
\item As a concrete example of universally coupled theories, we have considered $f(R)$ models, written in terms of $B(z)$, conventionally related to the first and second derivatives of $f(R)$ with respect to R. Results are compatible with \LCDM. Such theories assume that some screening mechanism is in place, in order to satisfy current bounds on baryonic physics at solar system scales.
\item Alternatively to screening mechanisms, one can assume that the coupling is not universal, such that baryons are still feeling standard gravitational attraction. As an example of this scenario, we have considered the case in which the dark matter evolution is coupled to the DE scalar field, feeling an effective fifth force stronger than gravity by a factor $\beta^2$. Constraints on coupled dark energy show a tension with \LCDM\ at the level of about 2.5 $\sigma$, slightly increasing when including polarization. The apparent tension, however, seems to hint at a dependence on priors, partly related to the degeneracy between the coupling and the slope of the background potential (and possibly others not identified here). Future studies will need to identify the source of tension and possibly disentangle background from perturbation effects.
\end{enumerate}

There are several ways in which the analysis can be extended. We have made an effort to (at least start) to put some order in the variety of theoretical frameworks discussed in literature. There are of course  scenarios not included in this picture that deserve future attention, such as additional cosmologies within the EFT (and Horndeski) framework, Galileons (see for example \cite{2014JCAP...08..059B}), other massive gravity models (see \cite{2014LRR....17....7D} for a recent review), general violations of Lorentz invariance as a way to modify GR \citep{Audren:2014hza}, non-local gravity (which, for some choices of the action, appears to fit \Planck\ 2013 data sets \citep{2014arXiv1411.7692D} as well as \LCDM, although there is no connection to a fundamental theory available at present); models of bigravity \citep{2012JHEP...02..126H} appear to be affected by instabilities in the gravitational wave sector \citep{2014arXiv1412.5979C} (see also \cite{2015PhLB..748...37A}) and are not considered in this paper.  In addition to extending the range of theories, which requires new numerical codes, future tests should verify whether all the assumptions (such as stability constraints, as pointed out in the text) in the currently available codes are justified. 
Further promising input may come from data sets such as WL and BAO/RSD, that allow to tighten considerably constraints on MG models in which perturbations are modified. With the data available at the time of this paper, there seems to be no significant trend (at more than 3 standard deviations) that compensates any possible tension in $\sigma_8$ or $H_0$ by favouring Modified Gravity; nevertheless, this issue will need to be further investigated in the future. 
 We also anticipate that these constraints will strengthen with future
releases of the \Planck\ data, including improved likelihoods for polarization and new likelihoods, not available at the time of this paper, such as
ISW, ISW-lensing and $B$-mode polarization, all of which can be used to further test MG scenarios.
\begin{acknowledgements}

It is a pleasure to thank
Luca Amendola,
Emilio Bellini,
Diego Blas,
Sarah Bridle,
Noemi Frusciante,
Catherine Heymans,
Alireza Hojjati,
Bin Hu,
Thomas Kitching,
Niall MacCrann,
Marco Raveri,
Ignacy Sawicki,
Alessandra Silvestri,
Fergus Simpson,
Christof Wetterich
and
Gongbo Zhao
for interesting discussions on theories, external data sets and numeri\-cal
codes. Part of the analysis for this paper
was run on the Andromeda and Perseus clusters of the University of Geneva and on WestGrid computers in Canada. 
We deeply thank Andreas Malaspinas for invaluable help with the Andromeda and Perseus Clusters.
This research was partly funded by the DFG TransRegio TRR33 grant on `The Dark Universe' and by the Swiss NSF.
The Planck Collaboration acknowledges the support of: ESA; CNES and CNRS/INSU-IN2P3-INP (France); ASI, CNR, and INAF (Italy); NASA and DoE (USA); STFC and UKSA (UK); CSIC, MINECO, JA, and RES (Spain); Tekes, AoF, and CSC (Finland); DLR and MPG (Germany); CSA (Canada); DTU Space (Denmark); SER/SSO (Switzerland); RCN (Norway); SFI (Ireland); FCT/MCTES (Portugal); ERC and PRACE (EU). A description of the Planck Collaboration and a list of its members, indicating which technical or scientific activities they have been involved in, can be found at \url{http://www.cosmos.esa.int/web/planck/planck-collaboration}.
\end{acknowledgements}

%Put these back into two lines as tex editor doesn't detect long line sets, arg.
\bibliography{Planck_bib,A16_Dark_energy_and_modified_gravity}{}
\bibliographystyle{aat}

\raggedright 
\end{document}